\documentclass[12pt,a4paper]{article}

\usepackage{lmodern}[lmr]

\usepackage[hidelinks]{hyperref}

\usepackage{amsfonts, amsmath, amssymb, amsthm, mathabx, color, epstopdf, enumitem, float, graphicx, lipsum, longtable, lscape, mathrsfs, mathtools, multirow, nccmath, nonfloat, pdflscape, rotating, tocloft}

\usepackage[margin=0.9in]{geometry}

\usepackage[round]{natbib}
\usepackage[doublespacing]{setspace}
\usepackage[table]{xcolor}

\usepackage{tikz}

\def\be{\begin{equation}}
\def\ee{\end{equation}}
\def\bea{\begin{eqnarray}}
\def\eea{\end{eqnarray}}

\author{}
\title{}

\DeclareMathOperator*{\argmin}{\arg\!\min}

\DeclareMathOperator*{\diag}{\normalfont\textrm{diag}}

\DeclareMathOperator*{\nn}{\mathcal{N}}
\DeclareMathOperator*{\win}{\normalfont\text{in}}
\DeclareMathOperator*{\wout}{\normalfont\text{out}}
\DeclareMathOperator*{\wmid}{\normalfont\text{mid}}

\begin{document}
\newcommand\blfootnote[1]{
\begingroup
\renewcommand\thefootnote{}\footnote{#1}
\addtocounter{footnote}{-1}
\endgroup
}

\newtheorem{corollary}{Corollary}
\newtheorem{definition}{Definition}
\newtheorem{lemma}{Lemma}
\newtheorem{proposition}{Proposition}
\newtheorem{remark}{Remark}
\newtheorem{theorem}{Theorem}
\newtheorem{example}{Example}
\newtheorem{assumption}{Assumption}

\numberwithin{corollary}{section}
\numberwithin{definition}{section}
\numberwithin{equation}{section}
\numberwithin{lemma}{section}
\numberwithin{proposition}{section}
\numberwithin{remark}{section}
\numberwithin{theorem}{section}

\allowdisplaybreaks[4]

\begin{titlepage}
\begin{center}
    {\large \bf Estimation and Inference for \\a Class of Generalized Hierarchical Models}
    \medskip

    { 
    {\sc Chaohua Dong$^{\ast}$, Jiti Gao$^{\dag}$, Bin Peng$^{\dag}$ and Yayi Yan$^\sharp$}
    \medskip

    $^{\ast}$Zhongnan University of Economics and Law\\ $^{\dag}$Monash University\\ $^\sharp$Shanghai University of Finance and Economics}

\bigskip

\today

\begin{abstract}

In this paper, we consider estimation and inference for the unknown parameters and function involved in a class of generalized hierarchical models. Such models are of great interest in the literature of neural networks (such as \citealp{BK2019}). We propose a rectified linear unit (ReLU)  based deep neural network (DNN) approach, and contribute to the design of DNN by i) providing more transparency for practical implementation, ii) defining different types of sparsity, iii) showing the differentiability, iv) pointing out the set of effective parameters, and v) offering a new variant of rectified linear activation function (ReLU), etc. Asymptotic properties are established accordingly, and a feasible procedure for the purpose of inference is also proposed. We conduct extensive numerical studies to examine the finite-sample performance of the estimation methods, and we also evaluate the empirical relevance and applicability of the proposed models and estimation methods to real data.

\bigskip
   

\medskip

\end{abstract}

\end{center}

\end{titlepage}

\section{Introduction}\label{Sec.1}

In recent decades, there has been a notable emphasis on deep neural networks (DNNs) in the literature of econometrics, (e.g., \citealp{cw1999,AC2003,SL2004,FLM2021}). Initially applied in machine learning, DNNs have since expanded into various fields, such as economics, finance, social sciences, among others. Related to the applications of DNN, \cite{LBH2015} offer a comprehensive overview of practical topics, while \cite{Athey2019} discusses its capacity in social science. Additionally, \cite{BMR2021} and \cite{FMZ2021}  offer reviews on recent methodological advancements. 

As the most important part of DNN, a variety of activation functions have been proposed and investigated theoretically and numerically (\citealp{DUBEY202292}). The rectified linear activation function (ReLU) sees its popularity due to its simplicity and partial linearity:
\begin{eqnarray*}
\sigma(x) =  x \vee 0 \quad\text{with}\quad x\in \mathbb{R}.
\end{eqnarray*}
ReLU is a piecewise linear function that will output the input directly if it is positive, otherwise, it will output zero. Compared to Sigmoid functions, ReLU has a low computational cost, which makes it efficient for large-scale neural networks practically. \cite{SH2020} and \cite{FG2022} for example establish some fundamental results with respect to using ReLU. 

However, there are still properties related to ReLU that remain unknown. To be more specific, we now define a simple DNN using ReLU for activation function, and then briefly review the relevant literature.

\begin{definition}[Simple DNN]\label{Def1}
For $\forall \mathbf{u},\mathbf{v} \in \mathbb{R}^n$, define the shifted activation function $\pmb{\sigma}_{\mathbf{v}}: \mathbb{R}^n \to \mathbb{R}^n$ as
\begin{eqnarray*}
\pmb{\sigma}_{\mathbf{v}} (\mathbf{u})=(\sigma (u_1 - v_1),\ldots, \sigma (u_n- v_n))^\top ,
\end{eqnarray*}
where $u_j$ and $v_j$ stand for the $j^{th}$ elements of $\mathbf{u}$ and $\mathbf{v}$ respectively. A simple DNN with $m$ hidden layers that realizes the mapping $\mathbb{U}\, (\subseteq \mathbb{R}^{c_1})\mapsto \mathbb{R}$ is defined as follows:

\begin{figure}[H]
\centering
\hspace*{-0.5cm}\begin{tikzpicture}

\tikzstyle{annot} = [text width=10em, text centered]

\node[rectangle, rounded corners, minimum size = 6mm, fill=orange!30] (Input-1) at (-0.15,-2) {$\mathbf{u} \in \mathbb{U}$};
 
\node[rectangle, rounded corners, minimum size = 6mm, fill=teal!10] (Hidden-1) at (2,-2) {$\mathbf{w}_1\,  \pmb{\sigma}_{\mathbf{v}_{1}}(\mathbf{u})$};

\node[rectangle, rounded corners, minimum size = 6mm, fill=teal!10] (Hidden-3) at (4,-2) {$\cdots$};

\node[rectangle, rounded corners, minimum size = 6mm, fill=teal!10] (Hidden-4) at (6,-2) {$\mathbf{w}_m\, \pmb{\sigma}_{\mathbf{v}_{m}}(\cdot)$};
 

\node[rectangle,  rounded corners, minimum size = 6mm, fill=purple!30] (Output-1) at (8.8,-2) {$\text{A scalar output}$};

\draw[->, shorten >=1pt] (Input-1) -- (Hidden-1);  
\draw[->, shorten >=1pt] (Hidden-1) -- (Hidden-3);   
\draw[->, shorten >=1pt] (Hidden-3) -- (Hidden-4);  
\draw[->, shorten >=1pt] (Hidden-4) -- (Output-1);  

\node[annot,above of=Input-1, node distance=1.1cm] (hl) {Input layer};
\node[annot,above of=Hidden-3, node distance=1.1cm] {Hidden layers};
\node[annot,above of=Output-1, node distance=1.1cm] {Output layer};
\end{tikzpicture}
\end{figure}
\noindent Mathematically, it is written as
\begin{eqnarray*}
\nn(\mathbf{u}\, |\, \mathbf{W}_{m})&:=&\mathbf{w}_m \, \pmb{\sigma}_{\mathbf{v}_m} \, \cdots \,\mathbf{w}_{1}\, \pmb{\sigma}_{\mathbf{v}_{1}}(\mathbf{u}),
\end{eqnarray*}
where $\mathbf{W}_{m}:= \{\mathbf{v}_1,\ldots, \mathbf{v}_m; \mathbf{w}_1,\ldots, \mathbf{w}_m \}$, and the weighting matrices and shift vectors have the following dimensions:
\begin{eqnarray*}
&&\mathbf{w}_j \text{ is }\left\{ 
\begin{array}{ll}
c_1\times c_1 & \text{for }j=1,\\
c_j\times c_{j-1} & \text{for } 2\le j \le m-1 , \\
1\times c_{m-1} & \text{for }j=m,
\end{array}
\right. \quad\text{and}\quad \mathbf{v}_j  \text{ is } \left\{ \begin{array}{ll}
c_1\times 1 & \text{for }j = 1,\\
c_{j-1}\times 1 & \text{for }j\ge 2.
\end{array}\right. 
\end{eqnarray*}
\end{definition}

The current literature agrees that ReLU is designed to offer sparsity, which leads to computational efficiency (e.g., \citealp{hoefler2021sparsity,glorot11a}). However, there have only been few efforts to explain how sparsity should be defined and why it occurs. To the best of our knowledge, there are already two types of sparsity involved: (1) non-active neurons and (2) parameters that are not effective. Additionally, the literature implicitly agrees that   $\mathbf{W}_{m}$ can be estimated through a minimization process. It is noteworthy that ReLU is piecewise linear, and it is not yet clear how to handle the accumulated (non-)differentiability through layers in both theory and practice. While the concern raised here actually exists in some well known software packages, to the best of our knowledge, no satisfactory treatment has been offered. We summarize a very brief survey about existing software packages in Appendix \ref{AppSurvey}.

Moving on to our discussion about modelling, when it comes to practical analysis using DNN based models and methods, the existing literature of model building primarily focuses on fully nonparametric models (e.g., \citealp{cw1999} and many extensions since then), with only a few mentions of semiparametric settings (e.g., \citealp{KK2017}, \citealp{BK2019}, and references therein).  It is not clear how to estimate and recover the parameters of interest for the  generalized hierarchical models mentioned in Definition 1 of \cite{BK2019}. There is a lack of investigations in this area of research, and it is also related to the (non-)differentiability of ReLU. As far as we know, these questions have not been thoroughly investigated. 

Meanwhile, the current literature heavily focuses on independent and identically distributed (i.i.d.) data (e.g., \citealp{BK2019, FLM2021}), while largely neglecting the implications of asymptotic properties when dealing with dependent data. This is especially significant for applications in the fields of finance and economics, such as those studied by \cite{Gu2020}, \cite{CLMZ2022},  and \cite{FALLAHGOUL2024105574}, where accounting for dependence can pose challenges in constructing inference. The literature on this topic dates at least back to \cite{NW1987}, with a comprehensive review provided by \cite{Shao2015}. In this paper, we aim to address this gap by training DNN with time series data, establishing asymptotic properties, and offering valid inference.

In what follows, in order to address the aforementioned concerns collectively, we follow the Definition 1 of \cite{BK2019} and consider a class of generalized hierarchical models of the form: for $1\leq t\leq T$,
\begin{eqnarray}\label{eq.md1}
y_t =  f_\star(\mathbf{z}_{1t}^\top \,\pmb{\theta}_{\star 1}, \ldots, \mathbf{z}_{rt}^\top\, \pmb{\theta}_{\star r}) + \varepsilon_t \equiv f_\star(x_{1t}, \cdots, x_{rt}) + \varepsilon_t, 
\end{eqnarray}
where $f_\star(\mathbf{x})$ is an unknown $r$-dimensional function of $\mathbf{x}=\left(x_1, \cdots, x_r\right)^{\top}$, $\pmb{\theta}_{\star j}$ is a vector of unknown  parameters of interest, each $\mathbf{z}_{jt}$ is a vector of $d_j\times 1$, for $d_j\ge 2$, observed time series, $\mathbf{z}_{jt}$ for $1\leq j\leq r$ may contain some common components, but $x_{jt}=\mathbf{z}_{jt}^\top\, \pmb{\theta}_{\star j}$ are all different for $1\leq j\leq r$, and $\varepsilon_t$ is an idiosyncratic error term.



When no misunderstanding arises, we write \eqref{eq.md1} as 

\begin{eqnarray}\label{eq.md2}
y_t  = f_\star(\mathbf{z}_t\, \pmb{\theta}_\star)+\varepsilon_t
\end{eqnarray} 
for notational simplicity, where $\mathbf{z}_t=\diag\{\mathbf{z}_{1t}^\top,\ldots,\mathbf{z}_{rt}^\top \}$ represents a block-wise diagonal matrix dataset of $r\times d$ dimensionality, and $\pmb{\theta}_\star=(\pmb{\theta}_{\star 1}^\top, \ldots, \pmb{\theta}_{\star r}^\top )^\top$ is a $d\times 1$ vector, in which $d=\sum_{j=1}^r d_j$. As discussed below Assumption \ref{As2}, meanwhile, a vector of lagged dependent variables of the form $\mathbf{Y}_{t-1} = \left(y_{t-1}, \cdots, y_{t-p}\right)^{\top}$ can be part of $\mathbf{z}_{t}$ when strict stationarity on $\mathbf{z}_{t}$ is imposed. 

Throughout the rest of this paper, we suppose that $d_j$'s and $r$ are finite, although both $r$ and $d_j$ may be very large. We also assign the script $\star$ to the true parameters and the true function.  For the purpose of identifiability and estimability of $\pmb{\theta}_\star$, we choose $\|\pmb{\theta}_{\star j} \|=1$ for all $j$'s, and assume that the first elements of $\pmb{\theta}_{\star j}$'s is positive. 

Note that model \eqref{eq.md2} is in the same spirit as Definition 1 of \cite{BK2019} and Eq (1.1) of \cite{Xia2002}. Note also that \cite{Xia2002} haven't imposed the same set of restrictions as ours due to a different purpose, and \cite{BK2019} haven't focused on such identifiability and estimability issues that we are interested in this paper.

To the best of our knowledge, there have been no attempts to infer $(\theta_{\star 1}, \cdots, \theta_{\star r})$ in the relevant DNN literature, although \cite{KK2017} and \cite{BK2019} draw to hierarchical models. The main goals of this paper are to estimate both $f_\star(\cdot)$ and $\pmb{\theta}_{\star j}$'s jointly by proposing a ReLU based DNN approach. Moreover, the proposed DNN based estimation method offers a unified way to deal with the case where the dimensionalities, $d_j$ for $1\leq j\leq r$, $r$ and $d$, can all be large, although being fixed (as reported in Table 2 of Section 5 below for the case of $r=8$ and $d=16$). By contrast, the existing nonparametric methods suffer from the so-called ``curse of dimensionality" issue when $r\geq 3$ and the sample size $T$ is not large enough. 

In summary, this paper adds the following main contributions to the relevant literature:

\begin{enumerate}
\item We enhance the design of DNN by i) providing more transparency for practical implementation, ii) defining different types of sparsity, iii) showing the differentiability, iv) pointing out the set of effective parameters, and v) offering a new variant of ReLU, etc. 

\item We investigate a class of generalized hierarchical models, and propose a ReLU based DNN approach for the unified estimation of both the parameters of interest and the unknown function.

\item We allow our models and methods to be applicable to dependent time series data and then establish a valid implementable procedure for the purpose of inference. 

\item We conduct extensive numerical results to validate the theoretical findings before we also demonstrate the empirical relevance and applicability of the proposed model and estimation method to real data.
\end{enumerate}

The remainder of this paper is structured as follows. Section \ref{Sec.2} presents the design of DNN, and establishes some basic results which can be applied to a wide class of nonparametric models. Section \ref{Sec.3} considers the estimation of model \eqref{eq.md1}, and derives the necessary asymptotic theory accordingly. In Section \ref{Sec.4}, we point out a few possible extensions. Section \ref{Sec.5} provides extensive numerical studies to examine the theoretical findings by both simulated and real datasets. We conclude in Section \ref{Sec.6} with a few remarks. Appendix \ref{SecAppA} presents some useful plots, notations, and preliminary lemmas. Due to the space limitation, we provide extra simulation results and plots in Appendix \ref{App.sim} and Appendix \ref{App.Plots} receptively, and present all the proofs of the main results in Appendix \ref{App.A3} of the online supplementary file.
\medskip

Before proceeding further, we introduce a few symbols which will be repeatedly used throughout this paper. 
\medskip

\textbf{Symbols \& basic operations} ---  For $\forall w\in \mathbb{R}$, we let $\lfloor w\rfloor$ and $\lceil w\rceil$ be the largest and smallest integers satisfying $\lfloor w\rfloor \le w$ and $\lceil w\rceil\ge w$ respectively. For $\forall n\in \mathbb{N}$, we let $\mathbf{I}_n$, $\mathbf{1}_n$, and $[n]$ be a $n\times n$ identify matrix, a $n\times 1$ vector of ones, and a set $\{1,2,\ldots,n\}$ respectively. For $\pmb{\alpha} \in \mathbb{N}_0^r$ with $\mathbb{N}_0 =0 \cup  \mathbb{N}$ and $\mathbf{x}\in  \mathbb{R}^r$, we let  
\begin{eqnarray*}
&& \pmb{\alpha}! =\prod_{i=1}^r \alpha_i !,\quad \mathbf{x}^{\pmb{\alpha}} =\prod_{i=1}^r x_i^{\alpha_i},\quad   \|\mathbf{x}\|_1 =\sum_{i=1}^r |x_i| ,\nonumber \\
&&\pmb{\ell}_{\mathbf{x}\, |\, {\pmb{\alpha}} } = ( x_1^{\alpha_1},\ldots, x_r^{\alpha_r}, \mathbf{1}_{q}^\top)^\top\quad\text{with}\quad q=\left\{\begin{array}{ll}
2^{\lceil \log_2 r \rceil} -r & \text{for }r\ge 2\\
1 & \text{for }r=1
\end{array} \right. .
\end{eqnarray*}

For a  matrix $\mathbf{A}$, we let $\|\mathbf{A} \|$ and $\| \mathbf{A}\|_2$ define its Frobenius norm and Spectral norm respectively. Throughout, we write
\begin{eqnarray*}
\mathbf{z}_t^\top \mathbf{1}_r :=\widetilde{\mathbf{z}}_t\quad \text{and}\quad I_{a,t} =I(\mathbf{z}_t\, \pmb{\theta}_\star\in [-a,a]^r)\quad \text{for}\quad t\in[T].
\end{eqnarray*}

\smallskip

\textbf{Function operations \& monomials} --- Let $f(\mathbf{x})$ be a sufficiently smooth function defined on $\mathbb{U}\subseteq \mathbb{R}^r$, and define  
\begin{eqnarray*}
&&\|f \|_{\infty}^{\mathbb{U}} =\sup_{\mathbf{x} \in  \mathbb{U}} |f(\mathbf{x})|, \quad f^{(\pmb{\alpha})}(\mathbf{x}) =\frac{\partial^{\|\pmb{\alpha}\|_1} f(\mathbf{x})}{\partial x_r^{\alpha_r} \cdots \partial x_1^{\alpha_1}} ,\nonumber \\
&& \mathbf{f}^{(1)}(\mathbf{x}) =\diag\Big \{\frac{\partial f (\mathbf{x})}{\partial x_1}\mathbf{I}_{d_1} ,\ldots, \frac{\partial f (\mathbf{x})}{\partial x_r}\mathbf{I}_{d_r} \Big\}.
\end{eqnarray*}
We define a space of monomials:
\begin{eqnarray*} 
\mathscr{P}_n =\left\{ \text{Linear span of $\mathbf{x}^{\pmb{\alpha}}$ with $0\le |\pmb{\alpha}|\le n$} \right\},
\end{eqnarray*}
of which the dimension is $\text{dim}\mathscr{P}_n =\binom{r+n}{r} :=r_n$ by direct calculation. Denote the basis of $\mathscr{P}_n$ by $\{\psi_1(\mathbf{x} ),\ldots, \psi_{r_n}(\mathbf{x})\}$, and let 
\begin{eqnarray*}
\pmb{\psi}_{r_n}(\mathbf{x}) = (\psi_1(\mathbf{x} ),\ldots, \psi_{r_n}(\mathbf{x}))^\top .
\end{eqnarray*}
For $\forall\mathbf{x}_0\in \mathbb{U}$, define the re-centred basis by $\{ \psi_1(\mathbf{x}\, |\, \mathbf{x}_0),\ldots, \psi_{r_n}(\mathbf{x}\, |\, \mathbf{x}_0) \}$, and let
\begin{eqnarray*}
\pmb{\psi}_{r_n}(\mathbf{x}\, |\, \mathbf{x}_0) = (\psi_1(\mathbf{x}\, |\, \mathbf{x}_0),\ldots, \psi_{r_n}(\mathbf{x}\, |\, \mathbf{x}_0))^\top .
\end{eqnarray*}

Having these notation and symbols in hand, we are now ready to start our investigation.

\section{DNN via ReLU}\label{Sec.2}

In this section, we present the design of DNN, and establish some basic results, which can be applied to a wider class of non- and semi-parametric models. Specifically, Section \ref{Sec.21} provides some basic definitions, while Section \ref{Sec.22} presents the detailed design.

\subsection{Basic Definitions}\label{Sec.21}

Recall that we have defined a simple DNN (i.e., $\nn(\cdot\, |\, \mathbf{W}_m)$) in Definition \ref{Def1}. Building on it, we further define a pair-wise hierarchical DNN (referred to as HDNN hereafter), which plays an important role in what follows.

\begin{definition}[HDNN]\label{Def2}
Define a mapping $\pmb{\ell}_{x,y}:(x,y)\mapsto \mathbb{U}$, where $x$ and $y$ are scalars. For $\forall\mathbf{u} =(u_1,\ldots, u_{2^q})^\top$ with $q\in \mathbb{N}$,  the  HDNN (written as $\nn_{\pmb{\ell}}  (\mathbf{u}\, |\,   \mathbf{W}_m)$) is implemented as follows:

\begin{enumerate}
\item[] Step 1 -- Divide $\mathbf{u}$ into pairs, and calculate $\nn (\pmb{\ell}_{u_1,u_2}  \, |\,  \mathbf{W}_m),\ldots,\nn ( \pmb{\ell}_{u_{2^q-1},u_{2^q}} \, |\,  \mathbf{W}_m)$;

\item[] Step $n\ (2\le n \le q)$ -- Apply $\nn  (\pmb{\ell}_{x,y} \, |\,  \mathbf{W}_m) $ to each pair of the outcomes from Step $n-1$. 
\end{enumerate}
\end{definition}

To better see Definition \ref{Def2}, we plot Figure \ref{HDNN} for the purpose of visualization.
\smallskip

In Figure \ref{HDNN} below, it is obvious that after $q$ steps, there is only one scalar left, which is the output of HDNN. Thus, the total number of hidden layers is $mq$.  Although $\nn_{\pmb{\ell}} (\mathbf{u}\, |\,   \mathbf{W}_m)$ may appear complex,  its parameters are entirely determined by $\mathbf{W}_m$ and $\pmb{\ell}_{x,y}$. Consequently, the number of effective parameters is significantly less than it appears. Finally, the ordering of the elements in $\mathbf{u}$ does not matter.

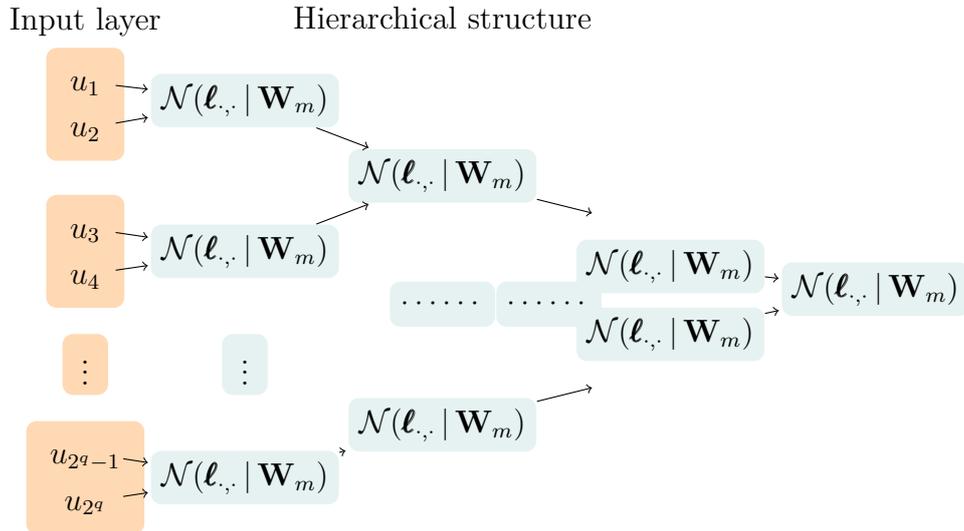
\begin{figure}[H]
\centering
\hspace*{-1cm}\begin{tikzpicture}

\tikzstyle{annot} = [text width=10em, text centered]

\node[rectangle, rounded corners, minimum size = 6mm, fill=orange!30] (Input-1) at (0.1,-2.55) {$\begin{array}{c}
u_1 \\ u_2
\end{array}$};

\node[rectangle, rounded corners, minimum size = 6mm, fill=orange!30] (Input-3) at (0.1,-4.5) {$\begin{array}{c}
u_3 \\ u_4
\end{array}$};
 
\node[rectangle, rounded corners, minimum size = 6mm, fill=orange!30] (Input-5) at (0.1,-6) {$\vdots$};

\node[rectangle, rounded corners, minimum size = 6mm, fill=orange!30] (Input-8) at (0.1,-7.5) {$\begin{array}{c}
u_{2^q-1} \\ u_{2^q}
\end{array}$};

\node[rectangle, rounded corners, minimum size = 6mm, fill=teal!10] (Hidden-11) at (2.2,-2.5) {$\nn (\pmb{\ell}_{\cdot,\cdot} \, |\, \mathbf{W}_m)$};
\node[rectangle, rounded corners, minimum size = 6mm, fill=teal!10] (Hidden-12) at (2.2,-4.5) {$\nn (\pmb{\ell}_{\cdot,\cdot} \, |\, \mathbf{W}_m)$};
\node[rectangle, rounded corners, minimum size = 6mm, fill=teal!10] (Hidden-13) at (2.2,-6) {$\vdots$};
\node[rectangle, rounded corners, minimum size = 6mm, fill=teal!10] (Hidden-15) at (2.2,-7.5) {$\nn (\pmb{\ell}_{\cdot,\cdot} \, |\, \mathbf{W}_m)$};

\node[rectangle, rounded corners, minimum size = 6mm, fill=teal!10] (Hidden-21) at (4.8,-3.5) {$\nn (\pmb{\ell}_{\cdot,\cdot} \, |\, \mathbf{W}_m)$};
\node[rectangle, rounded corners, minimum size = 6mm, fill=teal!10] (Hidden-22) at (6.2,-5.2) {$\cdots \cdots$};
\node[rectangle, rounded corners, minimum size = 6mm, fill=teal!10] (Hidden-24) at (4.8,-5.2) {$\cdots \cdots$};
\node[rectangle, rounded corners, minimum size = 6mm, fill=teal!10] (Hidden-23) at (4.8,-6.8) {$\nn (\pmb{\ell}_{\cdot,\cdot} \, |\, \mathbf{W}_m)$};

\node[rectangle, rounded corners, minimum size = 6mm, fill=teal!10] (Hidden-41) at (7.8,-4.7) {$\nn (\pmb{\ell}_{\cdot,\cdot} \, |\, \mathbf{W}_m)$};

\node[rectangle, rounded corners, minimum size = 6mm, fill=teal!10] (Hidden-42) at (7.8,-5.6) {$\nn (\pmb{\ell}_{\cdot,\cdot} \, |\, \mathbf{W}_m)$};

\node[rectangle, rounded corners, minimum size = 6mm, fill=teal!10] (Hidden-51) at (10.5,-5) {$\nn (\pmb{\ell}_{\cdot,\cdot} \, |\, \mathbf{W}_m)$};

\draw[->, shorten >=1pt] ++(0.5, -2.3) -- (Hidden-11);  
\draw[->, shorten >=1pt] ++(0.5, -2.8)  -- (Hidden-11);  

\draw[->, shorten >=1pt] ++(0.5, -4.25) -- (Hidden-12);  
\draw[->, shorten >=1pt] ++(0.5, -4.75)  -- (Hidden-12);  
 
\draw[->, shorten >=1pt] ++(0.6, -7.25) -- (Hidden-15);  
\draw[->, shorten >=1pt] ++(0.6, -7.75)  -- (Hidden-15);

\draw[->, shorten >=1pt] (Hidden-11) -- (Hidden-21);  
\draw[->, shorten >=1pt] (Hidden-12) -- (Hidden-21);  
 
\draw[->, shorten >=1pt] (Hidden-15) -- (Hidden-23);  

\draw[->, shorten >=1pt] (Hidden-21) -- ++(2, -0.5); 
\draw[->, shorten >=1pt] (Hidden-23) -- ++(2, 0.5); 

\draw[->, shorten >=1pt] (Hidden-41) -- (Hidden-51); 
\draw[->, shorten >=1pt] (Hidden-42) -- (Hidden-51); 

\node[annot,above of=Input-1, node distance=1.1cm] (hl) {Input layer};
\node[annot,above of=Hidden-21, node distance=2.1cm] {Hierarchical structure};
\end{tikzpicture}
\caption{HDNN of Definition \ref{Def2}}\label{HDNN}
\end{figure}

We then regulate the unknown function that is to be estimated. 

\begin{definition}[Smoothness]\label{Def3}
Let $p=\vartheta+s$ for some $\vartheta\in \mathbb{N}$ and $0<s\le 1$, and $\mathbb{U}\subseteq \mathbb{R}^r$. A function $f\,:\, \mathbb{U}\mapsto \mathbb{R}$ is called $(p, \mathscr{C})$-smooth, if for $\forall\boldsymbol{\alpha} \in \mathbb{N}_0^r$ with $\|\boldsymbol{\alpha}\|_1=\vartheta$ the partial derivative $f^{(\boldsymbol{\alpha})} (\mathbf{x})$ exists and satisfies that 
\begin{eqnarray*}
\|f^{(\boldsymbol{\alpha})} (\mathbf{x})-f^{(\boldsymbol{\alpha})} (\mathbf{z}) \|_{\infty}^{\mathbb{U}} \le \mathscr{C} \| \mathbf{x} - \mathbf{z}\|^s ,
\end{eqnarray*}
where $\mathscr{C}$ is a constant.
\end{definition}

Definition \ref{Def3} is adopted from \cite{BK2019}, and has different names in the literature (e.g., H{\"o}lder smoothness in \citealp{SH2020}).

\subsection{The Design}\label{Sec.22}

We present the design of neural network in this subsection, and establish some fundamental results regarding function approximation. First, we present the following lemma building on Definition \ref{Def1}.

\begin{lemma}\label{LM1}
For $\forall(x,y)\in [0,1]^2$, construct a DNN with $m+3$ hidden layers:
\begin{eqnarray*}
\nn (\pmb{\ell}^m_{x,y} \, |\, \mathbf{W}^{\star}_{m+3} )\quad\text{with}\quad \mathbf{W}^{\star}_{m+3} = \{\mathbf{v}_1^\star,\ldots, \mathbf{v}_{m+3}^\star; \mathbf{w}_1^\star,\ldots, \mathbf{w}_{m+3}^\star\},
\end{eqnarray*}
where
\begin{eqnarray*}
&&\pmb{\ell}_{x,y}^m=\frac{1}{2}\left( \frac{x-y+1}{2}, \ x-y+1, \ x+y+2^{-m} ,  \ \frac{x+y}{2}, \ x+y,\ \frac{1}{2} \right)^\top ,\nonumber \\
&& \mathbf{v}_k^\star =\left\{\begin{array}{ll}
\mathbf{1}_2\otimes (0,2^{1-2k},0)^\top & \text{for }k\in [m+1]\\
-1 & \text{for }k\in\{ m+2,m+3\}
\end{array}\right. ,\nonumber\\
&& \mathbf{w}_k^\star =\left\{\begin{array}{ll}  
\mathbf{I}_2\otimes \widetilde{\mathbf{w}}  & \text{for }k\in [m]\\
(-1,1,-1,1-1,1) &\text{for }k=m+1 \\
-1&\text{for }k=m+2\\
1 &\text{for }k=m+3
\end{array}\right.,\nonumber \\
&& \widetilde{\mathbf{w}} =   \begin{pmatrix}
\frac{1}{2} & -\frac{1}{2} & 0 \\
1 & -1 & 0 \\
1 & -1 & 1
\end{pmatrix}  .
\end{eqnarray*}
Here, $\nn (\pmb{\ell}^m_{x,y} \, |\, \mathbf{W}^{\star}_{m+3} )$ is piecewise linear in $x$ and $y$, and $\frac{\partial }{\partial x}  [\nn (\pmb{\ell}^m_{x,y} \, |\, \mathbf{W}^{\star}_{m+3} )]$ is defined accordingly. 

Let $\mathbb{C}_m=[0,1-2^{-m}]\times [0,1]$.  The following results hold:

\begin{enumerate}
\item $0\le \nn (\pmb{\ell}^m_{x,y} \, |\, \mathbf{W}^{\star}_{m+3} ) \le 1$ on $[0,1]^2$, and $\nn (\pmb{\ell}^m_{x,y} \, |\, \mathbf{W}^{\star}_{m+3} )=1$ at $(x,y)=(1,1)$,

\item $0\le \nn (\pmb{\ell}^m_{x,y} \, |\, \mathbf{W}^{\star}_{m+3} ) -xy \le 2^{-m}$ on $\mathbb{C}_m$,

\item $\left\|\frac{\partial }{\partial x}  [\nn (\pmb{\ell}^m_{x,y} \, |\, \mathbf{W}^{\star}_{m+3} )] -y \right\|_{\infty}^{\mathbb{C}_m} \le 2^{-m-1}$.
\end{enumerate}
\end{lemma}

In Lemma \ref{LM1}, $x$ and $y$ are symmetric in the sense that one can interchange $x$ and $y$ without violating the above results. In addition, Lemma \ref{LM1} actually offers a generic result. For example, one can replace $y$ with a generic function, say $f(y)$, and the result still holds with obvious modification. Thus, $\nn (\pmb{\ell}^m_{x,y} \, |\, \mathbf{W}^{\star}_{m+3} )$ offers a way to approximate different monomials practically, of which the space is the key to carry on nonparametric regression.  To see this numerically, we plot Figure \ref{FigNW3} in Section \ref{Sec.51}. Finally, it is worth mentioning that provided $\mathbb{C}_m$, $\nn (\pmb{\ell}^m_{x,y} \, |\, \mathbf{W}^{\star}_{m+3} )$ always approximates $xy$ from positive side.

Based on Lemma \ref{LM1}, we are then able to further estimate different monomials. Specifically, we provide the following lemma.

\begin{lemma}\label{LM.A7}
Using $\nn (\pmb{\ell}^m_{x,y} \, |\, \mathbf{W}^{\star}_{m+3})$ of Lemma \ref{LM1}, define $\mathcal{N}_{\pmb{\ell}^m}  (\pmb{\ell}_{\mathbf{x}\,|\, {\pmb{\alpha}}} \, |\, \mathbf{W}^{\star}_{m+3})$ according to Definition \ref{Def2}, where $r\ge 2$, $\mathbf{x}\in [0,h]^r$, $h\le 1 -\lceil\log_2 r\rceil \cdot 2^{-m}$, and $\pmb{\alpha} \in \mathbb{N}_0^r$. Then the following results hold:

\begin{enumerate}
\item $0\le \mathcal{N}_{\pmb{\ell}^m}  (\pmb{\ell}_{\mathbf{x}\,|\, {\pmb{\alpha}}} \, |\, \mathbf{W}^{\star}_{m+3})\le 1$ uniformly on $\mathbf{x}\in [0,h]^r$,

\item $0\le \mathcal{N}_{\pmb{\ell}^m}  (\pmb{\ell}_{\mathbf{x}\,|\, {\pmb{\alpha}}} \, |\, \mathbf{W}^{\star}_{m+3}) -\mathbf{x}^{\pmb{\alpha}} \le 3^{\lceil \log_2 r\rceil-1} 2^{-m} $ uniformly on $\mathbf{x}\in [0,h]^r$,

\item $\| \frac{\partial}{\partial x_i}[\mathcal{N}_{\pmb{\ell}^m}  (\pmb{\ell}_{\mathbf{x}\,|\, {\pmb{\alpha}}} \, |\, \mathbf{W}^{\star}_{m+3}) -\mathbf{x}^{\pmb{\alpha}}]\|_\infty^{[0,h]^r} \le 3^{\lceil \log_2 r\rceil-1} \cdot \|\pmb{\alpha}\|_1\cdot 2^{-m} $ for $\forall i\in [r]$.
\end{enumerate}
\end{lemma}

Lemma \ref{LM.A7} offers a specific range  (i.e., $[0,h]^r$) in which DNN can approximate monomials reasonably well. It is interesting to note that the range  includes non-negative quantities only, and $\mathcal{N}_{\pmb{\ell}^m}  (\pmb{\ell}_{\mathbf{x}\,|\, {\pmb{\alpha}}} \, |\, \mathbf{W}^{\star}_{m+3}) $ converges to $\mathbf{x}^{\pmb{\alpha}} $ from the positive side, which are consistent with Lemma \ref{LM1}. The construction of $\mathcal{N}_{\pmb{\ell}^m}  (\pmb{\ell}_{\mathbf{x}\,|\, {\pmb{\alpha}}} \, |\, \mathbf{W}^{\star}_{m+3})$ and Lemma \ref{LM.A7}.3 together offer a theoretical justification for \texttt{Adam} and \texttt{Keras} algorithms in which differentiability is required. In Section \ref{Sec.51}, we plot Figure \ref{FigHNN1} and Figure \ref{FigHNN2} for the purpose of demonstration.

\smallskip

We are now ready to consider the estimation of $f_\star (\mathbf{x})$. To proceed, we impose the following assumption to facilitate the development.

\begin{assumption}\label{As1}
Let $f_\star(\mathbf{x})$ of \eqref{eq.md1} be $(p, \mathscr{C})$-smooth on $\mathbf{x} \in [-a,a]^r$, where $a> 0$ and $r\geq 1$ are fixed.
\end{assumption}

As explained under Definition \ref{Def3}, Assumption \ref{As1} is commonly adopted in the literature. To proceed, we recall the notation and symbols defined at the end of Section \ref{Sec.1}, and present the following lemma that approximates $f_\star(\mathbf{x})$.

\begin{lemma}\label{LM2}
 Under Assumption 1,  for $\forall \mathbf{x}_0\in [-a,a)^r$, there exits $\pmb{\beta}_\star$ such that 
\begin{eqnarray*}
\| f_\star (\mathbf{x}) - \mathbf{N} (\mathbf{x} \, | \, \mathbf{x}_0)^\top \pmb{\beta}_{\star}  \|_{\infty}^{C_{\mathbf{x}_0,h}}=O(h^p +2^{-m}),
\end{eqnarray*}
where $ C_{\mathbf{x}_0,h} =\{\mathbf{x} \, |\, \mathbf{x} -\mathbf{x}_0\in [0,h]^r\}$ with $h\to 0$, $\pmb{\alpha}_{j}$'s are the power vectors of $\mathscr{P}_{r_\vartheta}$, and
\begin{eqnarray*}
 \mathbf{N} (\mathbf{x} \, | \, \mathbf{x}_0 ) &=& \begin{pmatrix}
\mathcal{N}_{\pmb{\ell}^m}(\pmb{\ell}_{\mathbf{x}-\mathbf{x}_0\, |\, \pmb{\alpha}_1 }\, |\, \mathbf{W}^\star_{m+3}) \\
\vdots \\
\mathcal{N}_{\pmb{\ell}^m}(\pmb{\ell}_{\mathbf{x}-\mathbf{x}_0\, |\, \pmb{\alpha}_{r_\vartheta}}\, |\, \mathbf{W}^\star_{m+3})
\end{pmatrix} .
\end{eqnarray*}
\end{lemma}

It should be understood that in Lemma \ref{LM2}, $\pmb{\beta}_\star$ varies with respect to $\mathbf{x}_0$. Although $h\to 0$ is not required in Lemma \ref{LM.A7}, it is essential to include the condition here from the perspective of function approximation. Building on Lemma \ref{LM2}, for $\forall M\in \mathbb{N}$, we can subdivide $[-a, a]^r$ into $M^r$ cubes of side length $h = \frac{2a}{M}$. For comprehensibility, we label these cubes by $C_{\mathbf{x}_{\mathbf{i}}}$ with $\mathbf{i}\in [M]^r$, where $\mathbf{x}_{\mathbf{i}}$ represents the point at the bottom left corner of each cube. Mathematically, $C_{\mathbf{x}_{\mathbf{i}}}$ is expressed as 
\begin{eqnarray}\label{def.cxi}
C_{\mathbf{x}_{\mathbf{i}}} = \{\mathbf{x}\, |\,  \mathbf{x} -\mathbf{x}_{\mathbf{i}}\in [0,h]^r\}.
\end{eqnarray}

Using the partition, we further present the first theorem of this paper.

\begin{theorem}\label{TM1}
Let Assumption \ref{As1} hold and $h\to 0$. There exists $\mathbf{B}_\star=\{\pmb{\beta}_{\star\mathbf{i}}\, |\, \mathbf{i}\in [M]^r\}$ such that
\begin{eqnarray*}
\| f_\star (\mathbf{x})-\mathscr{N} (\mathbf{x} \, | \, \mathbf{B}_\star)\|_{\infty}^{[-a,a]^r}=O(h^{p} +2^{-m}),
\end{eqnarray*}
where $\mathscr{N} (\mathbf{x} \, | \, \mathbf{B}_\star) =\sum_{\mathbf{i}} I(\mathbf{x}\in C_{\mathbf{x}_{\mathbf{i}}})\cdot \mathbf{N}  (\mathbf{x} \, | \, \mathbf{x}_{\mathbf{i}})^\top \pmb{\beta}_{\star \mathbf{i}}$.
\end{theorem}

Theorem \ref{TM1} shows that we can recover $f_\star(\mathbf{x})$ on $[-a,a]^r$ via HDNN. Having Theorem 1 in hand, we are ready to work on the estimation of \eqref{eq.md1} using data. In both theory and practice, the partial derivatives of an unknown function are also of great interest, as they allow us to further calculate marginal effects of some important variables. Along this line, we present some useful discussion and Corollary \ref{cor.nnvec} in Section \ref{Sec.4} later.

\section{Estimation and an Asymptotic Theory}\label{Sec.3}

In this section, we consider the estimation of \eqref{eq.md1}. To facilitate the development, we define a few more symbols. Let
\begin{eqnarray}\label{def.theta}
\pmb{\Theta} = \{ \pmb{\theta} = (\pmb{\theta}_1^\top,\ldots, \pmb{\theta}_r^\top)^\top\, | \, \pmb{\theta}_j\text{ is }d_j\times 1, \, \|\pmb{\theta}_j \|=1 , \, \theta_{j,1}>0, \,j\in [r] \},
\end{eqnarray}
where $\theta_{j,1}$ stands for the $1^{st}$ element of $\pmb{\theta}_j$. As in \eqref{eq.md2}, we write  $(\mathbf{z}_{1t}^\top \,\pmb{\theta}_1 ,\ldots, \mathbf{z}_{rt}^\top \, \pmb{\theta}_r)^\top = \mathbf{z}_t\, \pmb{\theta}$ when no misunderstanding arises.
 
Still, we partition $[-a,a]^r$ into $M^r$ cubes with side length $h =\frac{2a}{M}$, and work with $\{\mathbf{x}_{\mathbf{i}} \, | \, \mathbf{i}\in [M]^r\}$ as in \eqref{def.cxi}. To accommodate the structure of \eqref{eq.md1}, for $\forall\pmb{\theta}$ we map $\{\mathbf{x}_{\mathbf{i}} \, | \, \mathbf{i}\in [M]^r\}$ to $\{ \mathbf{z}_{\mathbf{i}}\, | \, \mathbf{z}_{\mathbf{i}}= \diag\{\mathbf{x}_{\mathbf{i}}\}\pmb{\theta}_{\diag}, \mathbf{i}\in [M]^r\}$, where $\pmb{\theta}_{\diag}=\diag\{\pmb{\theta}_1^\top, \ldots, \pmb{\theta}_r^\top\}$, and the dimension of $\mathbf{z}_{\mathbf{i}}$ is obviously consistent with $\mathbf{z}_t$. We then group $\mathbf{z}_t$'s using the following sets: 
\begin{eqnarray}\label{defcnew}
C_{\mathbf{i}} &=& \{\mathbf{z}\, |\, (\mathbf{z} - \mathbf{z}_{\mathbf{i}} )\pmb{\theta}_{\diag}^\top(\pmb{\theta}_{\diag}\,\pmb{\theta}_{\diag}^\top)^{-1}\mathbf{1}_r  \in[0,h]^r \}\nonumber \\
  &=& \{\mathbf{z}\, |\,  \mathbf{z}\,  \pmb{\theta}  -\mathbf{x}_{\mathbf{i}}  \in[0,h]^r \},
\end{eqnarray}
where the second equality follows from $\pmb{\theta}_{\diag}\,\pmb{\theta}_{\diag}^\top =\mathbf{I}_r$ and $\pmb{\theta}_{\diag}^\top \mathbf{1}_r =\pmb{\theta}$ by \eqref{def.theta}. Here, $C_{\mathbf{i}}$'s are equivalent to $C_{\mathbf{x}_{\mathbf{i}}}$'s of \eqref{def.cxi}.

With these symbols in hand, we conduct the following minimization procedure:
\begin{eqnarray}\label{def.est}
(\widehat{\pmb{\theta}}, \widehat{\mathbf{B}}) =\argmin Q_T (\pmb{\theta}, \mathbf{B}),
\end{eqnarray}
where $Q_T (\pmb{\theta}, \mathbf{B}) = \frac{1}{T}\sum_{t=1}^T [y_t - \mathscr{N} (\mathbf{z}_t\, \pmb{\theta}\, |\, \mathbf{B})  ]^2$, $\pmb{\theta}\in \pmb{\Theta}$, and $\mathbf{B} = \{\pmb{\beta}_{\mathbf{i}} \, |\, \| \pmb{\beta}_{\mathbf{i}}\|<\infty \}$. By Theorem \ref{TM1}, for $\forall \mathbf{x}_{0}\in [-a,a]^r$, $f_\star(\mathbf{x}_0)$ is naturally estimated by
\begin{eqnarray}\label{def.estmux}
\widehat{f}(\mathbf{x}_0) =\mathscr{N} (\mathbf{x}_0\, |\, \widehat{\mathbf{B}}).
\end{eqnarray}

To facilitate the theoretical development, we impose the following conditions.

\begin{assumption}\label{As2}
 
\item 
\begin{enumerate}

\item $\{(\mathbf{z}_t, \varepsilon_t )\, |\, t\in [T] \}$ are strictly stationary and $\alpha$-mixing with mixing coefficient
\begin{eqnarray*}
\alpha(k) = \sup_{A\in \mathcal{F}_{-\infty}^0, B\in  \mathcal{F}_k^\infty} |\Pr(A)\Pr(B) -\Pr(A\cap B)| \ \ \mbox{for $k\geq 1$}
\end{eqnarray*}
satisfying $\sum_{k= 1}^{\infty} \alpha(k)^{\nu/(2+\nu)} <\infty$ for some $\nu>0$, where $\mathcal{F}_{-\infty}^0$ and $\mathcal{F}_{k}^\infty$ are the $\sigma$-algebras generated by $\{(\mathbf{z}_{s},  \varepsilon_s )\, |\,  s \leq 0\}$ and $\{(\mathbf{z}_{s}, \varepsilon_s )\, |\,  s \geq t\}$, respectively. In addition, suppose that almost surely $E[\varepsilon_1\, |\,\mathbf{z}_1]=0$, $E[\varepsilon_1^2\, |\, \mathbf{z}_1]=\sigma_\varepsilon^2$, and $E[|\varepsilon_1|^{2+\nu}  \, |\, \mathbf{z}_1]<\infty$.

\item $E[f_\star(\mathbf{z}_1\,\pmb{\theta}_\star)-f_\star(\mathbf{z}_1\,\pmb{\theta})]^2$ is uniquely minimized on $\pmb{\Theta}$, and  $\inf_{ \pmb{\Theta}\times [-a, a]^r}\phi_{\pmb{\theta}} (\mathbf{w})\ge c_0>0$, where $\phi_{\pmb{\theta}} (\mathbf{w})$ defines the density function of $\mathbf{z}_1\,\pmb{\theta}$, and is Lipschitz continuous on $[-a, a]^r$.
\end{enumerate}
\end{assumption}

Assumption \ref{As2}.1 is standard (see, for example, Chapter 2 of \cite{FanYao}; Chapter 2 and Appendix A of \cite{gao2007}), and it requires strict stationarity. As alluded below model (\ref{eq.md2}), under the strict stationarity, model (\ref{eq.md2}) covers the dynamic case where a vector of lagged dependent variables of the form $\mathbf{Y}_{t-1} = \left(y_{t-1}, \cdots, y_{t-p}\right)^{\top}$ can be part of $\mathbf{z}_{t}$. In such dynamic cases, the relevant literature (see, \cite{Howell1990}; \cite{FanYao}; \cite{gao2007}; \cite{ttg2010}, for example) shows that $y_t$ of model (\ref{eq.md2}) can be strictly stationary when $f_{\star}(\cdot)$ satisfies $\lim_{\|\mathbf{x}\|\rightarrow \infty} \frac{\left|f_{\star}(\mathbf{x})\right|}{\|\mathbf{x}\|}<1$.

In Assumption \ref{As2}.2, the condition about $E[f_\star(\mathbf{z}_1\,\pmb{\theta}_\star)-f_\star(\mathbf{z}_1\,\pmb{\theta})]^2$ is necessary even in the case $f_\star(\cdot)$ is fully known. As $f_\star(\cdot)$ also needs to be estimated, we impose one more condition on $\phi_{\pmb{\theta}}$, which can be easily justified. For example, if $\mathbf{z}_1\,\pmb{\theta}$ follows a multivariate normal/$t$ distribution, the condition automatically holds. Notably, although we only infer $f_{\star}(\cdot)$ on $[-a,a]^r$, it does not mean that $\mathbf{z}_1\,\pmb{\theta}$ has to belong to a compact set. In Section \ref{Sec.6}, we discuss how to relax the condition on $a$ being finite.

Using Assumption \ref{As2}, we present the consistency in the following lemma.

\begin{lemma}\label{LM3}
Under Assumptions \ref{As1}-\ref{As2}, as $(h, Th^r)\to (0,\infty)$,

\begin{enumerate}
\item $ \frac{1}{M^r}\sum_{\mathbf{i}\in [M]^r}\|\mathbf{H}(\widehat{\pmb{\beta}}_{\mathbf{i}}-\pmb{\beta}_{\star\mathbf{i}} ) \|^2=o_P(1)$,

\item $\|\widehat{\pmb{\theta}}-\pmb{\theta}_\star\|=o_P(1)$,
\end{enumerate}
where $\mathbf{H}=\diag\{ h^{\|\pmb{\alpha}_1\|_1},\ldots, h^{\|\pmb{\alpha}_{r_\vartheta}\|_1} \}$ with $\pmb{\alpha}_j$'s being defined in Lemma \ref{LM2}.
\end{lemma}

After we have established Lemma \ref{LM3}, we recall the notation introduced in Section \ref{Sec.1} before we establish the following asymptotic distribution in the second theorem of this paper. 
\begin{theorem}\label{TM2}
Suppose that $\pmb{\Sigma}_{11} +\pmb{\Sigma}_{12}+\pmb{\Sigma}_{12}^\top$ and $\pmb{\Sigma}_{11}$ are positive definite, where  
\begin{eqnarray*} 
\pmb{\Sigma}_{11} &=& \sigma_\varepsilon^2 E [I_{a,1}\, \mathbf{f}^{(1)}_\star (\mathbf{z}_1\, \pmb{\theta}_\star) \, \widetilde{\mathbf{z}}_1 \widetilde{\mathbf{z}}_1^\top \mathbf{f}^{(1)}_\star(\mathbf{z}_1\, \pmb{\theta}_\star) ],\nonumber \\
\pmb{\Sigma}_{12}&=&\lim_{T} \sum_{t=1}^{T-1}(1-t/T ) E[\varepsilon_1 \varepsilon_{1+t} \,I_{a,1}I_{a,1+t}\, \mathbf{f}^{(1)}_\star (\mathbf{z}_1\, \pmb{\theta}_\star)\, \widetilde{\mathbf{z}}_1 \widetilde{\mathbf{z}}_{1+t}^\top \,\mathbf{f}^{(1)}_\star(\mathbf{z}_{1+t}\, \pmb{\theta}_\star) ],
\end{eqnarray*}
in which $\widetilde{\mathbf{z}}_t = \mathbf{z}_t^{\top} \mathbf{1}_r$.

Under Assumptions \ref{As1} and \ref{As2}, as $(h, Th^r)\to (0,\infty)$, for $\forall\mathbf{x}_0\in [-a, a]^r$
\be
\pmb{\Sigma}_{\mathbf{x}_{0}}^{-1/2}\mathbf{D}_T\left[\begin{pmatrix}
\widehat{\pmb{\theta}}-\pmb{\theta}_\star\\
\widehat{f}(\mathbf{x}_0) -f_\star(\mathbf{x}_0)
\end{pmatrix}+\mathbf{c}_{bias} +O_P( 2^{-m}) \right]\to_D N(\mathbf{0}_{d+1}, \mathbf{I}_{d+1}),
\label{jitibias}
\ee
where $\|\mathbf{c}_{\rm bias}\|=O_P(h^p)$ with its detailed form available from \eqref{defbias} of the supplementary appendix, $\mathbf{D}_T = \diag\{ \sqrt{T}\mathbf{I}_{d}, \sqrt{Th^r}\}$, and
\begin{eqnarray*}
\pmb{\Sigma}_{\mathbf{x}_{0}} &=&\diag\{\mathbf{I}_d , \pmb{\psi}_{r_\vartheta}(\mathbf{x}_0)^\top\mathbf{H}^{-1}\}\cdot\pmb{\Sigma}\cdot\diag\{\mathbf{I}_d , \mathbf{H}^{-1}\pmb{\psi}_{r_\vartheta}(\mathbf{x}_0)\} ,\nonumber \\
\pmb{\Sigma}&=&\diag\{ \pmb{\Sigma}_{11}^{-1}( \pmb{\Sigma}_{11} +\pmb{\Sigma}_{12}+\pmb{\Sigma}_{12}^\top)\pmb{\Sigma}_{11}^{-1}, \sigma_\varepsilon^2 \mathbf{I}_{r_{\vartheta}}\}.
\end{eqnarray*}
\end{theorem}

There are two bias terms involved in Theorem \ref{TM2}. The term $2^{-m}$ arises due to the use of ReLU, and it can be negligible as long as $m$ is sufficiently large. Also, $m$ is proportional to the number of layers, so it explains why increasing the layers of DNN can improve estimation accuracy exponentially. The term $\mathbf{c}_{\rm bias}$ comes from the nonparametric nature of DNN. It is pointed out that the bias terms remain with the estimate of $\pmb{\theta}_\star$ as well, because the approximation error term involved in the above Theorem 1 is not necessarily negligible asymptotically, unless a type of under-smoothing condition: $\sqrt{T} \left(h^p + 2^{-m}\right)=o(1)$, is imposed.

In addition to the involvement of two bias terms in equation (\ref{jitibias}), a long-run covariance matrix is involved in $\pmb{\Sigma}_{12}$. These terms make the asymptotic distribution in (\ref{jitibias}) infeasible in practice. Therefore, we propose a bootstrap procedure for the purpose of inference for $\pmb{\theta}_\star$.

\begin{enumerate}
\item For each bootstrap replication, we draw $\ell$-dependent time series $\{ \eta_t\, |\, t\in [T]\}$. where  $E[ \eta_t]=0$, $E[ \eta_t^2]=1$, $E[ \eta_t^4]<\infty$, $E[\eta_t\eta_s]=a\left(\frac{t-s}{\ell}\right)$, $(\frac{1}{\ell},\frac{\ell}{\sqrt{T}})\to (0,0)$, $a(\cdot)$ is a symmetric kernel defined on $[-1,1]$ satisfying that $a(0)=1$ and $K_a(x)=\int_{\mathbb{R}} a(u)e^{-iux}du\ge 0$ for $x\in \mathbb{R}$. 

\item Construct $y_t^*  = \mathscr{N} (\mathbf{z}_t\, \widehat{\pmb{\theta}} \, |\, \widehat{\mathbf{B}}) + \widehat{\varepsilon}_t \cdot \eta_{t}$, where $\widehat{\varepsilon}_t = y_t - \mathscr{N} (\mathbf{z}_t\, \widehat{\pmb{\theta}} \, |\, \widehat{\mathbf{B}})$. Conduct estimation using $\{y_t^*,\mathbf{z}_t \}$ as in \eqref{def.est} to obtain $\widehat{\pmb{\theta}}^*$.

\item Repeat Steps 1 and 2 $R$ times, where $R$ is sufficiently large.
\end{enumerate}

The condition $K_a(x)=\int_{\mathbb{R}} a(u)e^{-iux}du\ge 0$ for $x\in \mathbb{R}$ essentially regulates the kernel function, which together with other restrictions imposed on $a(\cdot)$ are satisfied by a few commonly used kernels, such as the Bartlett and Parzen kernels. More choices of the kernel function can be found in \cite{Andrews1991} and \cite{shao2010} for example. To get $\eta_{t}$'s, one may use $N(\mathbf{0}, \pmb{\Sigma}_{\eta})$ with $\pmb{\Sigma}_{\eta} =\{a\left(\frac{t-s}{\ell} \right) \}_{T\times T}$ for ease of implementation. 
 
\begin{corollary}\label{coroinf}
Let the conditions of Theorem \ref{TM2} hold. If, in addition, $\sqrt{T}(h^{p}+2^{-m})\to 0$ as $(T, m)\rightarrow (\infty, \infty)$, then we have  for $\forall \mathbf{x}_0\in [-a, a]^r$,
\begin{eqnarray*}
\sup_{\mathbf{w}}\left| \text{\normalfont Pr}\left\{ \sqrt{T}(\widehat{\pmb{\theta}}-\pmb{\theta}_\star)\le \mathbf{w} \right\} - \text{\normalfont Pr}^*\left\{ \sqrt{T}(\widehat{\pmb{\theta}}^*-\widehat{\pmb{\theta}})\le \mathbf{w}\right\} \right| =o_P(1),
\end{eqnarray*}
where $\text{\normalfont Pr}^*$ is the probability measure induced by the bootstrap procedure.
\end{corollary}

Because the biases come from diverse and intricate resources, we require a under-smoothing condition (i.e., $\sqrt{T}(h^{p}+2^{-m})\to 0$) in Corollary \ref{coroinf}. Up to this point, we have completed our theoretical development for model \eqref{eq.md1}.

\section{Extensions}\label{Sec.4}

Before we conduct simulation results and discuss real data analysis, we discuss about how to smooth  ReLU and also how to estimate marginal effects in this section.

\medskip

\textbf{Smoothed ReLU} --- In order to improve the finite sample performance of ReLU, many variants have been proposed, such as Leaky ReLU, parametric ReLU, Gaussian-error linear unit (GELU), etc. The rise of these variants is most likely due to the (non)--differentiability. See \cite{DUBEY202292} for a comprehensive review. Here, although it is not the main focus of the paper, we also offer a new variant, i.e., smoothed ReLU.

\medskip

We first introduce an assumption.

\begin{assumption}\label{As4}
Suppose that $\phi(\cdot)$ is nonnegative, $\int \phi(u)dv=1$, and $ \int |u| \phi(u)du<\infty$.
\end{assumption}

It is readily seen that conventional density functions satisfy this assumption. Then for positive integer $s$, we define 
\begin{equation}\label{b2}
\sigma_s(u)=\int \sigma(x)\phi_s(x-u)dx \quad\text{with}\quad \phi_s(u)=s\phi(su).
\end{equation}
The function sequence $\{\sigma_s(u)\}$ is called a regular sequence of $\sigma(u)$. For $\sigma_s(u)$, the following lemma holds.

\begin{lemma}\label{theorem44}
Let Assumption \ref{As4} hold.

\begin{enumerate}
\item $\sigma_s(u), \ s=1,2,\cdots,$ are smooth, and $$\|\sigma_s(u)-\sigma(u)\|_{\infty}^\mathbb{R}\le c \cdot s^{-1}$$ for some absolute constant $c$.

\item Furthermore, let $\phi(\cdot)$ be defined on $[-1,1]$, and be symmetric. Then we obtain that
\begin{eqnarray*}
\left\{\begin{array}{ll}
0\le \sigma_s(u)-\sigma(u) \le  O(1)s^{-1}, & \text{ for }|u|\le s^{-1},\\ \\
\sigma_s(u)-\sigma(u) =0, & \text{ for } |u|\ge s^{-1}.
\end{array} \right.
\end{eqnarray*}
\end{enumerate}
\end{lemma}

It is worth emphasizing that $\sigma_s(u)$'s of Lemma \ref{theorem44} are completely independent of the results in Sections \ref{Sec.2} and \ref{Sec.3}. Therefore, using triangle inequality and letting $s$ be sufficiently large. We can easily extend the results of Sections \ref{Sec.2} and \ref{Sec.3} via this lemma. Here, we basically trade smoothness with an additional approximation error due to the use of $\sigma_s(u)$'s. In Section \ref{Sec.51}, we further draw Figure \ref{FigSigm} to demonstrate this lemma.

\medskip

\textbf{Marginal Effects} --- Lemma \ref{LM1} shows the feasibility of derivatives, which may enable us to compute the marginal effects of different key variables. For example, we can have the following corollary immediately.  

\begin{corollary}\label{cor.nnvec}
Let $h\to 0$ and $p>r$. Under Assumption 1,  for $\forall \mathbf{x}_0\in [-a,a)^r$, there exits $\pmb{\beta}_\star$ such that for $\forall \pmb{\delta}=(\delta_{1},\ldots, \delta_{r})^\top$ with $\delta_{j}\in\{ 0,1\}$ and $j\in [r]$,
\begin{eqnarray*}
\| f_\star^{(\pmb{\delta})}(\mathbf{x}) - \mathbf{N}_{\pmb{\delta}} (\mathbf{x} \, | \, \mathbf{x}_0)^\top \pmb{\beta}_{\star}\|_{\infty}^{C_{\mathbf{x}_0,h}}=O(h^{p-\|\pmb{\delta}\|_1} +2^{-m}),
\end{eqnarray*}
where  
\begin{eqnarray*}
&&\mathbf{N}_{\pmb{\delta}}  (\mathbf{x} \, | \, \mathbf{x}_0 ) =\begin{pmatrix}
\pmb{\alpha}_1^{\pmb{\delta}}\, \mathcal{N}_{\pmb{\ell}^m}(\pmb{\ell}_{\mathbf{x}-\mathbf{x}_0,\, \overline{\pmb{\alpha}}_1 }\, |\, \mathbf{W}^\star_{m+3}) \\
\vdots \\
\pmb{\alpha}_{r_\vartheta}^{\pmb{\delta}}\, \mathcal{N}_{\pmb{\ell}^m}(\pmb{\ell}_{\mathbf{x}-\mathbf{x}_0, \, \overline{\pmb{\alpha}}_{r_\vartheta}}\, |\, \mathbf{W}^\star_{m+3})
\end{pmatrix} \quad \text{and}\quad\overline{\pmb{\alpha}}_{j} = \pmb{\sigma}_{\pmb{\delta}} (\pmb{\alpha}_j).
\end{eqnarray*}
\end{corollary}
 
In a similar fashion to the establishment of Theorem \ref{TM2}, we can estimate the marginal effects consistently.

\section{Finite-Sample Evaluation}\label{Sec.5}

In this section, we conduct extensive simulation studies to examine our theoretical findings in Section \ref{Sec.52} before we demonstrate the applicability and the finite-sample properties and advantages of the estimation method using a real dataset in Section \ref{Sec.53}.

\subsection{Simulation Results for Section \ref{Sec.3}}\label{Sec.52}

In this subsection, we conduct simulations to examine the theoretical findings for model \eqref{eq.md1}. Specifically, the data generating process is as follows. Let
\begin{eqnarray*}
y_t =  f_\star(\mathbf{x}_t) + \varepsilon_t = f_\star(\mathbf{z}_{1t}^\top \,\pmb{\theta}_{\star 1}, \ldots, \mathbf{z}_{rt}^\top\, \pmb{\theta}_{\star r}) + \varepsilon_t,
\end{eqnarray*}
where $\varepsilon_t = \rho_\varepsilon \varepsilon_{t-1}+v_t$ and $f_\star(\mathbf{x}) =\frac{2}{r}\left(\sum_{j\text{ is odd}} (5x_j+\sin(2x_j)) +\sum_{j\text{ is even}}\exp(2.5 x_j)\right)$ with $\mathbf{x} = \left(x_1, \cdots, x_r\right)^{\top}$. We consider the case of $r=2$ before we also have a look at the cases of $r=4,8$ in the second part of this simulation study. It is well understood that when $r\ge 4$, a typical nonparametric method such as kernel method will break down. When $r=8$, understandably, our NN method takes much longer time to compute than that used for the case of $r=2$.

That said, in what follows, we let
\begin{eqnarray*}
&& \pmb{\theta}_{\star j} =(0.6, 0.8)^\top \text{ with $j$ being odd},\quad \pmb{\theta}_{\star j} =(0.6, -0.8)^\top \text{ with $j$ being even},\nonumber \\
&& a=0.9,\quad \vartheta=2, \quad z_{it,j}\sim U(-1/1.4, 1/1.4),\quad \rho_\varepsilon=0.2,
\end{eqnarray*}
where $z_{it,j}$ stands for the $j^{th}$ element of $\mathbf{z}_{it}$. It is easy to know that as we set $a=0.9$, some observations will apparently be left out. Therefore, the design matches the framework of Sections \ref{Sec.2} and \ref{Sec.3}. 

For each generated dataset, we conduct estimation as in \eqref{def.est}. After $J$ replications, we report the following measures:
\begin{eqnarray*}
\text{RMSE}_{\pmb{\theta}} &=& \left\{\frac{1}{J}\sum_{j=1}^J \|\widehat{\pmb{\theta}}_j -\pmb{\theta}_\star\|^2 \right\}^{1/2}, \nonumber \\
\text{RMSE}_{f} &=& \left\{\frac{1}{JL}\sum_{j=1}^J\sum_{l=1}^L \|\widehat{f}_j(\mathbf{x}_l) -f_\star(\mathbf{x}_l) \|^2 \right\}^{1/2},\nonumber \\
\text{CR}_{\pmb{\theta}} &=&\frac{1}{Jd}\sum_{j=1}^J\sum_{s=1}^dI[(\widehat{\theta}_{j,s} - \theta_{\star,s})\in \text{CI}_{j,s}],
\end{eqnarray*}
where $\widehat{\pmb{\theta}}_j$ and $\widehat{f}_j(\cdot)$ stand for the estimates of $\pmb{\theta}_\star$ and $f_\star(\cdot)$ in the $j^{th}$ replication, and $\widehat{\theta}_{j,s} $ and $ \theta_{\star,s}$ are the $s^{th}$ elements of $\widehat{\pmb{\theta}}_j$ and $\pmb{\theta}_\star$. In addition, $\text{CI}_{j,s}$ is the 95\% confidence interval for the $s^{th}$ element in the $j^{th}$ replication based on the bootstrap draws, and $\{\mathbf{x}_l^{\ast}\, |\, l\in 0\cup [L]\}$ are some pre-determined points: $\mathbf{x}_l^{\ast}= (-a +  l\cdot 2a/L) \mathbf{1}_r.$ To implement the bootstrap procedure, we choose the Bartlett kernel (i.e., $y=(1-|x|)I(|x|\le 1)$), and follow the suggestion from such as \cite{PSU2011} to let $\ell =\lfloor 1.75T^{1/3} \rfloor$ for simplicity. We limit the number of bootstrap replications to 100 due to time and computation constraints. Also we let $T\in \{500, 1000, 2000\}$, $h\asymp T^{-1/4}$, and  $m\in\{3,4,5\}$, where $m$ controls the number of layers of in each hierarchy. Finally, to exam our argument under Lemma \ref{theorem44} we also repeat the above procedure using $\sigma_{s}(\cdot)$ by letting $s=32$, so $\sigma_{s}(\cdot)$ is sufficiently close to $\sigma(\cdot)$, and is smooth.

\begin{table}[H]
\caption{RMSE and Coverage Rate ($r=2$)}\label{TB.sim1}
\begin{tabular}{llrrrrrrr}
\hline\hline
 &  & \multicolumn{3}{c}{RMSE (via $\sigma$)} & \multicolumn{1}{l}{} & \multicolumn{3}{c}{RMSE (via $\sigma_{s}$)} \\
 & $T$ & $m=3$ & $m=4$ & $m=5$ &  & $m=3$ & $m=4$ & $m=5$ \\
$\text{RMSE}_{\pmb{\theta}}$ & 500 & 0.0212 & 0.0187 & 0.0177 &  & 0.0169 & 0.0142 & 0.0162 \\
 & 1000 & 0.0134 & 0.0137 & 0.0128 &  & 0.0101 & 0.0094 & 0.0078 \\
 & 2000 & 0.0097 & 0.0097 & 0.0092 &  & 0.0087 & 0.0076 & 0.0084 \\
$\text{RMSE}_{f}$ & 500 & 0.3929 & 0.4009 & 0.3969 &  & 0.4033 & 0.4396 & 0.3639 \\
 & 1000 & 0.2823 & 0.2765 & 0.2744 &  & 0.2636 & 0.2624 & 0.2681 \\
 & 2000 & 0.2003 & 0.2015 & 0.2000 &  & 0.1863 & 0.1839 & 0.1856 \\
 &  &  &  &  &  &  &  &  \\
 & \multicolumn{8}{c}{Coverage Rate} \\
$\text{CR}_{\pmb{\theta}}$ & 500 & 0.8875 & 0.9125 & 0.9150 &  & 0.8850 & 0.8750 & 0.9375 \\
 & 1000 & 0.9075 & 0.9225 & 0.9400 &  & 0.9075 & 0.9250 & 0.9550 \\
 & 2000 & 0.9275 & 0.9325 & 0.9375 &  & 0.9125 & 0.9600 & 0.9500 \\
  \hline\hline
\end{tabular}
\end{table}

As shown in Table \ref{TB.sim1}, it is clear that both $\text{RMSE}_{\pmb{\theta}}$ and $\text{RMSE}_{f}$ converge to 0, and $\text{RMSE}_{\pmb{\theta}}$ converges at a much faster rate. It is not surprising, as $\widehat{\pmb{\theta}}$ enjoys a parametric rate. Additionally, we see that as the sample size goes up, $\text{CR}_{\pmb{\theta}}$ converges to 95\% which is the nominal rate. Overall, $m=5$ offers better finite sample performance, which is also consistent with the finding of Section \ref{Sec.51}. Due to the complexity of DNN and the slow rate of $\text{RMSE}_{f}$, we need reasonably large sample size, which is also required by some of the existing simulation designs involving i.i.d. data (e.g., \citealp{DFLSP2021}). Last but not least, we see $\sigma_s(\cdot)$ and $\sigma(\cdot)$ offer equivalent finite sample performance. Although we see differences in the third digits, but in view of the number of replications of the Monte Carlo design, we should not emphasize these differences too much. Again, to understand the differences between $\sigma_s(\cdot)$ and $\sigma(\cdot)$, one should conduct a systematic comparison as in \cite{HG2023}.

As has been mentioned above, it is well understood that when $r\ge 4$, a typical nonparametric method such as kernel method will break down. Of course, nothing comes for free. When $r=4,8$, our NN method takes long time to compute due to large dimension involved in the calculation. Take $r=8$ for example, it takes a few hours to get one estimation done for $T\in \{1000, 2000\}$. As a result, it is impossible for us to calculate coverage rates using bootstrap method plus a large number of simulation replications. In practice, as one does not need to repeatedly implement the bootstrap draws, calculating confidence intervals is still feasible. That said, in what follows, we focus on bias and standard deviation, which are the focus of majority of DNN based studies anyway. Therefore, the following numerical results do not lose any generality. Additionally, we examine the case with $r=16$ in the online supplement Appendix \ref{App.sim}.

We now introduce some new measures as follows:
{\small
\begin{eqnarray}
\text{Bias}_{\pmb{\theta}}&=& \frac{1}{J}\sum_{j=1}^J\|\widehat{\pmb{\theta}}_j -\pmb{\theta}_\star \|_2/r,\ \ \text{Bias}_{f}=  \frac{1}{JL}\sum_{j=1}^J\sum_{l=1}^L |\widehat{f}_j(\mathbf{x}_l) - f_\star(\mathbf{x}_l)|,\nonumber\\
\text{Std}_{\pmb{\theta}}&=& \left\{\frac{1}{J}\sum_{j=1}^J\|\widehat{\pmb{\theta}}_j -\overline{\pmb{\theta}} \|_2/r\right\}^{1/2},\ \ \text{Std}_{f} = \left\{\frac{1}{JL}\sum_{j=1}^J\sum_{l=1}^L |\widehat{f}_j(\mathbf{x}_l) - \overline{f}_j(\mathbf{x}_l) |\right\}^{1/2},
\end{eqnarray}
where $\overline{\pmb{\theta}}= \frac{1}{J}\sum_{j=1}^J\widehat{\pmb{\theta}}_j$ and $ \overline{f}_j(\mathbf{x}_l)= \frac{1}{J}\sum_{j=1}^J\widehat{f}_j(\mathbf{x}_l)$}. Here $\mathbf{x}_l$'s are selected in the same way as in the main text. Note that ideally, we would like to select say 10 points on each dimension of $\mathbf{x}$ when evaluating the performance of $\widehat{f}$. However, it is not feasible practically, because $10^r$ with $r=8$ requires us to evaluate an extreme large number of points just for one simulation replication, which will definitely create lots of computation burden.

We summarize the results in Table \ref{extratable}. As we can see, the method presented in the paper still provides reasonable finite sample performance when $r=4, 8$. A few facts emerge: 1) It seems that $\text{Bias}_{\pmb{\theta}}$ has less bias via $\sigma(\cdot)$, and $\text{Std}_{\pmb{\theta}}$ is smaller via $\sigma_{32}(\cdot)$. Thus, introducing $\sigma_s$ seems to be balancing bias and standard deviation if we just focus on the estimate of $\pmb{\theta}_\star$. 2) The same pattern can be found for  the estimate of $f_\star$. In this case, the biases associated with $\sigma_{32}(\cdot)$ seem to be much larger, but they do not increase much when $r$ changes from 4 to 8. And 3) $m$ does not seem to play any serious role over all.

{\small
\begin{table}[h]
\caption{Extra Simulation Results}\label{extratable}
\begin{tabular}{lcrrrrrrrr}
\hline\hline
 & \multicolumn{1}{l}{} &  &  & \multicolumn{3}{c}{Bias} & \multicolumn{3}{c}{Std} \\
 & \multicolumn{1}{l}{} &  &  $T$ & $m=3$ & $m=4$ & $m=5$ & $m=3$ & $m=4$ & $m=5$ \\
$r=4$ & via $\sigma(\cdot)$ & $\pmb{\theta}$ & 500 & 0.0103 & 0.0080 & \multicolumn{1}{r|}{0.0062} & 0.0305 & 0.0309 & 0.0323 \\
 &  &  & 1000 & 0.0086 & 0.0072 & \multicolumn{1}{r|}{0.0099} & 0.0222 & 0.0211 & 0.0213 \\
 &  &  & 2000 & 0.0082 & 0.0085 & \multicolumn{1}{r|}{0.0114} & 0.0147 & 0.0150 & 0.0154 \\
 &  & $f$ & 500 & 0.2450 & 0.2646 & \multicolumn{1}{r|}{0.2694} & 0.3285 & 0.3289 & 0.3355 \\
 &  &  & 1000 & 0.2528 & 0.2616 & \multicolumn{1}{r|}{0.2571} & 0.2328 & 0.2281 & 0.2424 \\
 &  &  & 2000 & 0.2647 & 0.2675 & \multicolumn{1}{r|}{0.2630} & 0.1690 & 0.1598 & 0.1538 \\ \cline{5-10} 
 & via $\sigma_{32}(\cdot)$ & $\pmb{\theta}$ & 500 & 0.0042 & 0.0072 & \multicolumn{1}{r|}{0.0033} & 0.0392 & 0.0477 & 0.0495 \\
 &  &  & 1000 & 0.0044 & 0.0029 & \multicolumn{1}{r|}{0.0016} & 0.0385 & 0.0385 & 0.0326 \\
 &  &  & 2000 & 0.0016 & 0.0027 & \multicolumn{1}{r|}{0.0047} & 0.0296 & 0.0307 & 0.0322 \\
 &  & $f$ & 500 & 0.6926 & 0.7338 & \multicolumn{1}{r|}{0.5831} & 0.2821 & 0.2696 & 0.3010 \\
 &  &  & 1000 & 0.6816 & 0.7352 & \multicolumn{1}{r|}{0.5956} & 0.2151 & 0.1882 & 0.2048 \\
 &  &  & 2000 & 0.6821 & 0.7259 & \multicolumn{1}{r|}{0.5892} & 0.1409 & 0.1446 & 0.1352 \\ \cline{5-10} 
$r=8$ & via $\sigma(\cdot)$ & $\pmb{\theta}$ & 500 & 0.0114 & 0.0088 & \multicolumn{1}{r|}{0.0100} & 0.0630 & 0.0655 & 0.0625 \\
 &  &  & 1000 & 0.0108 & 0.0085 & \multicolumn{1}{r|}{0.0115} & 0.0436 & 0.0445 & 0.0429 \\
 &  &  & 2000 & 0.0090 & 0.0082 & \multicolumn{1}{r|}{0.0096} & 0.0313 & 0.0312 & 0.0299 \\
 &  & $f$ & 500 & 0.2976 & 0.2656 & \multicolumn{1}{r|}{0.2545} & 0.6052 & 0.6404 & 0.6536 \\
 &  &  & 1000 & 0.2539 & 0.2689 & \multicolumn{1}{r|}{0.2727} & 0.4445 & 0.4312 & 0.4159 \\
 &  &  & 2000 & 0.2580 & 0.2821 & \multicolumn{1}{r|}{0.2680} & 0.2811 & 0.3011 & 0.3073 \\ \cline{5-10} 
 & via $\sigma_{32}(\cdot)$ & $\pmb{\theta}$ & 500 & 0.0072 & 0.0060 & \multicolumn{1}{r|}{0.0070} & 0.0572 & 0.0536 & 0.0619 \\
 &  &  & 1000 & 0.0052 & 0.0037 & \multicolumn{1}{r|}{0.0039} & 0.0458 & 0.0402 & 0.0511 \\
 &  &  & 2000 & 0.0039 & 0.0026 & \multicolumn{1}{r|}{0.0022} & 0.0387 & 0.0348 & 0.0346 \\
 &  & $f$ & 500 & 0.7018 & 0.6016 & \multicolumn{1}{r|}{0.5444} & 0.4239 & 0.4467 & 0.4896 \\
 &  &  & 1000 & 0.7134 & 0.5710 & \multicolumn{1}{r|}{0.5389} & 0.2991 & 0.3290 & 0.3700 \\
 &  &  & 2000 & 0.7079 & 0.5698 & \multicolumn{1}{r|}{0.5268} & 0.2030 & 0.2105 & 0.2375 \\
 \hline\hline
\end{tabular}
\end{table}
}

\subsection{A Real Data Analysis}\label{Sec.53}

In this subsection, we examine the bond return predictability. Understanding the predictability of bond returns has always been a central topic in finance (e.g., \citealp{LN2009, AEMS2020,Betal2023}). The literature seems to agree that excess bond returns are forecastable by financial indicators such as yield spreads (\citealp{AEMS2020}), while also acknowledges the nonlinearity from modelling perspective (\citealp{Betal2023}). In addition, to avoid any variable selection problem, one often conducts principal component analysis (PCA) to collect some key factors from a large group of macro variables to be the key predictors of the forecasting model (\citealp{LN2009}). The above features naturally fit our example of Section \ref{Sec.4}.

We consider two benchmark models in this section:
\begin{eqnarray}
y_{t+1}&=& a + \varepsilon_{t+1},\label{emp.1} \\
y_{t+1}&=& a +\mathbf{x}_t^\top \pmb{\beta}+ \varepsilon_{t+1},\label{emp.2}
\end{eqnarray}
where $y_{t+1}$ denotes the 1-month log excess holding period return on 
a 2-month zero-coupon treasury bond. Here \eqref{emp.1} is a simple constant mean model, while \eqref{emp.2} is a typical one step ahead linear model including some key predictors. Specifically, $\mathbf{x}_t$ includes a set of variables adopted in \cite{Betal2023} and covers the period from December 1961 to December 2018: 1) yields spreads computed as the difference between the yield on a bond with 2 months to maturity and the implied yield on a one-month treasury bill obtained from CRSP;  2) forward spreads computed as the difference between the 2-month forward rate and the one-month yield; 3) the first principal component (PC) of yields obtained using 12, 24, 36, 48, and 60 month yields; 4) a linear combination of forward rates obtained from projecting 12, 24, 36, 48, and 60 month forward rates onto the mean excess bond return across the maturity spectrum; and 5) a linear combination of macroeconomic factors  obtained using the FRED-MD database  and estimated analogously to CP. As the first two predictors are observed directly, they form $\mathbf{z}_{1t}$. The last three predictors are obtained from estimation, so they form $\mathbf{z}_{2t}$.

Having the dataset ready, we first partition the data into a training set $S_{tr}$ (data from 1961 to 2011) and a test set $S_{te}$ (data from 2012 and 2018) (referred to as Partition I).  For each model, we run regression using training set only to obtain the parameter estimates, and calculate $\widehat{y}_{t}$ for the test set. For the DNN approach, we also let $m=3,4,5$ to examine the impact of  different layers practically. 

To evaluate the performance of different methods, we calculate
\begin{eqnarray}
\text{RMSE} =\left\{ \frac{1}{\sharp S_{te}}\sum_{t\in S_{te}} (\widehat{y}_{t}- y_{t})^2\right\}^{1/2}\quad\text{and}\quad
\text{CS} =  \frac{1}{\sharp S_{te}}\sum_{t\in S_{te}} I(\widehat{y}_{t}\cdot y_{t}>0 ),
\end{eqnarray}
where $\sharp S_{te}$ stands for the cardinality of $S_{te}$. The first measure is a typical root mean squared errors based on the test set, and the second measure examines the sign prediction and reports the percentage of correct sign prediction. There is a vast literature regarding sign prediction (e.g.,  \citealp{CD2006} and many extensions since then), which argues that return is not predictable in general, but one may better forecast the sign of return. Although predicting the sign is not the focus of the paper, it is interesting to see how DNN performs practically along this line of research. 

\begin{table}[h]
\caption{RMSE and CS}\label{Table.emp1}
\begin{tabular}{llccccc}
\hline\hline
 &  & \multirow{2}{*}{Model \eqref{emp.1}} & \multirow{2}{*}{Model \eqref{emp.2}} & \multicolumn{3}{c}{DNN} \\
 &  &  &  & $m=3$ & $m=4$ & $m=5$ \\
Partition I & RMSE & 0.2267 & 0.2209 & 0.2138 & 0.2047 & 0.2098 \\
 & CS & 0.5833 & 0.5833 & 0.5952 & 0.6429 & 0.6905 \\
Partition II & RMSE & 0.2274 & 0.2195 & 0.2025 & 0.2002 & 0.2095 \\
 & CS & 0.5833 & 0.5833 & 0.6905 & 0.5952 & 0.6429 \\
 \hline\hline
\end{tabular}
\end{table}

Alternatively, we consider a rolling window idea to partition the data (referred to as Partition II). The window size is always 50 years. For each window, we run regression using each model, and then use the estimated parameters to forecast the following year (i.e., the test set always includes data of one year). Again, we report RMSE and CS based on our forecasts. 

\begin{figure}[H] 
\hspace*{-1cm}\includegraphics[scale=0.18]{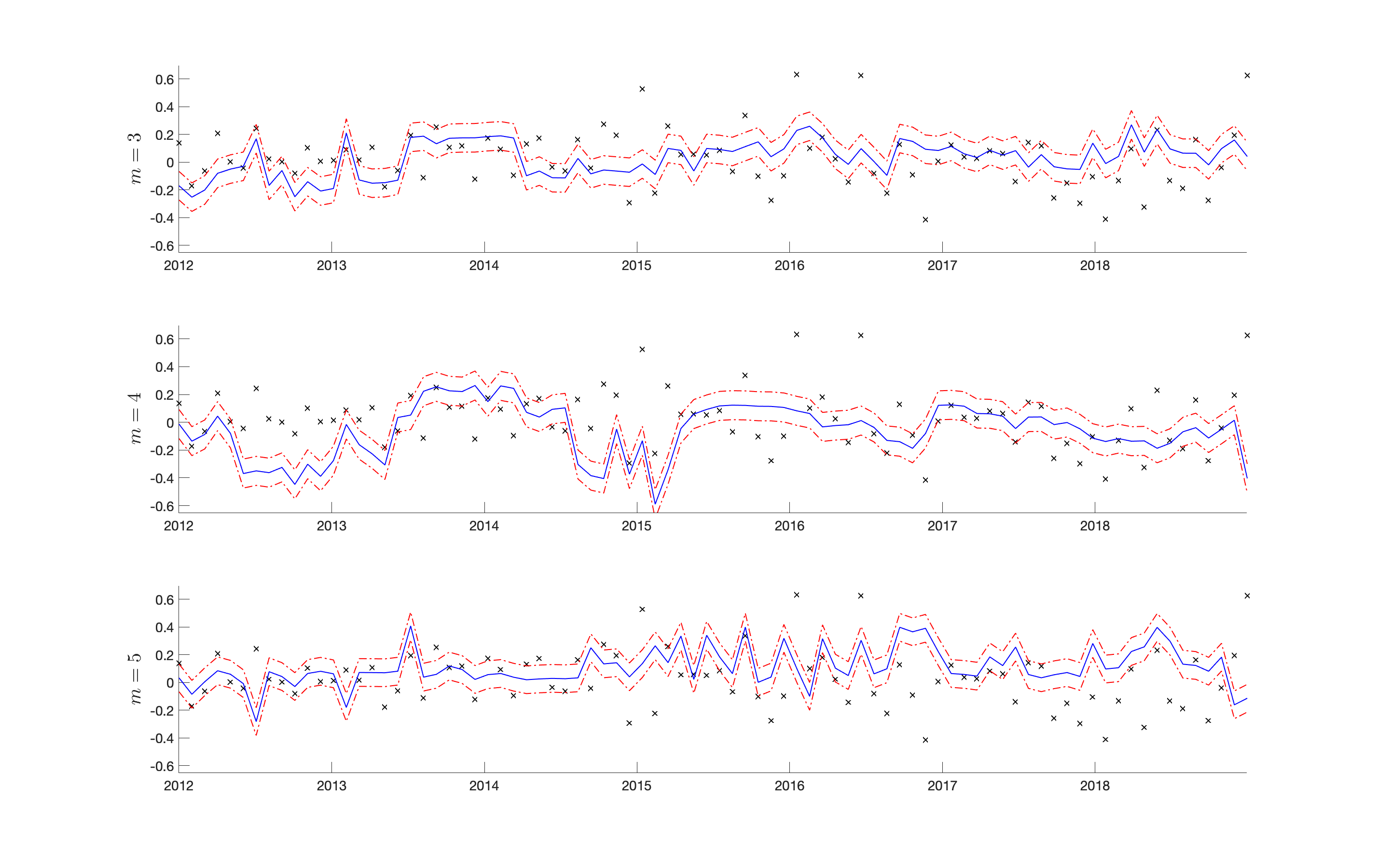}
\vspace*{-1cm}\caption{Examination using Test Set}\label{Fig.emp1}
\end{figure}

We summarize the results in Table \ref{Table.emp1}. As shown in the table, DNN approach in general outperforms Model \eqref{emp.1} and Model \eqref{emp.2} in terms of both RMSE and CS. Here, smaller RMSE implies less forecast errors, while higher CS means better  sign prediction. To further see the differences when using different layers (i.e., $m=3,4,5$) for DNN approach, we plot the point forecasts and their confidence intervals  using Partition I for example in Figure \ref{Fig.emp1}. In the figure, the actual $y_t$'s are marked using ``x", the estimated values are represented using blue solid lines, and the red dotted-hash lines stand for the 95\% confidence intervals. When $m=5$, DNN approach seems to offer a better coverage for the period from 2012 to 2015, while for the period from 2016 to 2018, it is quite hard to cover all the points regardless the value of $m$. This finding is consistent with \cite{Betal2023}, in which the authors argue that the predictability varies with respect to time and economic conditions. Certainly, one may conduct a more comprehensive investigation to be added to the relevant literature. We do not pursue these results further in order not to deviate from our main goal, but would like to leave this for future research.

\section{Conclusion with Discussion}\label{Sec.6}

Before concluding, we provide a few useful remarks.  Note that  we require $f_\star(\mathbf{x})$ to be defined on a compact set, but do not impose restriction on the range of $\{\mathbf{z}_t\}$. In fact, for time series data, it may make more sense to assume that $a$ is diverging, which is indeed achievable. Suppose that  $\mathbf{z}_t$ follows a sub-Gaussian distribution. In this case,  after some algebra we can relax the condition on $a$ to $\sqrt{\log (Th^r)} \cdot a\to \infty.$ Apparently, there is a price that we have to pay, which is the slow rate of convergence. A similar treatment has also been discussed in \cite{LTG2016} for example, so we do not further elaborate it here.

The total number of layers that we require for Theorem \ref{TM1} is 
\begin{eqnarray*}
\left\{\begin{array}{ll}
(m+3)\cdot \lceil \log_2 r \rceil  & \text{for }r\ge 2\\
m+3 & \text{for }r=1
\end{array} \right. .
\end{eqnarray*}
The width of most layers in the hierarchy is 6. The sparsity occurs naturally in view of Definition \ref{Def2} and Lemma \ref{LM1}.

\smallskip

\noindent \textbf{Connection with Some Existing Studies} --- Recently, \cite{DFLSP2021} and \cite{FLM2021} apply DNN to study treatment effects using micro datasets, and \cite{KN2020} apply DNN to study climate data. Our proposed model and method can be extendable to related topics, and the proposed estimation method offers a simple implementational procedure that complements existing studies. Specifically, we have a clear understanding of the nonlinear minimization process. 

\smallskip

To conclude,  we consider the estimation of both the parameters of interest and the unknown function involved in a class of hierarchical models in the literature of DNN. We contribute to the design of DNN by i) providing more transparency for practical implementation, ii) defining different types of sparsity, iii) showing the differentiability, iv) pointing out the set of effective parameters,  and v) offering a new variant of rectified linear activation function (ReLU), etc.   The asymptotic properties of the proposed estimates are derived accordingly, and they can be applied to a wider class of non- and semi-parametric models. Finally,  we conduct extensive numerical studies to examine the theoretical findings.

\section*{Acknowledgment}

The first author was supported by the National Natural Science Foundation of China (Grant 72073143) and the Fundamental Research Funds for the Central Universities at Zhongnan University of Economics and Law (2722022EG001). The second author was supported by Australian Research Council Discovery Project Program  under Grant Number  DP200102769. The third author was supported by Australian Research Council Discovery Project Program under Grant Number DP210100476. The fourth author was supported by the National Natural Science Foundation of China (Grant Number: 72303142) and Fundamental Research Funds for the Central Universities (Grant Numbers: 2022110877 \& 2023110099).

\appendix

\begin{appendix}

\renewcommand{\thelemma}{A\arabic{lemma}}
\setcounter{lemma}{0}

\renewcommand{\theequation}{A.\arabic{equation}}
\setcounter{equation}{0}

\renewcommand{\thesection}{A\arabic{section}}
\renewcommand{\thefigure}{A\arabic{figure}}
\renewcommand{\thetable}{A\arabic{table}}

\renewcommand{\theremark}{A\arabic{remark}}
\renewcommand{\thecorollary}{A\arabic{corollary}}

\setcounter{section}{0}
\setcounter{table}{0}
\setcounter{figure}{0}
\setcounter{remark}{0}

\section{Comments about Some Relevant Software Packages}\label{AppSurvey}

The following survey is also part of the motivation of this paper. The well known \texttt{neuralnet} in R does not even support the use of ReLU (\citealp{FG2010}). \texttt{PyTorch} does have ReLU and some of its variations included as the activation functions, but the explanation about the optimization process is very vague (\url{https://pytorch.org/docs/stable/optim.html}). \texttt{Keras} includes \texttt{Adam} algorithm and its variations (\url{https://keras.io/api/optimizers/adam/}), but \texttt{Adam} requires ``\textit{a stochastic scalar function that is differentiable w.r.t. parameters...}" (\citealp{KingBa15}), which does not apply to ReLU directly in an obvious manner. A comprehensive survey on the alternatives of ReLU is provided by \cite{DUBEY202292}, who comment on the pros and cons of different activation functions from the perspective of implementation. In this paper, we have settled some of these concerns in Lemmas \ref{LM1} and \ref{LM.A7} of Section \ref{Sec.2} about offering some generic results as well as in Section \ref{Sec.4} about smoothing ReLU.

\section{Extra Plots and Lemmas}\label{SecAppA}

\subsection{Some Useful Plots}\label{Sec.51}

We first provide justification for Lemma \ref{LM1}, Lemma \ref{LM.A7}, and Theorem \ref{theorem44} in Section \ref{Sec.51} below. Due to the space limitation, we concentrate on the results associated with $\sigma(\cdot)$ in this appendix, and provide extra plots to support our arguments about $\sigma_s(\cdot)$ in Appendix \ref{App.Plots} of the online supplementary document.

\begin{figure}[h] 
\hspace*{-2cm}\includegraphics[scale=0.25]{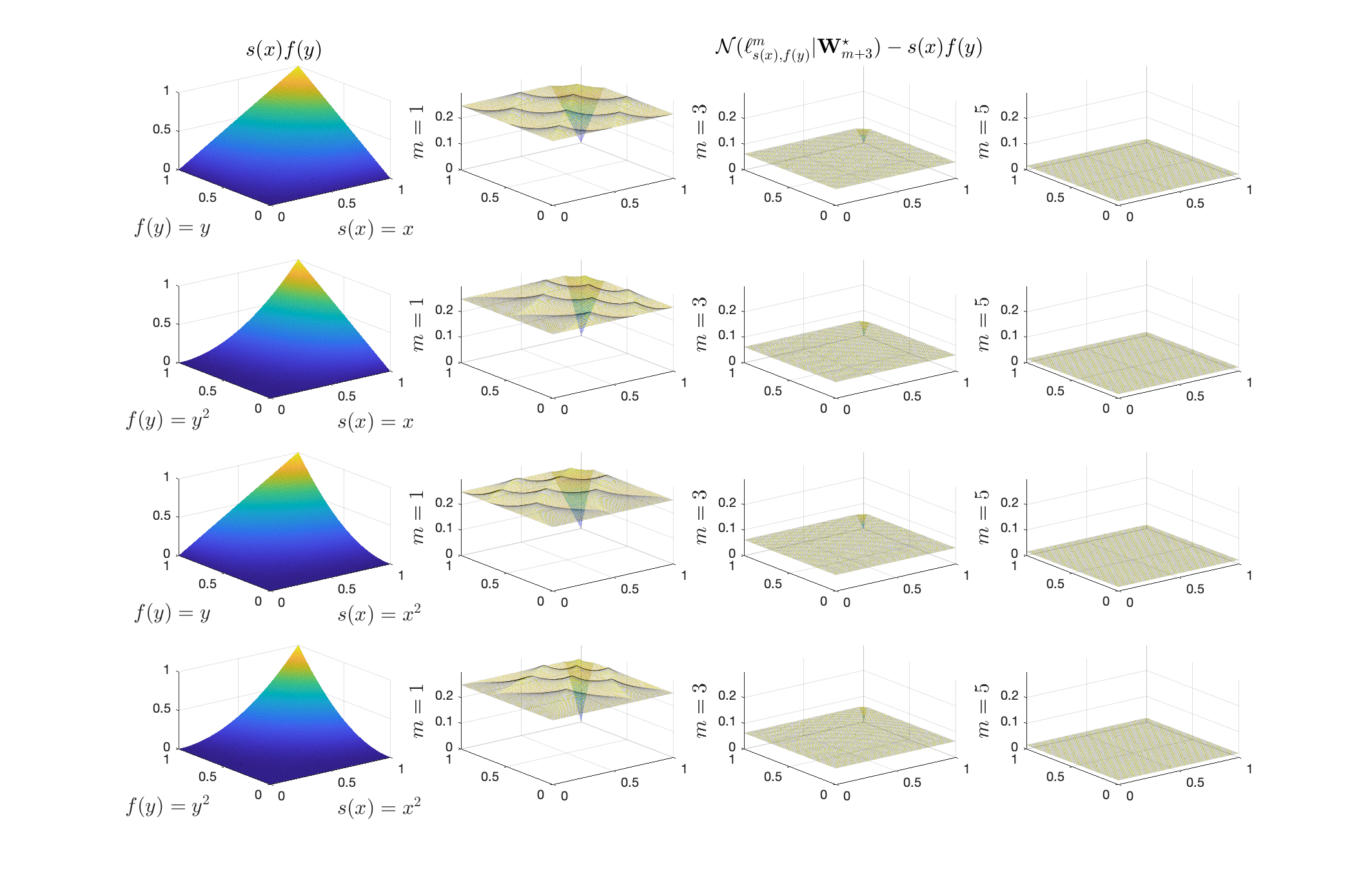}
\vspace*{-1cm}\caption{Illustration of Lemma \ref{LM1}}\label{FigNW3}
\end{figure}

We first show the validity of Lemma \ref{LM1} by approximating $s(x) f(y)$. Without loss of generality, we consider $s(x)\in \{ x,x^2\}$ and $f(y)\in \{y,y^2\}$. In Figure \ref{FigNW3}, the first subplot of each row represents the true function $s(x) f(y)$, while the rest subplots of the corresponding row show the differences 
\begin{eqnarray*}
\nn (\pmb{\ell}^m_{s(x),f(y)} \, |\, \mathbf{W}^{\star}_{m+3} ) -s(x) f(y) \quad\text{for}\quad m\in \{1,3,5\}.
\end{eqnarray*}
Regardless the shape of $s(x) f(y)$, the differences are always non-negative, and converge to 0 sufficiently fast as $m$ increases. More importantly, for $m=3,5$, the differences are very close to a flat surface, which also illustrates the third argument of Lemma \ref{LM1}.

\begin{figure}[h] 
\hspace*{-1cm}\includegraphics[scale=0.23]{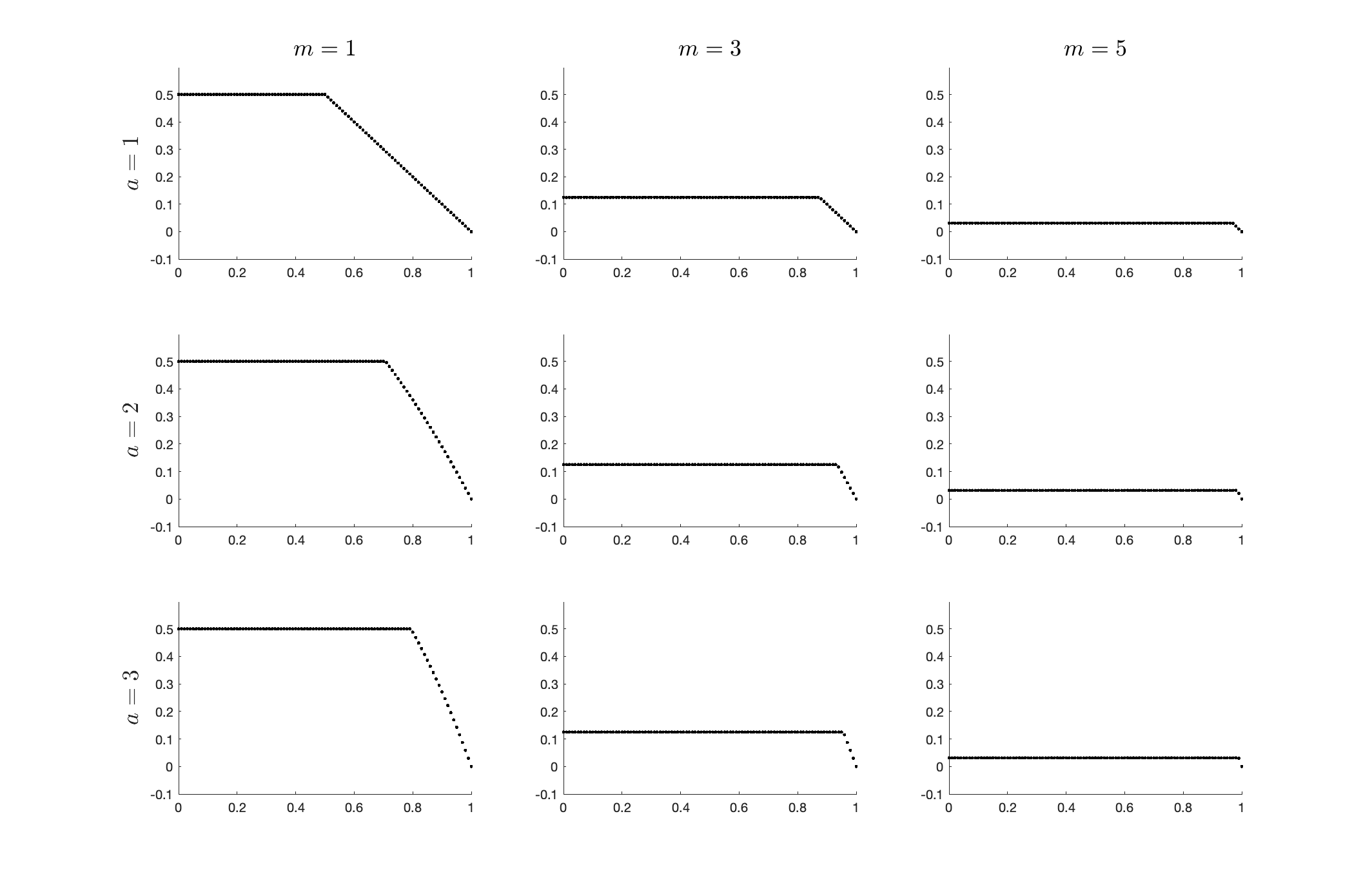}
\vspace*{-1cm}\caption{Plots of \eqref{plot1}}\label{FigHNN1}
\end{figure}

\begin{figure}[h] 
\hspace*{-1cm}\includegraphics[scale=0.23]{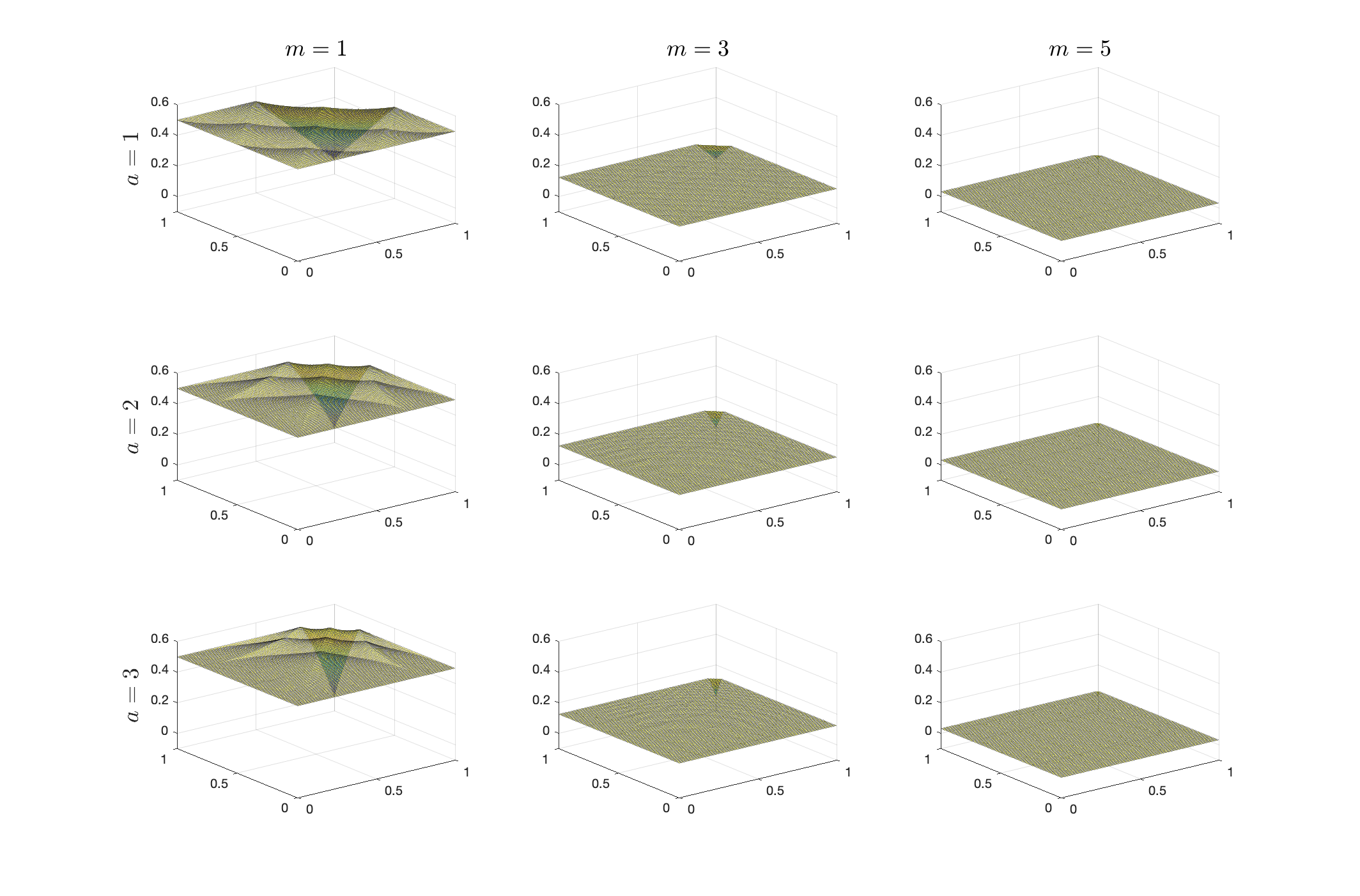}
\vspace*{-1cm}\caption{Plots of \eqref{plot2}}\label{FigHNN2}
\end{figure}

We then demonstrate Lemma \ref{LM.A7} using two plots.   In Figure \ref{FigHNN1}, we let $r=3$, $\pmb{\alpha} = (a,0,0)^\top$, and $\mathbf{x}= (x,1,1)^\top$. Therefore, $\mathcal{N}_{\pmb{\ell}^m}  (\pmb{\ell}_{\mathbf{x}\,|\, {\pmb{\alpha}}} \, |\, \mathbf{W}^{\star}_{m+3})$ should recover $x^a$. We thus plot
\begin{eqnarray}\label{plot1}
\mathcal{N}_{\pmb{\ell}^m}  (\pmb{\ell}_{\mathbf{x}\,|\, {\pmb{\alpha}}} \, |\, \mathbf{W}^{\star}_{m+3})-x^a\quad \text{for}\quad a\in[3],\ m\in\{1,3,5\}.
\end{eqnarray}
In Figure \ref{FigHNN2}, we let $r=3$, $\pmb{\alpha} = (a,a,0)^\top$, and $\mathbf{x}= (x,y,1)^\top$. Therefore, $\mathcal{N}_{\pmb{\ell}^m}  (\pmb{\ell}_{\mathbf{x}\,|\, {\pmb{\alpha}}} \, |\, \mathbf{W}^{\star}_{m+3})$ should recover $x^a y^a$. We thus plot
\begin{eqnarray}\label{plot2}
\mathcal{N}_{\pmb{\ell}^m}  (\pmb{\ell}_{\mathbf{x}\,|\, {\pmb{\alpha}}} \, |\, \mathbf{W}^{\star}_{m+3})-x^ay^a\quad \text{for}\quad a\in[3],\ m\in\{1,3,5\}.
\end{eqnarray}
Clearly the value of $\mathcal{N}_{\pmb{\ell}^m}  (\pmb{\ell}_{\mathbf{x}\,|\, {\pmb{\alpha}}} \, |\, \mathbf{W}^{\star}_{m+3}) -\mathbf{x}^{\pmb{\alpha}}$ is always non-negative. As $m$ increases, $\mathcal{N}_{\pmb{\ell}^m}  (\pmb{\ell}_{\mathbf{x}\,|\, {\pmb{\alpha}}} \, |\, \mathbf{W}^{\star}_{m+3}) -\mathbf{x}^{\pmb{\alpha}}$ converges to 0 in each figure from the positive side regardless the value of $a$. Furthermore, it is interesting to see that the differences in both \eqref{plot1} and \eqref{plot2} actually remain at a constant level, which is only affected by the value of $m$ (i.e., the number of layers of HDNN).

\begin{figure}[h] 
\hspace*{-1cm}\includegraphics[scale=0.23]{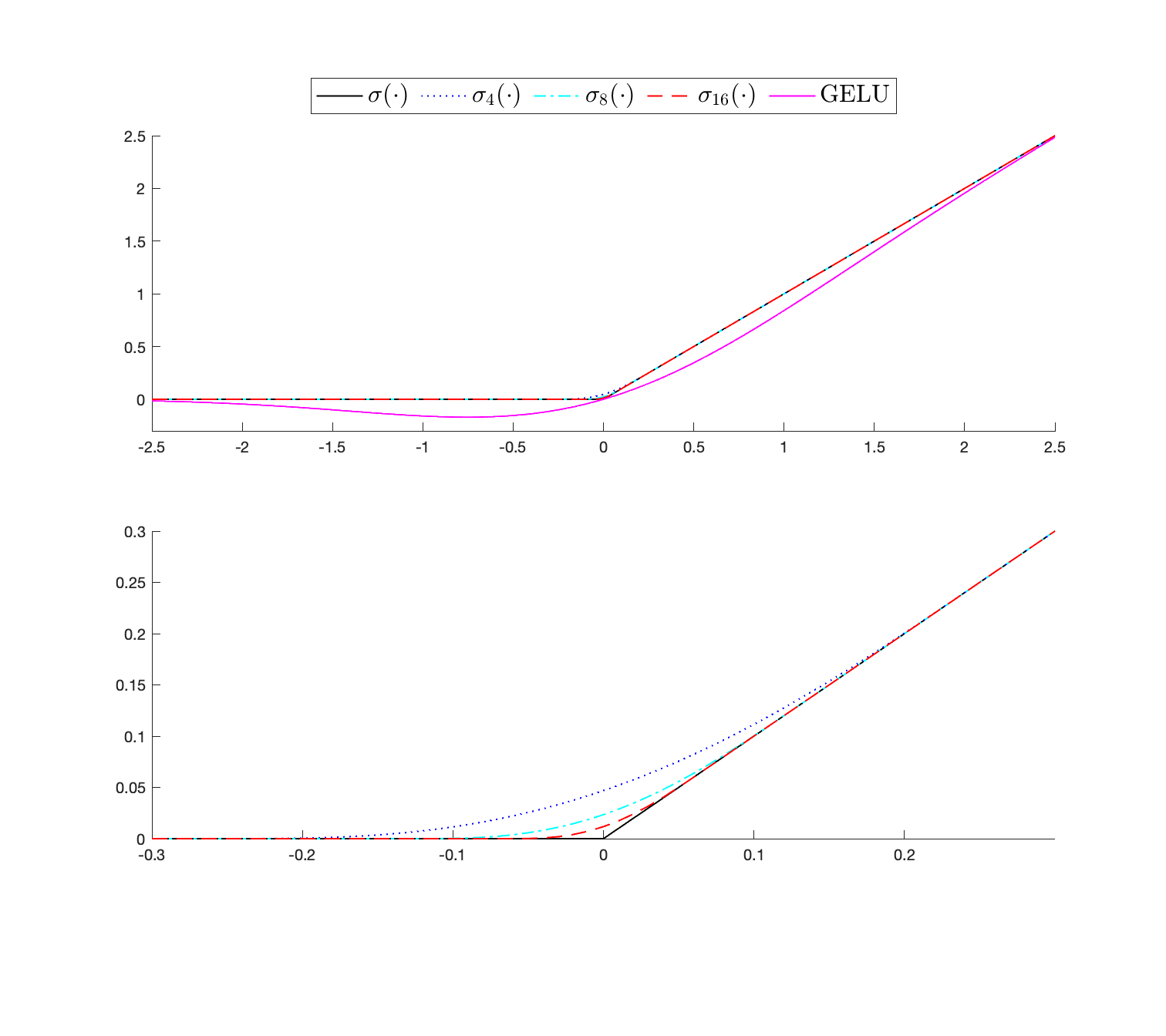}
\vspace*{-1.5cm}\caption{Plots of $\sigma_s(\cdot)$}\label{FigSigm}
\end{figure}

Finally, we plot $\sigma_s(\cdot)$ to justify Lemma \ref{theorem44}, and let $\phi(\cdot)$ be Epanechnikov kernel (i.e., $\phi(u)=0.75(1-u^2)I(|u|\le 1)$) without loss of generality. We first look at the top plot of Figure \ref{FigSigm}. Here, $\sigma(\cdot)$ is plotted using the black solid line, and GELU (i.e., $\sigma_G(u)=u\Phi(u)$ with $\Phi(u)$ being the CDF of $N(0,1)$) is plotted using the magenta solid line. The blue dotted lines stands for $\sigma_4(\cdot)$,  the cyan dashdot line stands for $\sigma_8(\cdot)$, and the red dashed line stands for $\sigma_{16}(\cdot)$. Obviously, $\sigma_s(\cdot)$ is able to mimic ReLU in a much better fashion, and the only difference happens at the origin. To see the difference in detail, we zoom in using the bottom plot of Figure \ref{FigSigm}. It is clear that $\sigma_s(\cdot)$ is quite smooth and moves towards $\sigma(\cdot)$ as $s$ increases. More importantly, unlike GELU, $\sigma_s(u)$ still acts as a linear term when $u\ge  s^{-1}$, so it remains at a low computational cost.

\subsection{Notation \& Preliminary Lemmas}\label{App.A2}

In this appendix, we first introduce extra notations which will be repeatedly used in the development, and then present the preliminary lemmas.

Throughout, we let $I_{\mathbf{i}}(\mathbf{x}) =I(\mathbf{x}\in C_{\mathbf{x}_{\mathbf{i}}}),$ where $C_{\mathbf{x}_{\mathbf{i}}}$ is defined in \eqref{def.cxi}. For $\forall k\in \mathbb{N}$, denote a mapping $T^k:[0, 2^{2-2k}] \mapsto [0, 2^{-2k}]$ as follows:
\begin{eqnarray}\label{def.Tk1}
T^k (x)= \frac{x}{2} \wedge \left(2^{1-2k}-\frac{x}{2}\right) .
\end{eqnarray}
To see the validity of \eqref{def.Tk1}, we write
\begin{eqnarray}\label{def.Tk2}
T^k (x) & = & \frac{x}{2} +0 \wedge \left(2^{1-2k} -x \right) = \frac{x}{2} -0 \vee \left( x-2^{1-2k} \right) 
\nonumber\\
& = & \sigma\left(\frac{x}{2} \right) - \sigma\left( x-2^{1-2k} \right) ,
\end{eqnarray}
where the third equality follows from the fact that $\sigma\left(\frac{x}{2} \right) = \frac{x}{2}$ because of $x$ being defined on $[0, 2^{2-2k}]$. In view of \eqref{def.Tk2}, we partition $[0, 2^{2-2k}]$ as $[0, 2^{1-2k}] \cup [2^{1-2k}, 2^{2-2k}]$, and immediately obtain that
\begin{eqnarray*}
T^k (x)  =\left\{ \begin{array}{ll}
\frac{1}{2}x & \text{for }x \in [0, 2^{1-2k}]\\ 
2^{1-2k} -\frac{1}{2}x & \text{for } x\in [2^{1-2k}, 2^{2-2k}]
\end{array}
\right. ,
\end{eqnarray*}
of which either expression on the right hand side fulfils $T^k (x)\in [0, 2^{-2k}]$. Thus, we conclude $T^k:[0, 2^{2-2k}] \to [0, 2^{-2k}]$.

\smallskip

\begin{figure}[h] 
\hspace*{-2cm}\includegraphics[scale=0.3]{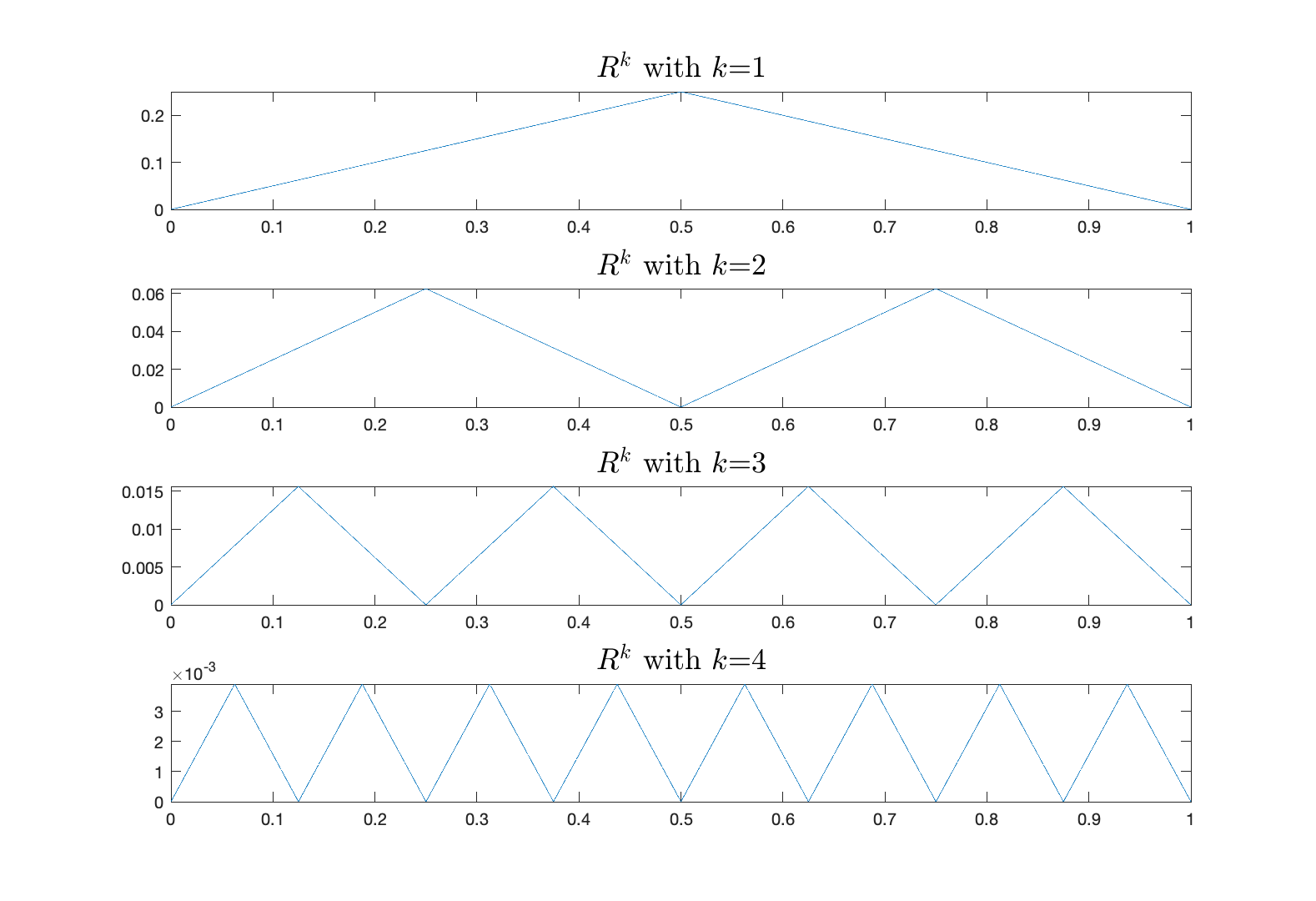}
\vspace*{-1cm}\caption{Plots of $R^k(x)$ with $k\in [4]$}\label{FigRk}
\end{figure}

We then define $R^k:[0,1]\mapsto [0, 2^{-2k}]$ as follows:
\begin{eqnarray}\label{def.Rk}
R^k (x)= T^k\circ T^{k-1}\circ \cdots \circ T^1(x),
\end{eqnarray}
and let further $\mathbf{R}_m(x) = (R^1(x),\ldots, R^m(x))^\top$. In Figure \ref{FigRk}, we plot $R^k(x)$ with $k\in [4]$ for the purpose of demonstration. It is easy to see that $R^k(x)$ is piece wise linear, and the value of $R^k(x)$ shrinks towards 0 as $k$ increases.

Next, we provide a few preliminary lemmas.

\begin{lemma}\label{LM.A2}
Let $f\,:\, [-a, a]^r\to \mathbb{R}$ be a $(p,\mathscr{C})$-smooth function, where $p=\vartheta+s$. For $\forall \mathbf{x}_0\in [-a, a]^r$, let $p_{\vartheta}(\mathbf{x}\, |\, \mathbf{x}_0) =\sum_{\|\mathbf{J}\|_1\le \vartheta} \frac{1}{\mathbf{J}!} f^{(\mathbf{J})}(\mathbf{x}_0)  (\mathbf{x}-\mathbf{x}_0)^{\mathbf{J}}$. Then for $\forall\pmb{\delta} =(\delta_{1},\ldots, \delta_{r})^\top$ satisfying $\|\pmb{\delta}\|_1\le \vartheta$,
\begin{eqnarray*}
\|f^{(\pmb{\delta})}(\mathbf{x})- p_\vartheta^{(\pmb{\delta})}(\mathbf{x}\, |\, \mathbf{x}_0) \|_{\infty}^{[-a, a]^r}\le O(1)\| \|\mathbf{x}-\mathbf{x}_0 \|^{p-\|\pmb{\delta}\|_1}\|_{\infty}^{[-a,a]^r},
\end{eqnarray*}
where $O(1)$ depends on $r$ and $\vartheta$ only.
\end{lemma}

\begin{lemma}\label{LM.A3}
For $\forall m\in \mathbb{N}$, $\|\mathbf{R}_{m} (x)\|_1 $ admits a DNN representation:
\begin{eqnarray*}
\|\mathbf{R}_{m} (x)\|_1&=& \mathbf{w}_{\wout}^\top  \pmb{\sigma}_{\widetilde{\mathbf{v}}_{m}} \widetilde{\mathbf{w}}  \, \pmb{\sigma}_{\widetilde{\mathbf{v}}_{m-1}}\cdots  \widetilde{\mathbf{w}} \, \pmb{\sigma}_{\widetilde{\mathbf{v}}_{1}} (  \mathbf{w}_{\win}\cdot x): =\nn ( \mathbf{w}_{\win} \cdot x \, | \, \mathbf{W}_{\mathbf{R}_{m}}),
\end{eqnarray*}
where $\widetilde{\mathbf{w}}$ is defined in Lemma \ref{LM1} already, and for $k\in [m]$,
\be
 \mathbf{w}_{\win}  =\left(\frac{1}{2}, 1, 0\right)^{\top}, \  \mathbf{w}_{\wout}=(1, -1, 1)^{\top} \ \ \mbox{and} \ \  \widetilde{\mathbf{v}}_{k}  =\left(0, 2^{1-2k}, 0\right)^{\top}.
 \nonumber
\ee
\end{lemma}

\noindent According to Definition \ref{Def1}, the value of $\mathbf{W}_{\mathbf{R}_{m}}$ is obvious. It is noteworthy that $\widetilde{\mathbf{w}}$ is a $3\times 3$ square matrix, but only has rank 2.

\begin{lemma}\label{LM.A4}
For $\forall m\in \mathbb{N}$, $\|\mathbf{R}_m(x) \|_1$ is Lipschitz continuous on $[0,1]$, and is piecewise linear on the following intervals:
\begin{eqnarray*}
&&[\ell \cdot 2^{-m},(\ell+1)\cdot 2^{-m})  \text{ for }\ell =0\cup[2^m-2],\nonumber \\
&&[\ell \cdot 2^{-m},(\ell+1)\cdot 2^{-m}] \text{ for }\ell = 2^m-1.
\end{eqnarray*}
Let $\frac{\partial (\|\mathbf{R}_{m} (x)\|_1)}{\partial x}$ be defined accordingly. For $g(x) = x(1-x)$, we obtain the following results:
\be
\|g(x) - \|\mathbf{R}_m(x) \|_1 \|_{\infty}^{[0,1]} \le 2^{-m} \ \ \ \mbox{and} \ \ \ \left\|\frac{\partial (\|\mathbf{R}_{m} (x)\|_1)}{\partial x} - \frac{\partial g(x)}{\partial x}\right\|_{\infty}^{[0,1]}\le 2^{-m+1}.
\nonumber
\ee
\end{lemma}

\begin{figure}[h] 
\hspace*{-2cm}\includegraphics[scale=0.25]{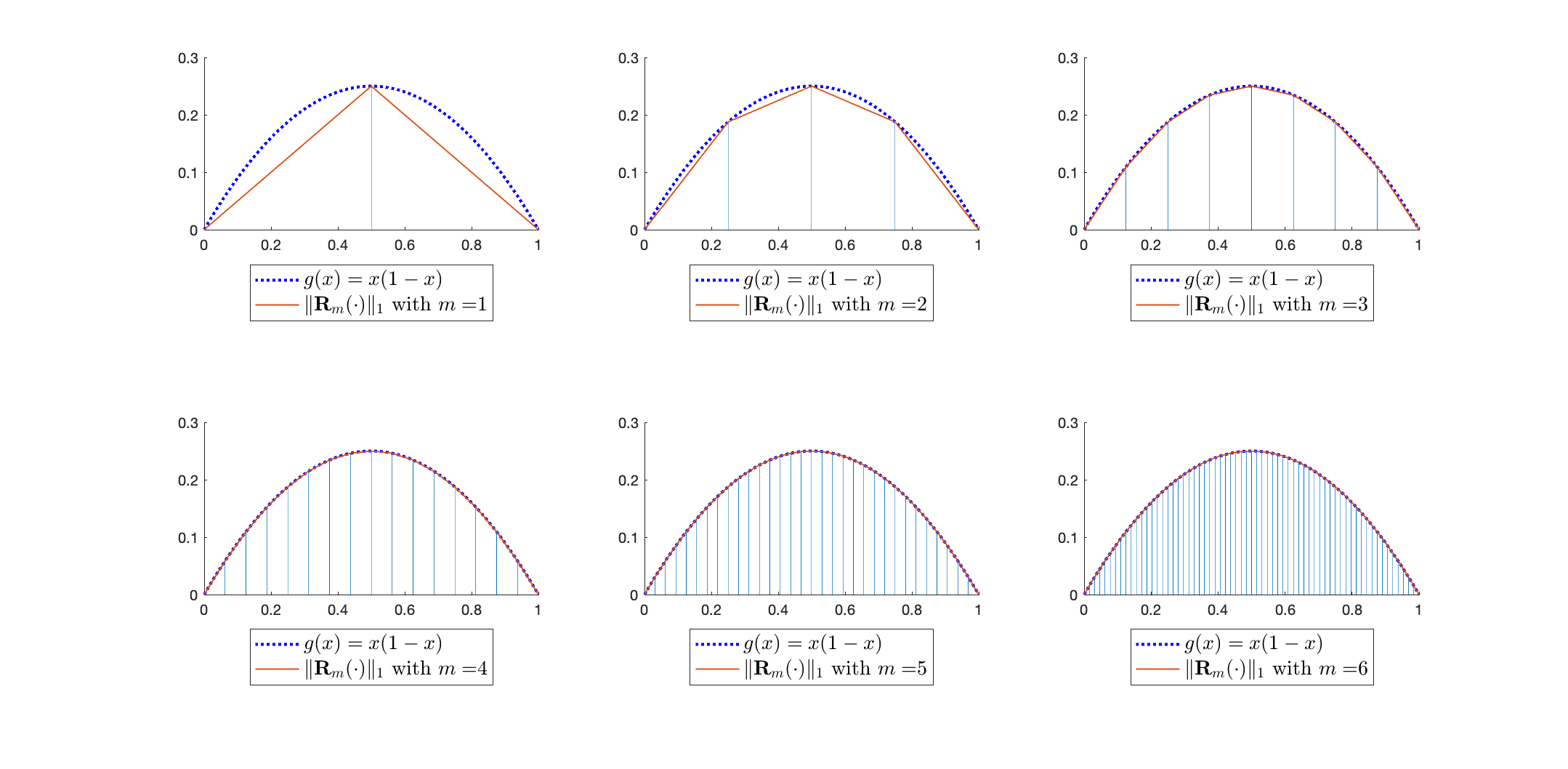}
\vspace*{-1cm}\caption{Illustration of Lemma \ref{LM.A4}}\label{FigRm}
\end{figure}

\noindent In Figure \ref{FigRm}, we give some plots to illustrate our statements in Lemma \ref{LM.A4}. The difference between $\|\mathbf{R}_m(x) \|_1$ and $g(x)$ becomes visually negligible when $m\ge 3$.

We now recall the mapping $\pmb{\ell}_{\mathbf{x}\, |\,\pmb{\alpha}} $ defined in the end of Section \ref{Sec.1}, which gives
\begin{eqnarray}\label{defx1}
\pmb{\ell}_{\mathbf{x}\, |\, {\mathbf{1}_r}}= (\mathbf{x}^\top, \mathbf{1}_q^\top)^\top\quad \text{with}\quad q=\left\{\begin{array}{ll}
2^{\lceil \log_2 r \rceil} -r & \text{for }r\ge 2\\
1 & \text{for }r=1
\end{array} \right. .
\end{eqnarray}
We are then able to present the following lemma.

\begin{lemma}\label{LM.A6}
Using $\nn (\pmb{\ell}^m_{x,y} \, |\, \mathbf{W}^{\star}_{m+3})$ of Lemma \ref{LM1}, define $\mathcal{N}_{\pmb{\ell}^m}  (\pmb{\ell}_{\mathbf{x}\,|\, {\mathbf{1}_r}} \, |\, \mathbf{W}^{\star}_{m+3}) $ according to Definition \ref{Def2}, where $r\ge 2$, $\mathbf{x}\in [0,h]^r$, and $h\le 1 -\lceil\log_2 r\rceil \cdot 2^{-m}$. Then the following results hold:

\begin{enumerate}
\item $0\le \mathcal{N}_{\pmb{\ell}^m}  (\pmb{\ell}_{\mathbf{x}\,|\, {\mathbf{1}_r}} \, |\, \mathbf{W}^{\star}_{m+3}) \le 1$ uniformly on $\mathbf{x}\in [0,h]^r$,

\item $0\le \mathcal{N}_{\pmb{\ell}^m}  (\pmb{\ell}_{\mathbf{x}\,|\, {\mathbf{1}_r}} \, |\, \mathbf{W}^{\star}_{m+3}) -\mathbf{x}^{\mathbf{1}_r} \le  3^{\lceil \log_2 r\rceil-1} 2^{-m} $ uniformly on $\mathbf{x}\in [0,h]^r$,

\item $ \|\frac{\partial}{\partial x_i}[ \mathcal{N}_{\pmb{\ell}^m}  (\pmb{\ell}_{\mathbf{x}\,|\, {\mathbf{1}_r}} \, |\, \mathbf{W}^{\star}_{m+3})-\mathbf{x}^{\mathbf{1}_r}] \|_{\infty}^{[0,h]^r}\le 3^{\lceil \log_2 r\rceil-1} 2^{-m}$ for $\forall i\in [r]$.
\end{enumerate}
\end{lemma}

\begin{lemma}\label{LM.A8}
Suppose that $\pmb{\Sigma}_{11} +\pmb{\Sigma}_{12}+\pmb{\Sigma}_{12}^\top$ is positive definite, where $\pmb{\Sigma}_{11}$ and $\pmb{\Sigma}_{12}$ are defined in Theorem \ref{TM2}. Under Assumptions \ref{As1}-\ref{As2}, as $(h,Th^r)\to (0,\infty)$,

\begin{enumerate}
\item $\frac{1}{\sqrt{T}}\sum_{t=1}^T  \varepsilon_t\, \mathbf{f}^{(1)}_\star(\mathbf{z}_t\, \pmb{\theta}_\star)\, \widetilde{\mathbf{z}}_t I_{a,t}\to_D N(\mathbf{0}, \pmb{\Sigma}_{11} +\pmb{\Sigma}_{12}+\pmb{\Sigma}_{12}^\top)$,

\item for $\forall \mathbf{x_0}\in [-a, a]^r$, $\frac{1}{\sqrt{Th^r}}\sum_{t=1}^T\sum_{\mathbf{i}\in [M]^2} \varepsilon_t I_{\mathbf{i},t,\mathbf{x}_0} \mathbf{H}^{-1}\pmb{\psi}_{r_\vartheta} (\mathbf{z}_t\, \pmb{\theta}_\star \, | \, \mathbf{x}_{\mathbf{i}}) \to_D N(\mathbf{0}, \pmb{\Sigma}_{2,\mathbf{x}_0} )$,
\end{enumerate}
where $I_{\mathbf{i},t,\mathbf{x}_0}= I(\mathbf{x}_0\in C_{\mathbf{x}_\mathbf{i}})I_{\mathbf{i}} (\mathbf{z}_t\,\pmb{\theta}_\star)$ and $\pmb{\Sigma}_{2,\mathbf{x}_0} =\sigma_\varepsilon^2\phi_{\pmb{\theta}_\star}(\mathbf{x}_0)\int_{[0,1]^r}\pmb{\psi}_{r_\vartheta}(\mathbf{w})\pmb{\psi}_{r_\vartheta}(\mathbf{w})^\top \mathrm{d}\mathbf{w}.$
\end{lemma}

\end{appendix}

{\footnotesize
\bibliography{Refs}

@article{Xia2002,
 author = {Yingcun Xia and Howell Tong and W. K. Li and Li-Xing Zhu},
 journal = {Journal of the Royal Statistical Society. Series B (Statistical Methodology)},
 number = {3},
 pages = {363--410},
 title = {An Adaptive Estimation of Dimension Reduction Space},
 volume = {64},
 year = {2002}
}

@article{cw1999,
title={Improved Rates and Asymptotic Normality for Nonparametric Neural Network Estimators},
author={Xiaohong Chen and Halbert White},
journal={IEEE Transactions on Information Theory},
volume={45},
number={6},
pages={682--691},
year={1999}}

@ARTICLE{SL2004,
title = {Nonparametric neural network estimation of Lyapunov exponents and a direct test for chaos},
author = {Shintani and Linton},
year = {2004},
journal = {Journal of Econometrics},
volume = {120},
number = {1},
pages = {1-33}
}

@unpublished{CLMZ2022,
	title={Casual inference of General Treatment Effects using Neural Networks with a Diverging Number of Confounders},
	author={Xiaohong Chen and Ying Liu and Shujie Ma and Zheng Zhang},
	note={Available at \url{https://arxiv.org/abs/2009.07055v5}},
	year={2022},
}

@article{hoefler2021sparsity,
  title={Sparsity in deep learning: Pruning and growth for efficient inference and training in neural networks},
  author={Hoefler, Torsten and Alistarh, Dan and Ben-Nun, Tal and Dryden, Nikoli and Peste, Alexandra},
  journal={The Journal of Machine Learning Research},
  volume={22},
  number={1},
  pages={10882--11005},
  year={2021},
  publisher={JMLRORG}
}

@ARTICLE {BK2019,
    author  = "Benedikt Bauer and Michael Kohler",
    title   = "{On Deep Learning as a Remedy for the Curse of Dimensionality in Nonparametric Regression}",
    journal = "The Annals of Statistics",
    year    = "2019",
    volume  = "47",
    number  = "4",
    pages   = "2261-2285"
}

@book{ttg2010,
	title={Modelling Nonlinear Economic Time Series},
	author={Timo Ter\"{a}svirta and Dag Tj{\o}stheim and Clive W. J. Granger},
	year={2010},
	publisher={Oxford University Press, London}
	}

@book{Howell1990,
	title={Nonlinear Time Series: A Dynamical System Approach},
	author={Howell Tong},
	year={1990},
	publisher={Oxford University Press, London}
	}

@book{gao2007,
	title={Nonlinear Time Series: Semiparametric and Nonparametric Methods},
	author={Jiti Gao},
	volume={108},
	year={2007},
	publisher={Chapman \& Hall/CRC Monographs on Statistics and Applied Probability, London}
	}

@article{KK2017,
 author = {Michael Kohler and Adam Krzy\'{z}ak},
 journal = {IEEE Transactions on Information Theory},
 number = {3},
 Pages = {341-356},
 title = {Nonparametric regression based on hierarchical interaction models},
 volume = {63},
 year = {2017}
}

@BOOK {FanYao,
    author    = "J. Fan and Q. Yao",
    title     = "Nonlinear Time Series: {N}onparametric and Parametric Methods",
    publisher = "Springer-Verlag",
    year      = "2003"
}

@article{AC2003,
author = {Ai, Chunrong and Chen, Xiaohong},
title = {Efficient Estimation of Models with Conditional Moment Restrictions Containing Unknown Functions},
journal = {Econometrica},
volume = {71},
number = {6},
pages = {1795-1843},
year = {2003}
}

@ARTICLE {SH2020,
    author  = "Johannes Schmidt-Hieber",
    title   = "{Nonparametric Regression Using Deep Neural Networks with ReLU Activation Function}",
    journal = "The Annals of Statistics",
    year    = "2020",
    volume  = "48",
    number  = "4",
    pages   = "1875-1897"
}

@article{CGL2012ET,
 title={A NEW DIAGNOSTIC TEST FOR CROSS-SECTION UNCORRELATEDNESS IN NONPARAMETRIC PANEL DATA MODELS}, 
 volume={28}, 
 number={5}, 
 journal={Econometric Theory}, 
 author={Chen, Jia and Gao, Jiti and Li, Degui}, 
 year={2012}, 
 pages={1144-1163}
 }

@article{hansen_1991,
 title={Strong Laws for Dependent Heterogeneous Processes},
 volume={7},
 number={2},
 journal={Econometric Theory},
 author={Hansen, Bruce E.},
 year={1991},
 pages={213-221}
 }

@article{LBH2015,
 title={Deep learning},
 volume={521},
 journal={Nature},
 author={Y. {LeCun} and Y. Bengio and G. Hinton},
 year={2015},
 pages={436–444}
 }

@INCOLLECTION {Athey2019,
    author    = "Susan Athey",
    title     = "The Impact of Machine Learning on Economics",
    booktitle = "The Economics of Artificial Intelligence: {A}n Agenda",
    year      = "2019",
    editor    = "Ajay Agrawal, Joshua Gans, and Avi Goldfarb",
    pages     = "507-547"
}

@article{BMR2021, 
title={Deep learning: {A} statistical viewpoint}, 
volume={30},
journal={Acta Numerica}, 
author={Bartlett, Peter L. and Montanari, Andrea and Rakhlin, Alexander}, 
year={2021}, 
pages={87201}}

@article{FMZ2021,
author = {Jianqing Fan and Cong Ma and Yiqiao Zhong},
title = "A Selective Overview of Deep Learning",
volume = {36},
journal = {Statistical Science},
number = {2},
pages = {264-290},
year = {2021}
}

@article{FALLAHGOUL2024105574,
title = {Asset pricing with neural networks: {S}ignificance tests},
journal = {Journal of Econometrics},
volume = {238},
number = {1},
pages = {105574},
year = {2024},
author = {Hasan Fallahgoul and Vincentius Franstianto and Xin Lin}
}

@unpublished{FG2022,
	title={Factor Augmented Sparse Throughput Deep ReLU Neural Networks for High Dimensional Regression},
	author={Jianqing Fan and Yihong Gu},
	note={Available at \url{https://doi.org/10.48550/arXiv.2210.02002}},
	year={2022},
}

@article{FLM2021,
author = {Farrell, Max H. and Liang, Tengyuan and Misra, Sanjog},
title = {Deep Neural Networks for Estimation and Inference},
journal = {Econometrica},
volume = {89},
number = {1},
pages = {181-213},
year = {2021}
}

@InProceedings{glorot11a,
  title = 	 {Deep Sparse Rectifier Neural Networks},
  author = 	 {Glorot, Xavier and Bordes, Antoine and Bengio, Yoshua},
  booktitle = 	 {Proceedings of the Fourteenth International Conference on Artificial Intelligence and Statistics},
  pages = 	 {315-323},
  year = 	 {2011},
  editor = 	 {Gordon, Geoffrey and Dunson, David and Dudík, Miroslav},
  volume = 	 {15},
  series = 	 {Proceedings of Machine Learning Research}
}

@article{DUBEY202292,
title = {Activation Functions in Deep Learning: {A} Comprehensive Survey and Benchmark},
journal = {Neurocomputing},
volume = {503},
pages = {92-108},
year = {2022},
issn = {0925-2312},
author = {Shiv Ram Dubey and Satish Kumar Singh and Bidyut Baran Chaudhuri},
}

@article{NW1987,
 author = {Whitney K. Newey and Kenneth D. West},
 journal = {Econometrica},
 number = {3},
 pages = {703-708},
 title = {A Simple, Positive Semi-Definite, Heteroskedasticity and Autocorrelation Consistent Covariance Matrix},
 volume = {55},
 year = {1987}
}

@article{Shao2015,
author = {Xiaofeng Shao},
title = {Self-Normalization for Time Series: {A} Review of Recent Developments},
journal = {Journal of the American Statistical Association},
volume = {110},
number = {512},
pages = {1797-1817},
year  = {2015}
}

@article{Gu2020,
    author = {Gu, Shihao and Kelly, Bryan and Xiu, Dacheng},
    title = "Empirical Asset Pricing via Machine Learning",
    journal = {The Review of Financial Studies},
    volume = {33},
    number = {5},
    pages = {2223-2273},
    year = {2020},
}

@article{FG2010,
author = {G{\"u}nther, Frauke and Fritsch, Stefan},
year = {2010},
month = {06},
pages = {30-38},
title = {Neuralnet: {T}raining of Neural Networks},
volume = {2},
journal = {R Journal}
}

@InProceedings{KingBa15,
  author    = {Kingma, Diederik and Ba, Jimmy},
  booktitle = {International Conference on Learning Representations (ICLR)},
  title     = {Adam: A Method for Stochastic Optimization},
  year      = {2015},
  address   = {San Diega, CA, USA},
  optmonth  = {12},
}

@ARTICLE {shao2010,
    author    = "Shao, Xiaofeng",
    title     = "The dependent wild bootstrap",
    journal   = "Journal of the American Statistical Association",
    year      = "2010",
    volume    = "105",
    number    = "489",
    pages     = "218-235",
}

@article{Andrews1991,
 author = {Donald W. K. Andrews},
 journal = {Econometrica},
 number = {3},
 pages = {817-858},
 title = {Heteroskedasticity and Autocorrelation Consistent Covariance Matrix Estimation},
 volume = {59},
 year = {1991}
}

@article{Hansen1992,
 author = {Bruce E. Hansen},
 journal = {Econometrica},
 number = {4},
 pages = {967-972},
 title = {Consistent Covariance Matrix Estimation for Dependent Heterogeneous Processes},
 volume = {60},
 year = {1992}
}

@ARTICLE {LTG2016,
    author  = "D. Li and D. Tj{\o}stheim and J. Gao",
    title   = "Estimation in nonlinear regression with {H}arris recurrent {M}arkov chains",
    journal = "The Annals of Statistics",
    year    = "2016",
    volume  = "44",
    number  = "5",
    pages   = "1957-1987"
}

@unpublished{DFLSP2021,
	title={Dimension-Free Average Treatment Effect Inference with Deep Neural Networks},
	author={Xinze Du and Yingying Fan and Jinchi Lv and Tianshu Sun and Patrick Vossler},
	note={Available at \url{https://doi.org/10.48550/arXiv.2112.01574}},
	year={2021},
}

@article{KN2020,
    author = {Keane, Michael and Neal, Timothy},
    title = "{Comparing deep neural network and econometric approaches to predicting the impact of climate change on agricultural yield}",
    journal = {The Econometrics Journal},
    volume = {23},
    number = {3},
    pages = {S59-S80},
    year = {2020}
}

@ARTICLE{PSU2011,
title = {Cross-sectional dependence robust block bootstrap panel unit root tests},
author = {Palm, Franz and Smeekes, Stephan and Urbain, Jean-Pierre},
year = {2011},
journal = {Journal of Econometrics},
volume = {163},
number = {1},
pages = {85-104}
}

@article{AEMS2020,
    author = {Andreasen, Martin M and Engsted, Tom and M{\o}ller, Stig V and Sander, Magnus},
    title = "{The Yield Spread and Bond Return Predictability in Expansions and Recessions}",
    journal = {The Review of Financial Studies},
    volume = {34},
    number = {6},
    pages = {2773-2812},
    year = {2020}
}

@article{Betal2023,
author = {Borup, Daniel and Eriksen, Jonas N. and Kj\ae{}r, Mads M. and Thyrsgaard, Martin},
title = {Predicting Bond Return Predictability},
journal = {Management Science, forthcoming},
year = {2023}
}

@article{LN2009,
    author = {Ludvigson, Sydney C. and Ng, Serena},
    title = "{Macro Factors in Bond Risk Premia}",
    journal = {The Review of Financial Studies},
    volume = {22},
    number = {12},
    pages = {5027-5067},
    year = {2009},
    month = {10}
}

@ARTICLE {CD2006,
    author  = "Christoffersen, Peter F. and Diebold, Francis X.",
    title   = "Financial Asset Returns, Direction-of-Change Forecasting, and Volatility Dynamics",
    journal = "Management Science",
    year    = "2006",
    volume  = "52",
    number  = "8",
    pages   = "1273-1287"
}

@BOOK{rudin2004,
  title = {{Principles of Mathematical Analysis}},
  publisher = {McGraw-Hill Companies, Inc.},
  year = {2004},
  author = {Walter Rudin},
  series = {},
  address = {New York}
}

@unpublished{HG2023,
	title={Gaussian Error Linear Units ({GELU}s)},
	author={Dan Hendrycks and Kevin Gimpel},
	note={Available at \url{https://doi.org/10.48550/arXiv.1606.08415}},
	year={2023},
}
}

\newpage
\setcounter{page}{1}

{\small

\begin{center}
    { \large \bf Online Supplementary Appendices to \\``Estimation and Inference for \\a Class of Generalized Hierarchical Models"}
    \medskip

    { 
    {\sc Chaohua Dong$^{\ast}$, Jiti Gao$^{\dag}$, Bin Peng$^{\dag}$ and Yayi Yan$^\sharp$}
    \medskip

    $^{\ast}$Zhongnan University of Economics and Law\\ $^{\dag}$Monash University\\ $^\sharp$Shanghai University of Finance and Economics} 
\end{center}

\renewcommand{\theequation}{B\arabic{section}.\arabic{equation}}
\renewcommand{\thesection}{B\arabic{section}}
\renewcommand{\thefigure}{B\arabic{figure}}
\renewcommand{\thetable}{B\arabic{table}}
\renewcommand{\thelemma}{B\arabic{lemma}}
\renewcommand{\theremark}{B\arabic{remark}}
\renewcommand{\thecorollary}{B\arabic{corollary}}

\setcounter{equation}{0}
\setcounter{lemma}{0}
\setcounter{section}{0}
\setcounter{table}{0}
\setcounter{figure}{0}
\setcounter{remark}{0}
\setcounter{corollary}{0}
 
In the appendices, we provide additional simulation results in Appendix \ref{App.sim}, present some extra plots in Appendix \ref{App.Plots}, and then give the proofs in Appendix \ref{App.A3}.

\medskip

\section{Extra Simulation Results}\label{App.sim}

In this section, we examine the case with $r=16$, and Table \ref{extratabler16} is organized in the same way as Table \ref{extratable} of the main text. As $r$ is reasonably large now,  typical nonparametric regression approaches will definitely fail. To obtain reasonable bias and standard deviations, we increase the sample size to $2000$ and $4000$, which are still acceptable in view of the numerical exercises such as those in \cite{BK2019} and \cite{FLM2021}. The findings are consistent with those presented in the main text, so we no longer repeat them here. Overall, we conclude our approach is robust in terms of the value of $r$.

{\small
\begin{table}[h]
\caption{Extra Simulation Results (with $r=16$)}\label{extratabler16}
\begin{tabular}{crrrrrrrr}
\hline\hline
\multicolumn{1}{l}{} &  &  & \multicolumn{3}{c}{Bias} & \multicolumn{3}{c}{Std} \\
\multicolumn{1}{l}{} &  &  $T$ & $m=3$ & $m=4$ & $m=5$ & $m=3$ & $m=4$ & $m=5$ \\
via $\sigma(\cdot)$ & $\pmb{\theta}$ & 2000 & 0.0471 & 0.0517 & 0.0502 & 0.0522 & 0.0638 & 0.0611 \\
 &  & 4000 & 0.0374 & 0.0366 & 0.0354 & 0.0438 & 0.0397 & 0.0442 \\
 & $f$ & 2000 & 0.5691 & 0.4700 & 0.5241 & 0.5734 & 0.4634 & 0.4794 \\
 &  & 4000 & 0.3487 & 0.4787 & 0.4242 & 0.3343 & 0.5078 & 0.4163 \\ \cline{4-9}
via $\sigma_{32}(\cdot)$ & $\pmb{\theta}$ & 2000 & 0.0238 & 0.0298 & 0.0265 & 0.0369 & 0.0417 & 0.0383 \\
 &  & 4000 & 0.0227 & 0.0060 & 0.0216 & 0.0350 & 0.0137 & 0.0301 \\
 & $f$ & 2000 & 0.7491 & 0.7563 & 0.7704 & 0.3158 & 0.4165 & 0.3654 \\
 &  & 4000 & 0.7337 & 0.7312 & 0.7331 & 0.2502 & 0.2726 & 0.2671 \\
 \hline\hline
\end{tabular}
\end{table}
}

\section{Extra Plots}\label{App.Plots}

In this appendix, we provide some extra plots using $\sigma_s(\cdot)$. As in Figure \ref{FigNW3} of the main text, we show that Lemma \ref{LM1} still holds when $\sigma_s(\cdot)$ is adopted. The setting is the same as those in Figure \ref{FigNW3}, but we replace $\sigma(\cdot)$ with $\sigma_s(\cdot)$ in which $s =16, \, 32$, and $\phi(\cdot)$ is chosen to be Epanechnikov kernel (i.e., $\phi(u)=0.75(1-u^2)I(|u|\le 1)$) without loss of generality.

As shown in Figure \ref{FigNW3}, Figures \ref{FigNW32} and \ref{FigNW33}, the differences are overall at the same magnitude, and they all shrink towards 0 as $m$ increases. As $s$ increases, the differences of Figure \ref{FigNW33} look more similar to those presented in \ref{FigNW3}.  

In addition, we re-plot \eqref{plot1} and \eqref{plot2} of the main text using using $\sigma_s(\cdot)$. Again, we replace $\sigma(\cdot)$ with $\sigma_s(\cdot)$ in which $s =16, \, 32$, and $\phi(\cdot)$ is chosen to be Epanechnikov kernel. The findings are almost the same as aforementioned.  As $s$ increases, the differences in Figures \ref{FigHNN13} and \ref{FigHNN23} very much similar to those in Figures \ref{FigHNN1} and \ref{FigHNN2} of the main text. It is worth mentioning that in Figure \ref{FigHNN1}, there are some obvious non-smooth changing points, which become much smoother in Figures \ref{FigHNN12} and \ref{FigHNN13} due to the use of $\sigma_s(\cdot)$.

Overall, we can conclude that the main results developed in Sections \ref{Sec.2} and \ref{Sec.3} remain valid by replacing $\sigma(\cdot)$ with $\sigma_s(\cdot)$.

\begin{figure}[h]
\hspace*{-1cm}\includegraphics[scale=0.25]{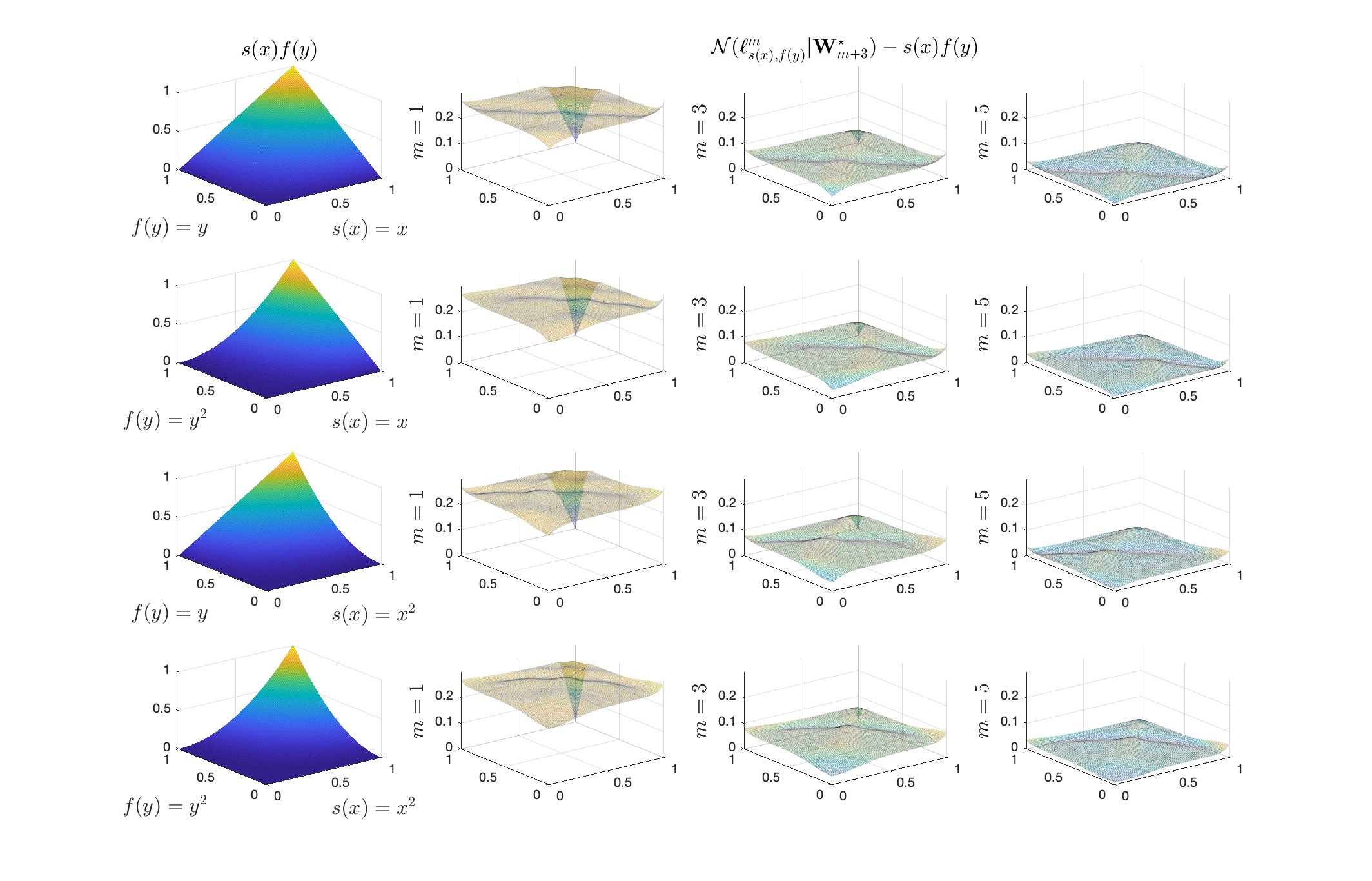}
\vspace*{-1cm}\caption{Illustration of Lemma \ref{LM1} using $\sigma_{16}(\cdot)$}\label{FigNW32}
 
\hspace*{-1cm}\includegraphics[scale=0.25]{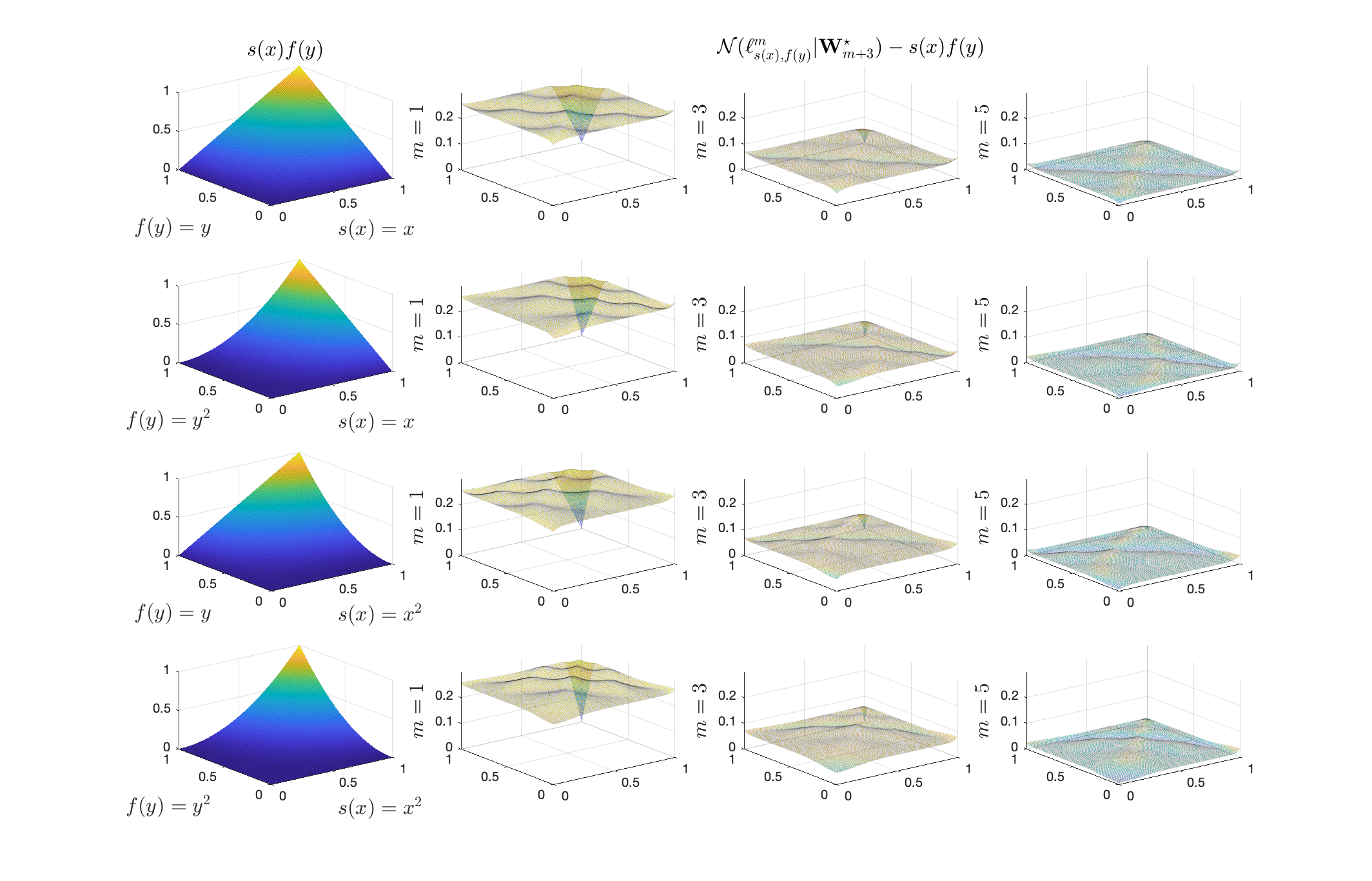}
\vspace*{-1cm}\caption{Illustration of Lemma \ref{LM1} using $\sigma_{32}(\cdot)$}\label{FigNW33}
\end{figure}

\begin{figure}[h] 
\hspace*{-1cm}\includegraphics[scale=0.23]{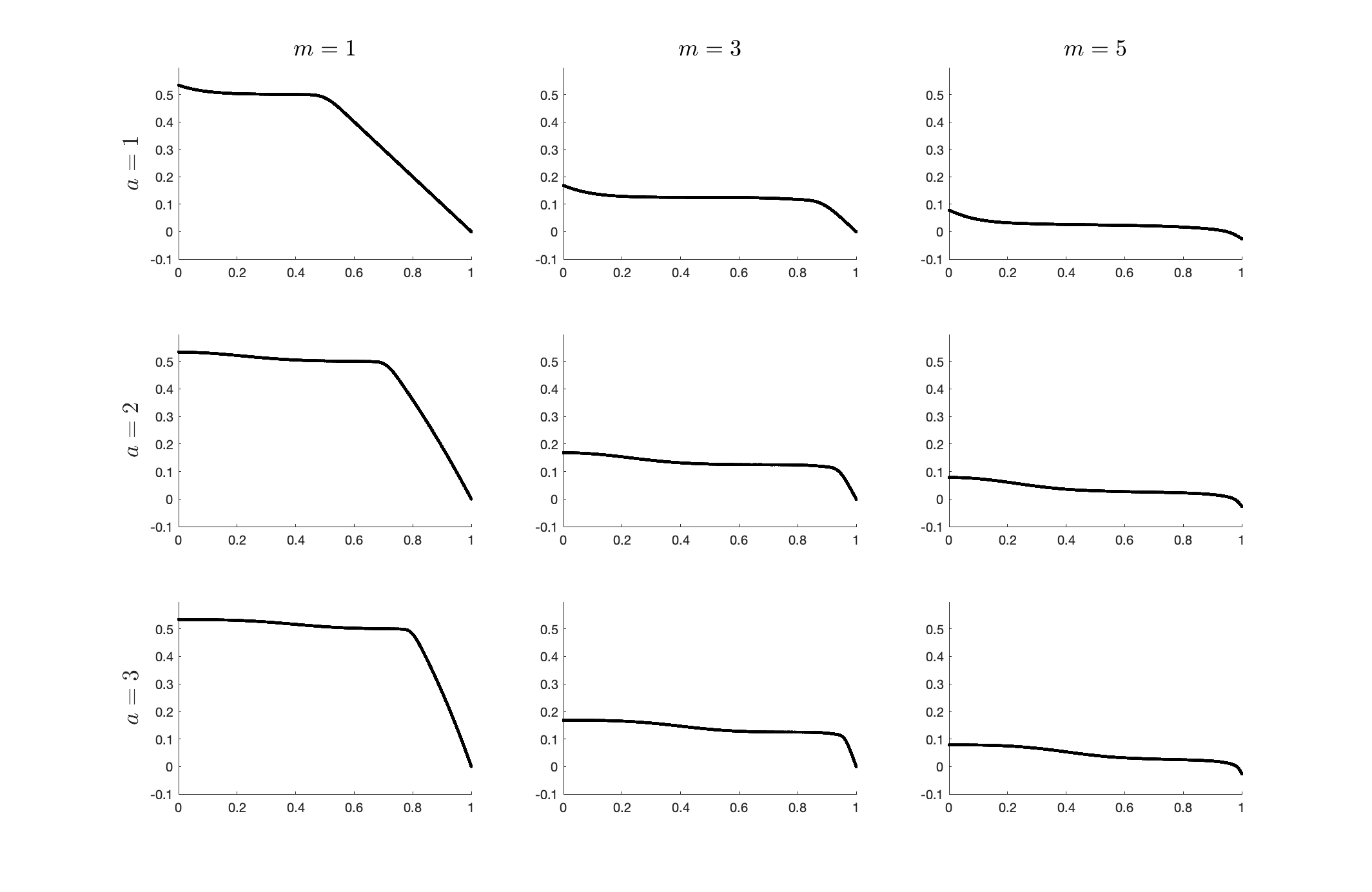}
\vspace*{-1cm}\caption{Plots of \eqref{plot1} using $\sigma_{16}(\cdot)$}\label{FigHNN12}
 
\hspace*{-1cm}\includegraphics[scale=0.23]{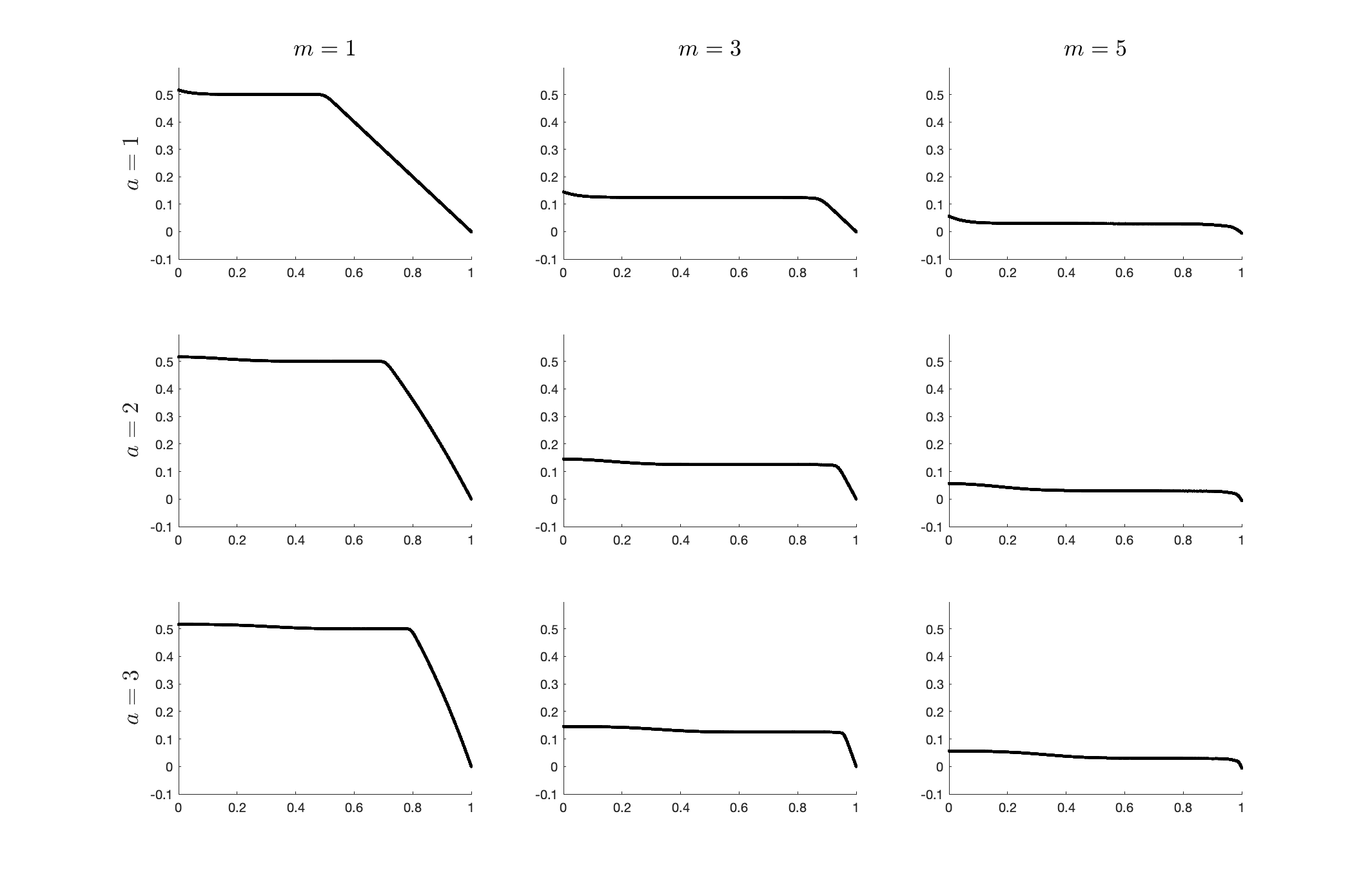}
\vspace*{-1cm}\caption{Plots of \eqref{plot1} using $\sigma_{32}(\cdot)$}\label{FigHNN13}
\end{figure}

\begin{figure}[h] 
\hspace*{-1cm}\includegraphics[scale=0.23]{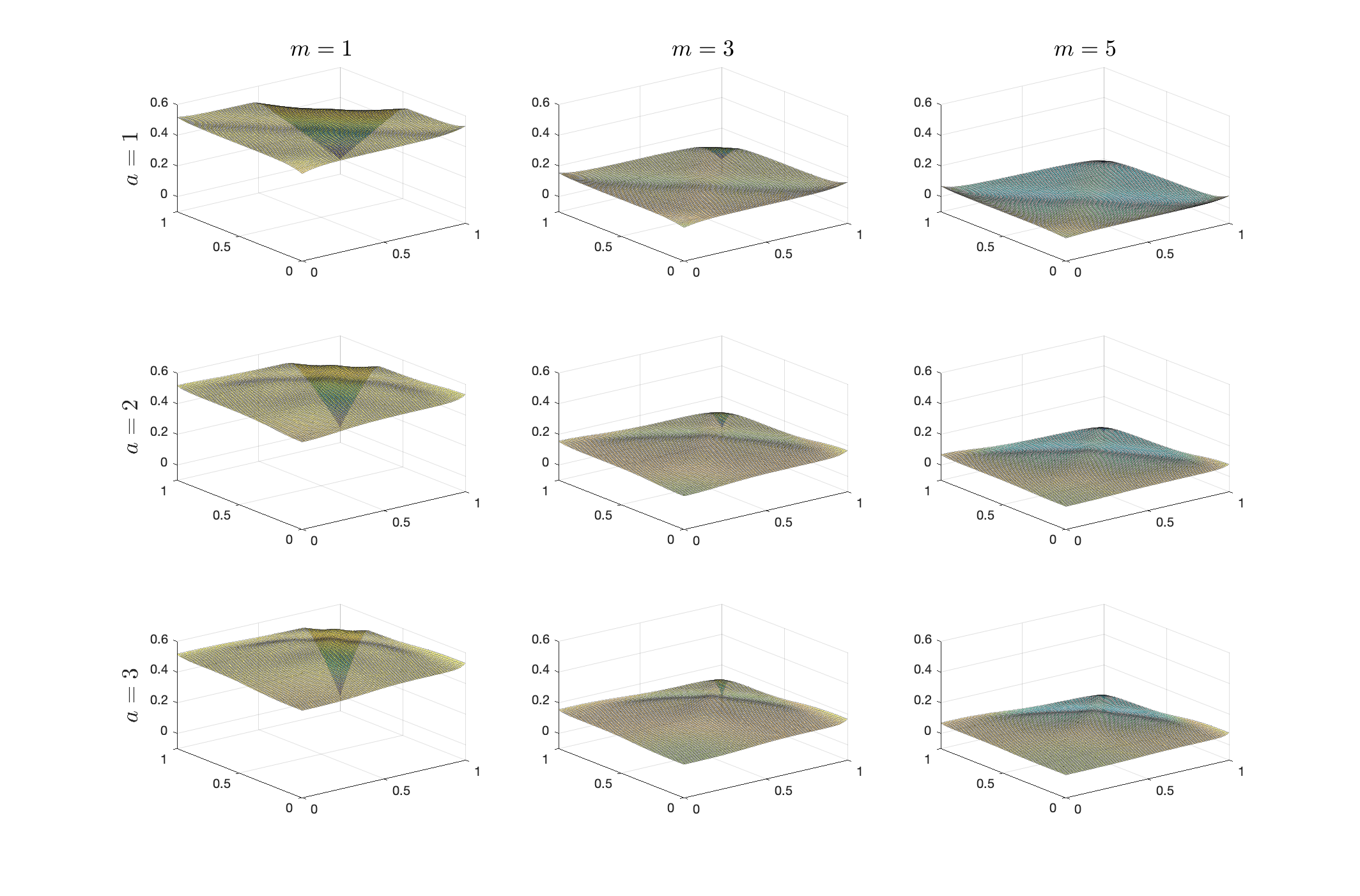}
\vspace*{-1cm}\caption{Plots of \eqref{plot2} using $\sigma_{16}(\cdot)$}\label{FigHNN22}

\hspace*{-1cm}\includegraphics[scale=0.23]{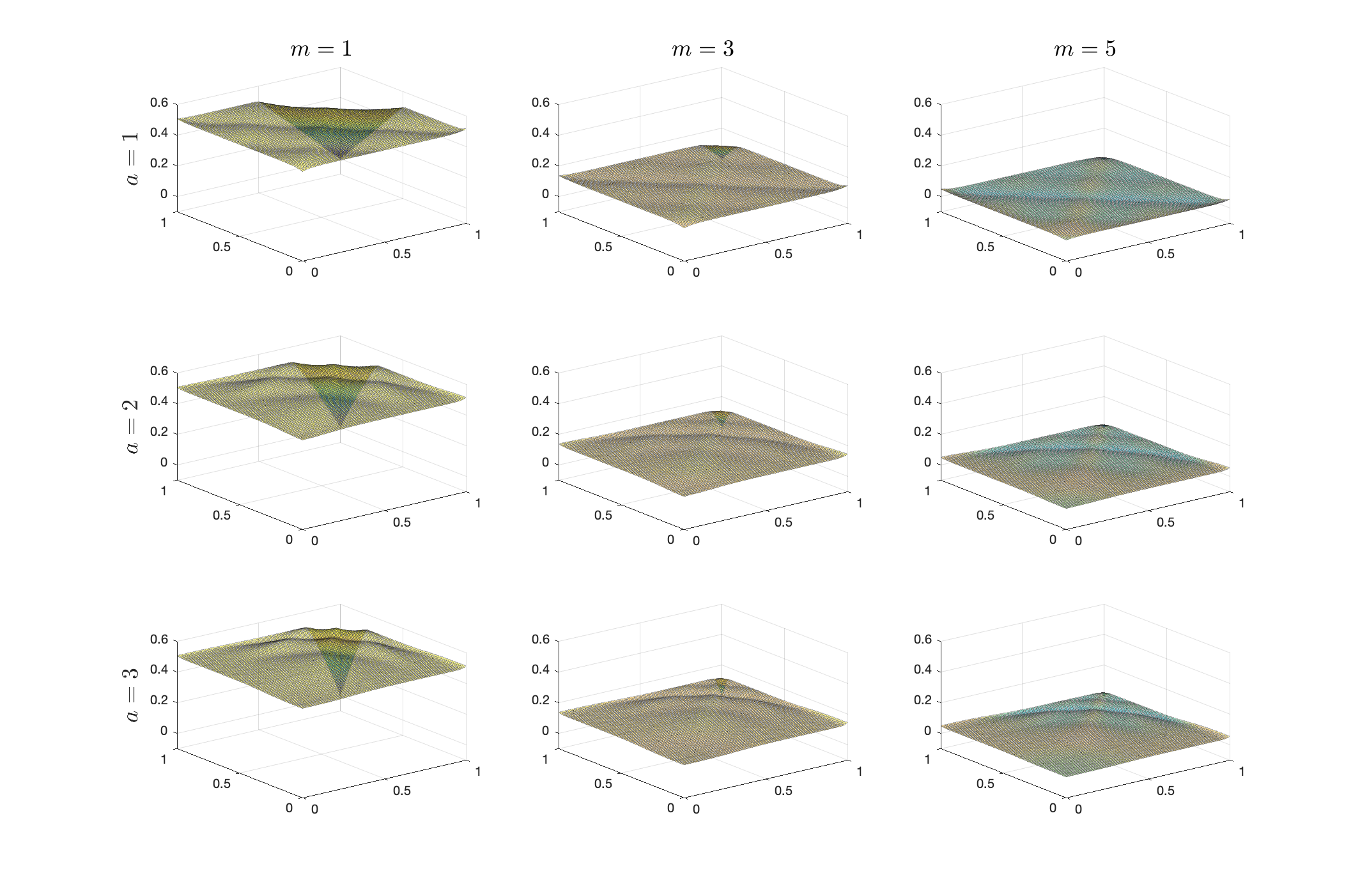}
\vspace*{-1cm}\caption{Plots of \eqref{plot2} using $\sigma_{32}(\cdot)$}\label{FigHNN23}
\end{figure}

\section{Proofs}\label{App.A3}

This appendix provides all the proofs. First, we denote

\begin{eqnarray}\label{defbias}
\mathbf{c}_{bias}&=&(\mathbf{c}_{\pmb{\theta}}^\top, c_{f})^\top,\\
\mathbf{c}_{\pmb{\theta}}&=&  \sum_{\mathbf{i}}\int_{C_{\mathbf{i}}}\phi(\mathbf{z})\sum_{\|\mathbf{J}\|_1=\vartheta}\frac{\vartheta (\mathbf{z}\,\pmb{\theta}_\star-\mathbf{x}_{\mathbf{i}})^{\mathbf{J}}}{\mathbf{J}!} F_{\mathbf{i}}(\mathbf{z}\, \pmb{\theta}_\star)\mathbf{f}^{(1)}_\star (\mathbf{z}\, \pmb{\theta}_\star)\cdot \mathbf{z}^\top \mathbf{1}_r  d\mathbf{z},\nonumber \\
c_{f} &=&\sum_{\mathbf{i}}\frac{I(\mathbf{x}_0\in C_{\mathbf{X}_\mathbf{i}})}{h^r}\int_{C_{\mathbf{X}_\mathbf{i}}}\phi_{\pmb{\theta}}(\mathbf{x})\sum_{\|\mathbf{J}\|_1=\vartheta}\frac{\vartheta (\mathbf{x}-\mathbf{x}_{\mathbf{i}})^{\mathbf{J}}}{\mathbf{J}!} F_{\mathbf{i}}(\mathbf{x})\mathbf{H}^{-1}\pmb{\psi}_{r_\vartheta} (\mathbf{x} \, | \, \mathbf{x}_{\mathbf{i}}) d\mathbf{x},\nonumber 
\end{eqnarray}
where $\phi(\cdot)$ denotes the density function of $\mathbf{z}_t$, $\phi_{\pmb{\theta}}(\cdot)$ is defined in Assumption \ref{As2}, and

\begin{eqnarray*}
  F_{\mathbf{i}}(\mathbf{x} ) =\int_{0}^1\left[ (1-w)^{\vartheta-1} f_\star^{(\mathbf{J})}(\mathbf{x}_{\mathbf{i}} + w(\mathbf{x}-\mathbf{x}_{\mathbf{i}})) - f_\star^{(\mathbf{J})}(\mathbf{x}_{\mathbf{i}})   (\mathbf{x}-\mathbf{x}_{\mathbf{i}})^{\mathbf{J}} \right] \mathrm{d}w.
\end{eqnarray*}
Due to the use of $(p,\mathscr{C})$-smooth, the term $F_{\mathbf{i}}(\mathbf{x} )$ remains in $\mathbf{c}_{bias}$. See the proof of Lemma \ref{LM.A2} for details.

\medskip

\noindent \textbf{Proof of Lemma \ref{LM.A2}:}

Start with the case where $\|\pmb{\delta}\|_1=0$. Write

\begin{eqnarray*}
&&f(\mathbf{x}) -p_{\vartheta}(\mathbf{x}\, |\, \mathbf{x}_0)\nonumber \\
&=&  f(\mathbf{x}) -p_{\vartheta-1}(\mathbf{x}\, |\, \mathbf{x}_0) -\sum_{\|\mathbf{J}\|_1=\vartheta}\frac{1}{\mathbf{J}!} f^{(\mathbf{J})}(\mathbf{x}_0) (\mathbf{x}-\mathbf{x}_0)^{\mathbf{J}} \nonumber \\
&=& \sum_{\|\mathbf{J}\|_1=\vartheta}\frac{\vartheta}{\mathbf{J}!} \int_{0}^1 (1-w)^{\vartheta-1}  f^{(\mathbf{J})}(\mathbf{x}_0 + w(\mathbf{x}-\mathbf{x}_0))   (\mathbf{x}-\mathbf{x}_0)^{\mathbf{J}} \mathrm{d}w\nonumber \\
&& -\sum_{\|\mathbf{J}\|_1=\vartheta}\frac{\vartheta}{\mathbf{J}!} \int_{0}^1 (1-w)^{\vartheta-1}  f^{(\mathbf{J})}(\mathbf{x}_0) (\mathbf{x}-\mathbf{x}_0)^{\mathbf{J}} \mathrm{d}w  \nonumber \\
&=& \sum_{\|\mathbf{J}\|_1=\vartheta}\frac{\vartheta (\mathbf{x}-\mathbf{x}_0)^{\mathbf{J}}}{\mathbf{J}!} \int_{0}^1\left[ (1-w)^{\vartheta-1} f^{(\mathbf{J})}(\mathbf{x}_0 + w(\mathbf{x}-\mathbf{x}_0)) - f^{(\mathbf{J})}(\mathbf{x}_0)   (\mathbf{x}-\mathbf{x}_0)^{\mathbf{J}} \right] \mathrm{d}w,
\end{eqnarray*}
where the first equality follows from the definition of $p_{\vartheta}(\mathbf{x}\, |\, \mathbf{x}_0)$, and the second equality follows from using the integral form of the remainder of Taylor expansion for $ f(\mathbf{x}) -p_{\vartheta-1}(\mathbf{x}\, |\, \mathbf{x}_0)$ and the fact that $-\vartheta \int_0^1 (1-w)^{\vartheta-1}\mathrm{d}w =1$.

Therefore, we can further write

\begin{eqnarray*}
\|f(\mathbf{x}) -p_{\vartheta}(\mathbf{x}\, |\, \mathbf{x}_0)\|_{\infty}^{[-a,a]^r}&\le &  O(1) \| \| \mathbf{x}-\mathbf{x}_0\|^{\vartheta+s} \|_{\infty}^{[-a,a]^r} \nonumber \\
&=&  O(1)\|\| \mathbf{x}-\mathbf{x}_0\|^{p}\|_{\infty}^{[-a,a]^r},
\end{eqnarray*}
where the inequality follows from the property of $(p,\mathscr{C})$-smooth, and $p=\vartheta+s$ is by the definition of $(p,\mathscr{C})$-smooth.

Similarly, we can prove the results for $0<\|\pmb{\delta}\|_1\le\vartheta$. The proof is now completed. \hspace*{\fill}{$\blacksquare$}

\bigskip

\noindent \textbf{Proof of Lemma \ref{LM.A3}:}

First, we show that $R^k $ admits a few representations, which will facilitate the development. By \eqref{def.Tk2}, we note that for $k\ge 2$

\begin{eqnarray*}
T^k \circ T^{k-1}(x) &=&  \sigma \left(\frac{1}{2} T^{k-1}(x)  \right) - \sigma \left( T^{k-1}(x)-2^{1-2k} \right) \nonumber \\
&=&\overline{\mathbf{w}}^\top \left( \begin{array}{c}
 \sigma \left(\frac{1}{2} T^{k-1}(x) \right) \\
 \sigma \left( T^{k-1}(x)-2^{1-2k} \right)
\end{array}\right) \nonumber \\
&=&\overline{\mathbf{w}} ^\top \pmb{\sigma}_{\overline{\mathbf{v}}_k} (\underline{\mathbf{w}}\cdot T^{k-1}(x)),
\end{eqnarray*}
where $\overline{\mathbf{w}} = (1,-1)^\top$, $\underline{\mathbf{w}} = (\frac{1}{2},1)^\top$, and $\overline{\mathbf{v}}_k = (0, 2^{1-2k})^\top$. It then yields that for $k\ge 3$

\begin{eqnarray*}
T^k \circ T^{k-1} \circ T^{k-2}(x) &=&\overline{\mathbf{w}}^\top \pmb{\sigma}_{\overline{\mathbf{v}}_k}( \underline{\mathbf{w}} \cdot T^{k-1} \circ T^{k-2}(x))\nonumber \\
&=&\overline{\mathbf{w}}^\top \pmb{\sigma}_{\overline{\mathbf{v}}_k} (\underline{\mathbf{w}}  \cdot \overline{\mathbf{w}}^\top \pmb{\sigma}_{\overline{\mathbf{v}}_{k-1}} (\underline{\mathbf{w}} \cdot T^{k-2}(x))) \nonumber \\
&=& \overline{\mathbf{w}}^\top \pmb{\sigma}_{\overline{\mathbf{v}}_k} (\underline{\overline{\mathbf{w}}}  \cdot \pmb{\sigma}_{\overline{\mathbf{v}}_{k-1}} (\underline{\mathbf{w}} \cdot T^{k-2}(x))),
\end{eqnarray*}
where $\underline{\overline{\mathbf{w}}} = \underline{\mathbf{w}}  \cdot \overline{\mathbf{w}}^\top $. Finally, we can write for $k\ge 2$

\begin{eqnarray}\label{defRknew1}
R^k(x)   = \overline{\mathbf{w}}^\top \pmb{\sigma}_{\overline{\mathbf{v}}_k} \, \underline{\overline{\mathbf{w}}} \, \pmb{\sigma}_{\overline{\mathbf{v}}_{k-1}} \cdots  \,  \underline{\overline{\mathbf{w}}} \, \pmb{\sigma}_{\overline{\mathbf{v}}_{1}}(\underline{\mathbf{w}}\cdot x)    ,
\end{eqnarray} 
and 

\begin{eqnarray}\label{defRknew2}
R^k(x)  &=&  \overline{\mathbf{w}}^\top \pmb{\sigma}_{\overline{\mathbf{v}}_k}  (\underline{\mathbf{w}}\cdot R^{k-1}(x) )  .
\end{eqnarray} 

\medskip

We now start proving the main result of the lemma. For $m=1,2$, it is easy to see the validity of the lemma. Without loss of generality, we suppose that $m\ge 3$ in what follows. Note that by \eqref{defRknew1} and \eqref{defRknew2}, we can obtain that

\begin{eqnarray}\label{defRknew3}
\|\mathbf{R}_{m} (x)\|_1  &=& \overline{\mathbf{w}}^\top \pmb{\sigma}_{\overline{\mathbf{v}}_{m}}  (\underline{\mathbf{w}} \cdot R^{m-1}(x) )+\|\mathbf{R}_{m-1} (x)\|_1\nonumber \\
&=& \mathbf{w}_{\wout}^\top \left( \begin{matrix}
\pmb{\sigma}_{\overline{\mathbf{v}}_{m}}  (\underline{\mathbf{w}} \cdot R^{m-1}(x) )\\
\|\mathbf{R}_{m-1} (x)\|_1
\end{matrix}\right) \nonumber \\
&=&\mathbf{w}_{\wout}^\top  \pmb{\sigma}_{\widetilde{\mathbf{v}}_{m}}  \left( \begin{matrix}
 \underline{\mathbf{w}} \cdot R^{m-1}(x) \\
\|\mathbf{R}_{m-1} (x)\|_1
\end{matrix}\right) \nonumber \\
&=& \mathbf{w}_{\wout}^\top   \pmb{\sigma}_{\widetilde{\mathbf{v}}_{m}}   \widetilde{\mathbf{w}}   \left( \begin{matrix}
\pmb{\sigma}_{\overline{\mathbf{v}}_{m-1}}  (\underline{\mathbf{w}}\cdot R^{m-2}(x) ) \\ 
\|\mathbf{R}_{m-2} (x)\|_1
\end{matrix}\right)  \nonumber \\
&=&  \mathbf{w}_{\wout}^\top   \pmb{\sigma}_{\widetilde{\mathbf{v}}_{m}}    \widetilde{\mathbf{w}}  \, \pmb{\sigma}_{\widetilde{\mathbf{v}}_{m-1}}  \left( \begin{matrix}
\underline{\mathbf{w}}\cdot R^{m-2}(x)  \\ 
\|\mathbf{R}_{m-2} (x)\|_1
\end{matrix}\right)  \nonumber \\
&=&  \mathbf{w}_{\wout}^\top  \pmb{\sigma}_{\widetilde{\mathbf{v}}_{m}}  \widetilde{\mathbf{w}}  \, \pmb{\sigma}_{\widetilde{\mathbf{v}}_{m-1}}\cdots  \widetilde{\mathbf{w}} \, \pmb{\sigma}_{\widetilde{\mathbf{v}}_{1}} (  \mathbf{w}_{\win}\cdot x),
\end{eqnarray}
where $\widetilde{\mathbf{v}}_{k} = ( \overline{\mathbf{v}}_{k}^\top,0)^\top$ for $k\in [m]$, $\mathbf{w}_{\wout}=(\overline{\mathbf{w}}^\top, 1)^\top$, $ \mathbf{w}_{\win}=(\underline{\mathbf{w}}^\top ,0)^\top$, 

\begin{eqnarray*}
\widetilde{\mathbf{w}}  = \begin{pmatrix}
\underline{\mathbf{w}}  \cdot \overline{\mathbf{w}}^\top & \mathbf{0}_2\\
\overline{\mathbf{w}}^\top & 1
\end{pmatrix} =\begin{pmatrix}
\underline{\mathbf{w}} & \mathbf{0}_2\\
1 & 1
\end{pmatrix} \cdot \begin{pmatrix}
\overline{\mathbf{w}}^\top & 0\\
\mathbf{0}_2^\top & 1
\end{pmatrix},
\end{eqnarray*}
and the last line of \eqref{defRknew3} follows from repeating the procedure from the third equality to the fifth equality.

The proof is now completed. \hspace*{\fill}{$\blacksquare$}

\bigskip

\noindent \textbf{Proof of Lemma \ref{LM.A4}:}

(1). Recall that we let $g(x) = x(1-x)$ for notational simplicity in the body of this lemma. We first show that

\begin{eqnarray}\label{L11}
\text{$R^k(\cdot)$ is piecewise linear on $ [\ell\cdot 2^{-k}, (\ell+1)\cdot 2^{-k} ]$ for $\ell =0\cup[2^k-1]$}
\end{eqnarray}
with endpoints 

\begin{eqnarray}\label{L12}
\left\{ \begin{array}{ll}
R^k \left( \frac{\ell}{2^k}\right) =2^{-2k} & \text{$\ell$ is odd} \\
R^k \left( \frac{\ell}{2^k}\right) =0 & \text{$\ell$ is even}
\end{array}\right. .
\end{eqnarray}
In view of Figure \ref{FigRk}, the argument of \eqref{L11} can easily be proved by using induction, so we omit the details here.

\medskip

In what follows, we show that for $\forall m\ge 1$,

\begin{eqnarray}\label{A4}
g(\ell\cdot 2^{-m}) =\|\mathbf{R}_m(\ell \cdot 2^{-m})\|_1 \quad \text{for}\quad \ell  \in 0 \cup[2^m]
\end{eqnarray}
using induction over $m$. For $m=1$, we have

\begin{eqnarray*}
\left\{ \begin{array}{ll}
g(\ell \cdot 2^{-1})=0 & \ell \in \{0, 2\}, \\
g(\ell \cdot 2^{-1})=2^{-2} & \ell =1 .
\end{array}\right.
\end{eqnarray*}
Apparently, \eqref{A4} holds, which can also be verified in view of Figure \ref{FigRk}. 

For the inductive step, we now suppose that the claim holds for $m$, and consider two cases: (1). $\ell$ being even, and (2). $\ell$ being odd respectively.  We start with Case (1). If $\ell$ is even, we have 

\begin{eqnarray}\label{A5}
R^{m+1} (\ell \cdot 2^{-m-1}) =0
\end{eqnarray}
according to \eqref{L12}. Therefore,

\begin{eqnarray*}
g(\ell \cdot 2^{-m-1})&=&g(\ell/2\cdot 2^{-m}) \nonumber \\
&=&\|\mathbf{R}_m(\ell/2 \cdot 2^{-m}) \|_1\nonumber \\
&=&\|\mathbf{R}_m(\ell \cdot 2^{-m-1})\|_1 \nonumber \\
&=& \|\mathbf{R}_{m+1}(\ell \cdot 2^{-m-1})\|_1,
\end{eqnarray*}
where the second equality follows from \eqref{A4} and $\ell$ being even, and the fourth equality follows from \eqref{A5}.

It thus remains to consider Case (2), i.e., $\ell$ being odd. By \eqref{L11}, it is not hard to see that given $\ell$ being odd, $x\mapsto \| \mathbf{R}_m(x)\|_1$  is linear on 

\begin{eqnarray}\label{deflinear}
&&[(\ell-1)/2\cdot 2^{-m}, (\ell+1)/2 \cdot 2^{-m}]\nonumber \\
&=&[(\ell-1)\cdot 2^{-m-1}, (\ell+1) \cdot 2^{-m-1}].
\end{eqnarray} 

In addition, by the definition of $g(x)$, for $\forall w$

\begin{eqnarray*}
g(x) - \frac{g(x+w) +g(x-w)}{2} =w^2.
\end{eqnarray*}
Therefore, for $x=\ell \cdot 2^{-m-1}$ and $w= 2^{-m-1}$, we have

\begin{eqnarray}\label{A6}
 2^{-2m-2} &=&g(\ell \cdot 2^{-m-1}) -\frac{g(\ell \cdot 2^{-m-1}+2^{-m-1}) +g(\ell \cdot 2^{-m-1}-2^{-m-1})}{2} \nonumber \\
 &=&g(\ell \cdot 2^{-m-1}) -\frac{g((\ell+1)/2 \cdot 2^{-m} ) +g((\ell-1)/2 \cdot 2^{-m} )}{2} \nonumber \\
 &=&g(\ell \cdot 2^{-m-1}) -\frac{1}{2}\left(\|\mathbf{R}_m((\ell+1)/2 \cdot 2^{-m})\|_1+\|\mathbf{R}_m((\ell-1)/2 \cdot 2^{-m})\|_1 \right) \nonumber \\
 &=&g(\ell \cdot 2^{-m-1}) -\frac{1}{2}\left(\|\mathbf{R}_m((\ell+1)  \cdot 2^{-m-1})\|_1+\|\mathbf{R}_m((\ell-1)  \cdot 2^{-m-1})\|_1 \right) \nonumber \\
 &=&g(\ell \cdot 2^{-m-1}) - \|\mathbf{R}_m(\ell \cdot 2^{-m-1})\|_1,
\end{eqnarray}
where the third equality follows from \eqref{A4}, and the fifth equality follows from the fact that $x\mapsto \|\mathbf{R}_m(x)\|_1$  is linear on $[(\ell-1)\cdot 2^{-m-1}, (\ell+1)\cdot 2^{-m-1}]$ as mentioned in \eqref{deflinear}. In connection with the fact $R^{m+1}(\ell \cdot 2^{-m-1}) =2^{-2m-2}$ by \eqref{L12}, \eqref{A6} yields that

\begin{eqnarray*}
g(\ell \cdot 2^{-m-1}) = 2^{-2m-2} + \|\mathbf{R}_m(\ell \cdot 2^{-m-1}) \|_1= \|\mathbf{R}_{m+1}(\ell \cdot 2^{-m-1})\|_1.
\end{eqnarray*}
Putting the development of Case (1) and Case (2) together completes the inductive step.

\medskip

So far we have proved that $R^*(x)\equiv \|\mathbf{R}_m(x)\|_1$ interpolates $g(x)$ at the points $\ell \cdot 2^{-m}$ and is linear on the intervals $[\ell \cdot 2^{-m},(\ell+1)\cdot 2^{-m}]$. Therefore, we have for $x\in [\ell \cdot 2^{-m},(\ell+1)\cdot 2^{-m}]$

\begin{eqnarray*}
\frac{R^*(x) - R^*(\ell \cdot 2^{-m})}{x - \ell \cdot 2^{-m}} = \frac{R^*((\ell+1) \cdot 2^{-m})-R^*(x) }{(\ell+1)\cdot 2^{-m}-x} ,
\end{eqnarray*}
which yields that

\begin{eqnarray*}
R^*(x) &=&(2^m x-\ell) R^*((\ell+1) \cdot  2^{-m}) + (\ell+1-2^m x)R^*(\ell \cdot 2^{-m})\nonumber \\
&=&(2^m x-\ell) g((\ell+1) \cdot 2^{-m}) + (\ell+1-2^m x)g(\ell \cdot 2^{-m}).
\end{eqnarray*}
Thus, for any $x$, there exists an $\ell$ such that

\begin{eqnarray}\label{rateBP1}
\left|g(x) -\|\mathbf{R}_m(x) \|_1 \right| &=& \left|g(x) -(2^m x-\ell) g((\ell+1)\cdot 2^{-m}) - (\ell+1-2^m x)g(\ell \cdot 2^{-m}) \right| \nonumber \\
&=&  \left|g(x) -g(\ell \cdot 2^{-m})-(2^m x-\ell) [g((\ell+1) \cdot 2^{-m}) - g(\ell \cdot 2^{-m})] \right| \nonumber \\
&\le &  |x-\ell \cdot 2^{-m}|+2^{m}|( x-\ell \cdot 2^{-m}) [g((\ell+1) \cdot 2^{-m}) - g(\ell \cdot 2^{-m})]  | \nonumber \\
&\le &  |x-\ell \cdot 2^{-m}|+2^{m}|x-\ell \cdot 2^{-m} | 2^{-m}\nonumber \\
&=&2 |x-\ell\cdot 2^{-m}|\le 2^{-m},
\end{eqnarray}
where the second inequality follows from the fact that $g$ is Lipschitz continuous with Lipschitz constant one, and the last steps follows from assuming $x$ is closer to $\ell \cdot 2^{-m}$. If $x$ is closer to $(\ell+1) 2^{-m}$, one can easily modify the above step to ensure  $ |g(x) -\|\mathbf{R}_m(x) \|_1 | \le 2^{-m}$ as well.

In connection with Lemma \ref{LM.A3}, the proof of the first result is now completed.

\medskip

(2). Next, we consider the derivative of $\|\mathbf{R}_{m} (x)\|_1$, and recall that we have shown that $\|\mathbf{R}_{m} (x)\|_1$ is piecewise linear on $[\ell \cdot 2^{-m},(\ell+1)\cdot 2^{-m}]$ for $\ell =0\cup[2^m-1]$. Therefore, we now partition $\|\mathbf{R}_{m} (x)\|_1$ using the following intervals:
 
\begin{eqnarray*}
&&[\ell \cdot 2^{-m},(\ell+1)\cdot 2^{-m})  \text{ for }\ell =0\cup[2^m-2],\nonumber \\
&&[\ell \cdot 2^{-m},(\ell+1)\cdot 2^{-m}] \text{ for }\ell = 2^m-1.
\end{eqnarray*}

Similarly, we define $\frac{\partial (\|\mathbf{R}_{m} (x)\|_1)}{\partial x}$ on the same set of intervals.
 
Note that

\begin{eqnarray}\label{rateBP2}
\frac{\partial (\|\mathbf{R}_{m} (x)\|_1)}{\partial x} &=&  \frac{\partial g(x)}{\partial x} + \frac{\partial (\|\mathbf{R}_{m} (x)\|_1-g(x))}{\partial x} .
\end{eqnarray}
In what follows, we focus on $\frac{\partial (\|\mathbf{R}_{m} (x)\|_1-g(x))}{\partial x}$ below. As in \eqref{rateBP1}, for any $x$, there exists an $\ell$ so that we can write

\begin{eqnarray}\label{rateBP3}
&&\left| \frac{\partial (\|\mathbf{R}_{m} (x)\|_1-g(x))}{\partial x}\right| \nonumber \\
&=&\left| \frac{\partial}{\partial x}\left( g(\ell \cdot 2^{-m})+(2^m x-\ell) [g((\ell+1) \cdot 2^{-m}) - g(\ell \cdot 2^{-m})]-g(x) \right) \right|\nonumber \\
&=&\left| 2^m   [g((\ell+1) \cdot 2^{-m}) - g(\ell \cdot 2^{-m})]  - g^{(1)}(x) \right| \nonumber \\
&=&\left| 2^m [(\ell+1) \cdot 2^{-m} -\ell \cdot 2^{-m}]  \cdot g^{(1)}(\tilde{x} )  - g^{(1)}(x) \right|\nonumber \\
&=&\left| g^{(1)}(\tilde{x} )  - g^{(1)}(x) \right| \le 2^{-m+1},
\end{eqnarray}
where the first equality follows from the second equality of \eqref{rateBP1}, the second equality follows from Mean Value theorem with $\tilde{x}$ in between $(\ell+1) \cdot 2^{-m} $ and $\ell \cdot 2^{-m}$, and the last inequality follows from both $\tilde{x}$ and $x$ are in between $(\ell+1) \cdot 2^{-m} $ and $\ell \cdot 2^{-m}$ and the fact that $g^{(1)}(x)=1-2x$.

By \eqref{rateBP2} and \eqref{rateBP3}, we conclude that

\begin{eqnarray*}
\left\|\frac{\partial (\|\mathbf{R}_{m} (x)\|_1)}{\partial x} - \frac{\partial g(x)}{\partial x}\right\|_{\infty}^{[0,1]}\le 2^{-m+1}.
\end{eqnarray*}
The proof of the second result is now completed. \hspace*{\fill}{$\blacksquare$}

\bigskip

\noindent \textbf{Proof of Lemma \ref{LM1}:}

Let $g(x)=x(1-x)$ for notational simplicity. Simple algebra shows that

\begin{eqnarray}\label{gxy}
g\left( \frac{x-y+1}{2} \right) - g\left( \frac{x+y}{2} \right) + \frac{x+y}{2} - \frac{1}{4} =xy.
\end{eqnarray}
Additionally, let $H:[0,1]\mapsto [0,\infty)$ be a generic non-negative function.

\medskip

(1). We now start our investigation. First, we note that by the development of Lemma \ref{LM.A3}, it is easy to see that for $\forall z_1, z_2\in [0,1]$

\begin{eqnarray}\label{defMNN}
&&\|\mathbf{R}_{m+1}(z_1)\|_1  +H(z_2) \nonumber \\
&=&  \mathbf{w}_{\wout}^\top \pmb{\sigma}_{\widetilde{\mathbf{v}}_{m+1}} \widetilde{\mathbf{w}}\, \pmb{\sigma}_{\widetilde{\mathbf{v}}_{m}}\cdots  \, \widetilde{\mathbf{w}}\, \pmb{\sigma}_{\widetilde{\mathbf{v}}_{1}} (\mathbf{z} ) \nonumber \\
&=& \nn (\mathbf{z}  \, | \,  \mathbf{W}_{\mathbf{R}_m}),
\end{eqnarray}
where $\mathbf{z} =  (\frac{1}{2} z_1, z_1, H(z_2))^\top$, and $\nn (\cdot \, | \,  \mathbf{W}_{\mathbf{R}_m})$ is defined in Lemma \ref{LM.A3}.

\medskip

Next, we show that there is a DNN with $m+3$ hidden layers that computes the function

\begin{eqnarray}\label{defNNxy}
&&(x,y)\mapsto  \sigma\left(\big\|\mathbf{R}_{m+1}\big( \frac{x-y+1}{2} \big)\big\|_1 -\big\|\mathbf{R}_{m+1}\big( \frac{x+y}{2}\big) \big\|_1+ \frac{x+y+2^{-m}}{2} - \frac{1}{4} \right)\wedge 1 . \nonumber \\ 
\end{eqnarray}
In order to do so, we apply \eqref{defMNN} by replacing $\mathbf{z}$ with

\begin{eqnarray*}
&& \widetilde{\pmb{\ell}}_{x,y}^m=\left( \frac{x-y+1}{4},  \frac{x-y+1}{2}, \frac{x+y+2^{-m}}{2} \right)^\top ,\nonumber
\end{eqnarray*}
and 
\begin{eqnarray*}
&&\overline{\pmb{\ell}}_{x,y}=\left( \frac{x+y}{4}, \frac{x+y}{2}, \frac{1}{4} \right)^\top\nonumber
\end{eqnarray*}
respectively. It then gives a combination of two parallel DNNs:

\begin{eqnarray}\label{cbNN1}
&&\begin{pmatrix}
 \nn (\widetilde{\pmb{\ell}}_{x,y}^m \, | \,  \mathbf{W}_{\mathbf{R}_m}) \\
 \nn (\overline{\pmb{\ell}}_{x,y} \, | \,  \mathbf{W}_{\mathbf{R}_m})
\end{pmatrix} = \begin{pmatrix}
\mathbf{w}_{\wout}^\top \pmb{\sigma}_{\widetilde{\mathbf{v}}_{m+1}} \widetilde{\mathbf{w}}\, \pmb{\sigma}_{\widetilde{\mathbf{v}}_{m}}\cdots \, \widetilde{\mathbf{w}}\, \pmb{\sigma}_{\widetilde{\mathbf{v}}_{1}} (\widetilde{\pmb{\ell}}_{x,y}^m) \\
\mathbf{w}_{\wout}^\top \pmb{\sigma}_{\widetilde{\mathbf{v}}_{m+1}} \widetilde{\mathbf{w}}\, \pmb{\sigma}_{\widetilde{\mathbf{v}}_{m}}\cdots  \,\widetilde{\mathbf{w}}\, \pmb{\sigma}_{\widetilde{\mathbf{v}}_{1}} (\overline{\pmb{\ell}}_{x,y})
\end{pmatrix} \nonumber \\
&=& (\mathbf{I}_2\otimes \mathbf{w}_{\wout}^\top ) \, \pmb{\sigma}_{\mathbf{1}_2\otimes \widetilde{\mathbf{v}}_{m+1}} (\mathbf{I}_2\otimes \widetilde{\mathbf{w}}) \, \pmb{\sigma}_{\mathbf{1}_2\otimes\widetilde{\mathbf{v}}_{m}}\cdots   (\mathbf{I}_2\otimes  \widetilde{\mathbf{w}}) \, \pmb{\sigma}_{\mathbf{1}_2\otimes\widetilde{\mathbf{v}}_{1}} (\pmb{\ell}^m_{x,y})\nonumber \\
&:=& \widetilde{\mathbf{w}}_{\wout}^{\top} \, \pmb{\sigma}_{ \pmb{\nu}_{m+1}} \widetilde{\mathbf{w}}_{\wmid} \, \pmb{\sigma}_{\pmb{\nu}_{m}}\cdots  \, \widetilde{\mathbf{w}}_{\wmid} \, \pmb{\sigma}_{\pmb{\nu}_{1}} (\pmb{\ell}^m_{x,y}),
\end{eqnarray}
where $\pmb{\ell}^m_{x,y}$ has been defined in the body of this lemma, and

\begin{eqnarray*}
&&\widetilde{\mathbf{w}}_{\wout} =\mathbf{I}_2\otimes \mathbf{w}_{\wout},\nonumber \\
&&\widetilde{\mathbf{w}}_{\wmid}=\mathbf{I}_2\otimes \widetilde{\mathbf{w}},\nonumber \\
&&\pmb{\nu}_k= \mathbf{1}_2\otimes \widetilde{\mathbf{v}}_{k}\text{ for }k\in [m+1].
\end{eqnarray*}

It is easy to see that \eqref{cbNN1} admits the following mapping:

\begin{eqnarray}\label{defMNN2}
(x,y)\mapsto \begin{pmatrix}
\|\mathbf{R}_{m+1}\left( (x-y+1)/2 \right)\|_1 + \frac{x+y+2^{-m}}{2} \\
\|\mathbf{R}_{m+1}\left( (x+y)/2\right) \|_1+ \frac{1}{4}
\end{pmatrix}  .
\end{eqnarray}
Below, we further apply to the output of \eqref{defMNN2} the two hidden layer network

\begin{eqnarray}\label{defMNN3}
(u, v)\mapsto \sigma_{-1}( -\sigma_{-1}(-\overline{\mathbf{w}}^\top\mathbf{v}^*)) &=&\sigma (1 -\sigma(1- (u-v)) )  \nonumber \\
&=&\sigma (1 -\max(1- u+v, 0)) \nonumber \\
&=&\sigma (1 -1- \max(- u+v, -1)) \nonumber \\
&=&\sigma(\min(u-v, 1))\nonumber \\
&=& \sigma(u-v) \wedge 1,
\end{eqnarray}
where $\overline{\mathbf{w}}=(1,-1)^\top$, $\mathbf{v}^* =(u, v)^\top$, and the above calculation should be straightforward. 

The network formed by \eqref{defMNN2} and \eqref{defMNN3} must compute \eqref{defNNxy}, of which the output always belongs to $[0,1]$ in view of the right hand side of \eqref{defMNN3}. Specifically, the DNN has the following representation:

\begin{eqnarray}\label{defMNN4}
 \sigma_{-1}(-\sigma_{-1}\, \overline{ \pmb{\omega}} \,  \pmb{\sigma}_{ \pmb{\nu}_{m+1}} \widetilde{\mathbf{w}}_{\wmid} \, \pmb{\sigma}_{\pmb{\nu}_{m}}\cdots  \, \widetilde{\mathbf{w}}_{\wmid} \, \pmb{\sigma}_{\pmb{\nu}_{1}} (\pmb{\ell}^{\star}_{x,y})):=\nn (\pmb{\ell}^m_{x,y} \, |\, \mathbf{W}^{\star}_{m+3} ),
\end{eqnarray}
where  

\begin{eqnarray*}
\overline{ \pmb{\omega}} =-\overline{\mathbf{w}}^\top \widetilde{\mathbf{w}}_{\wout}^{\top}=-(1,-1)  (\mathbf{I}_2\otimes \mathbf{w}_{\wout}^\top) =(-\mathbf{w}_{\wout}^\top,  \mathbf{w}_{\wout}^\top ).
\end{eqnarray*}

Moreover, by direct calculation, we have $\nn (\pmb{\ell}^m_{x,y} \, |\, \mathbf{W}^{\star}_{m+3} )=1$ at $(x,y)=(1,1)$. The first result then follows.

\medskip

(2). Below, we let again $g(w) =w(1-w)$, and consider $\nn (\pmb{\ell}^m_{x,y} \, |\, \mathbf{W}^{\star}_{m+3} )$ defined in \eqref{defMNN4} on the following set:

\begin{eqnarray*}
(x,y)\in[0,1-2^{-m}]\times [0,1]:=\mathbb{C}_m.
\end{eqnarray*}

For $(x,y)\in\mathbb{C}_m$, we write

\begin{eqnarray}\label{rateBP5}
&& \|\mathbf{R}_{m+1}\left( (x-y+1)/2 \right)\|_1 -\|\mathbf{R}_{m+1}\left( (x+y)/2\right) \|_1+ \frac{x+y+2^{-m}}{2} - \frac{1}{4}    \nonumber \\
&\le &   g((x-y+1)/2) -\|\mathbf{R}_{m+1}\left( (x+y)/2\right) \|_1+ \frac{x+y+2^{-m}}{2} - \frac{1}{4}   \nonumber \\
&=& \frac{-(x-y)^2+2x+2y}{4} -\|\mathbf{R}_{m+1}\left( (x+y)/2\right) \|_1 +2^{-m-1} \nonumber \\
&\le & \frac{x+y+2^{-m} }{2} \le 1 ,
\end{eqnarray}
where  the first inequality follows from the fact that $\|\mathbf{R}_{m+1}\left(w \right)\|_1\le g(w)$ (see Figure \ref{FigRm} for example), the first equality follows from the fact that

\begin{eqnarray*}
 g((x-y+1)/2)  + \frac{x+y}{2} - \frac{1}{4}=\frac{-(x-y)^2+2x+2y}{4},
\end{eqnarray*}
and the third inequality follows from the fact that for $\forall (x,y)\in \mathbb{C}_m$

\begin{eqnarray*}
\frac{x+y+2^{-m} }{2} \le \frac{1-2^{-m}+1+2^{-m}}{2}=1.
\end{eqnarray*} 
Additionally,

\begin{eqnarray}\label{rateBP52}
&&\|\mathbf{R}_{m+1}\left( (x-y+1)/2 \right)\|_1 -\|\mathbf{R}_{m+1}\left( (x+y)/2\right) \|_1 + \frac{x+y+2^{-m}}{2} - \frac{1}{4} \nonumber \\
&\ge &  g((x-y+1)/2)-2^{-m-1} -\|\mathbf{R}_{m+1}\left( (x+y)/2\right) \|_1  + \frac{x+y+2^{-m}}{2} - \frac{1}{4} \nonumber \\
&\ge &  g((x-y+1)/2) -2^{-m-1} -g((x+y)/2) + \frac{x+y+2^{-m}}{2} - \frac{1}{4} \nonumber \\
&=&xy-2^{-m-1}+2^{-m-1}\geq 0,
\end{eqnarray}
where the first inequality follows  from using Lemma \ref{LM.A4}, the second inequality follows from the fact that $\|\mathbf{R}_{m+1}\left(w \right)\|_1\le g(w)$, the equality follows from \eqref{gxy}, and the last inequality follows from $xy\ge 0$ on $\mathbb{C}_m$.

By \eqref{defMNN3}, \eqref{rateBP5}, and \eqref{rateBP52}, we obtain that on $\mathbb{C}_m$,

\begin{eqnarray}\label{BP6}
&&\nn (\pmb{\ell}^m_{x,y} \, |\, \mathbf{W}^{\star}_{m+3} )\nonumber \\
&=&\|\mathbf{R}_{m+1}\left( (x-y+1)/2 \right)\|_1 -\|\mathbf{R}_{m+1}\left( (x+y)/2\right) \|_1+ \frac{x+y+2^{-m}}{2} - \frac{1}{4},
\end{eqnarray}
which falls in $[0,1]$ uniformly in $(x,y)\in\mathbb{C}_m$.  

Further, we can obtain that

\begin{eqnarray}\label{rateBP53}
&&\|\mathbf{R}_{m+1}\left( (x-y+1)/2 \right)\|_1 -\|\mathbf{R}_{m+1}\left( (x+y)/2\right) \|_1+ \frac{x+y+2^{-m}}{2} - \frac{1}{4}    \nonumber \\
&\le & g( (x-y+1)/2 )-g((x+y)/2)+2^{-m-1}+ \frac{x+y+2^{-m}}{2} - \frac{1}{4} \nonumber \\
&= &xy+2^{-m},
\end{eqnarray}
where the inequality follows from using Lemma \ref{LM.A4} and the fact that $\|\mathbf{R}_{m+1}\left(w \right)\|_1\le g(w)$, and the equality follows from \eqref{gxy}. 

By \eqref{rateBP52}, \eqref{BP6} and \eqref{rateBP53}, we obtain that on $\mathbb{C}_m$,

\begin{eqnarray*}
0\le \nn (\pmb{\ell}^m_{x,y} \, |\, \mathbf{W}^{\star}_{m+3} )-xy\le 2^{-m},
\end{eqnarray*}
which is the second result of this lemma.

\medskip

(3). First, note that by Lemma \ref{LM1}, $\nn (\pmb{\ell}^m_{x,y} \, |\, \mathbf{W}^{\star}_{m+3} )$ is piecewise linear on $\mathbb{C}_m$. By \eqref{BP6}, we can write

\begin{eqnarray}
&&\left\|\frac{\partial }{\partial x}  [\nn (\pmb{\ell}^m_{x,y} \, |\, \mathbf{W}^{\star}_{m+1} ) -xy] \right\|_{\infty}^{\mathbb{C}_m} \nonumber \\
&=&\Big\|\frac{\partial }{\partial x} \Big[ \|\mathbf{R}_{m+1}\left( (x-y+1)/2 \right)\|_1 -\|\mathbf{R}_{m+1}\left( (x+y)/2\right) \|_1+ \frac{x+y+2^{-m}}{2} - \frac{1}{4} \nonumber \\
&&-xy\Big] \Big\|_{\infty}^{\mathbb{C}_m} \nonumber \\
&=&\Big\|\frac{\partial }{\partial x}  \Big[ \|\mathbf{R}_{m+1}\left( (x-y+1)/2 \right)\|_1 -\|\mathbf{R}_{m+1}\left( (x+y)/2\right) \|_1\nonumber \\
&& -g((x-y+1)/2)+g((x+y)/2)\Big] \Big\|_{\infty}^{\mathbb{C}_m}\nonumber \\
&\le &\Big\|\frac{\partial }{\partial x}  \Big[ \|\mathbf{R}_{m+1}\left( (x-y+1)/2 \right)\|_1 -g((x-y+1)/2) \Big] \Big\|_{\infty}^{\mathbb{C}_m} \nonumber \\
&& + \Big\|\frac{\partial }{\partial x}  \Big[\mathbf{R}_{m+1}\left( (x+y)/2\right) \|_1- g((x+y)/2)\Big] \Big\|_{\infty}^{\mathbb{C}_m}\nonumber \\
&\le &2\cdot 2^{-m-1}\cdot \frac{1}{2}=2^{-m-1},
\end{eqnarray}
where the first equality follows \eqref{BP6}, the second equality follows from \eqref{gxy}, and the second inequality follows from Lemma \ref{LM.A4}.2 and the chain rule.

The proof is now completed. \hspace*{\fill}{$\blacksquare$}

\bigskip

\noindent \textbf{Proof of Lemma \ref{LM.A6}:}

Before proceeding further, recall that we have defined $\nn (\pmb{\ell}^m_{x,y} \, |\, \mathbf{W}^{\star}_{m+3} ) $ in Lemma \ref{LM1}, and have defined $\pmb{\ell}_{\mathbf{x}\, | \, {\mathbf{1}_r}} $ in \eqref{defx1}.  By Definition \ref{Def2} about a pair-wise HDNN, we define

\begin{eqnarray}\label{defx1r}
\mathcal{N}_{\pmb{\ell}^m}  (\pmb{\ell}_{\mathbf{x}\, | \, {\mathbf{1}_r}} \, |\, \mathbf{W}^{\star}_{m+3})  , 
\end{eqnarray}
which will be repeatedly used below.

By the construction of \eqref{defx1r}, we need to invoke Lemma \ref{LM1} $ \lceil \log_2 r\rceil -1$ times for $r\ge 2$, so we require

\begin{eqnarray}\label{defh}
&&h+(\lceil\log_2 r\rceil -1)\cdot 2^{-m}\le 1-2^{-m}\nonumber \\
&\Rightarrow & h\le 1 -\lceil\log_2 r\rceil \cdot 2^{-m}.
\end{eqnarray}
As a consequence, the output still falls in the range $[0, 1-2^{-m}]$ after using Lemma \ref{LM1}.2 $ \lceil \log_2 r\rceil -1$ times. 

Just in the proof of this lemma, when no misunderstanding arises below, we write 

\begin{eqnarray}\label{defX2}
\pmb{\ell}_{\mathbf{x}\, | \,{\mathbf{1}_r}}:=\pmb{\ell}
\end{eqnarray}
for notational simplicity. Accordingly, for any two given positive integers $j_1\le j_2$, we let $\pmb{\ell}_{j_1:j_2}$ be a column vector including the  elements  from  $j_1^{th}$ position to $j_2^{th}$ position  of $\pmb{\ell}$. When $j_1=j_2$, we write $\pmb{\ell}_{j_1}$ for simplicity.  Also, it is worth mentioning that

\begin{eqnarray}\label{def11}
\nn (\pmb{\ell}_{1,1}^m \, |\, \mathbf{W}^{\star}_{m+3}  ) =1
\end{eqnarray}
by Lemma \ref{LM1}.1.

\medskip

We are now ready to start the investigation. 

(1). The first result follows immediately by Lemma \ref{LM1}.1 and the construction of $\mathcal{N}_{\pmb{\ell}^m}  (\pmb{\ell}_{\mathbf{x}\, | \, {\mathbf{1}_r}} \, |\, \mathbf{W}^{\star}_{m+3})$ according to Definition \ref{Def2}.

\medskip

(2). Note that if $(a,b)\in \mathbb{C}^m$ and $(c,d)\in \mathbb{C}^m$, we have

\begin{eqnarray*}
&&\nn (\pmb{\ell}_{a,b}^m \, |\, \mathbf{W}^{\star}_{m+3} )-cd  \nonumber \\
&= & \nn (\pmb{\ell}_{a,b}^m \, |\, \mathbf{W}^{\star}_{m+3}  ) -ab  +  ab-cd \nonumber \\
&= & \nn (\pmb{\ell}_{a,b}^m \, |\, \mathbf{W}^{\star}_{m+3}  ) -ab  +  b \cdot (a-c)+c\cdot (b-d).
\end{eqnarray*}
Using Lemma \ref{LM1}.2 and the facts that $(a,b)\in \mathbb{C}^m$ and $(c,d)\in \mathbb{C}^m$, we obtain that

\begin{eqnarray}\label{defineq1}
0\le \nn (\pmb{\ell}_{a,b}^m \, |\, \mathbf{W}^{\star}_{m+3} )-cd \le 2^{-m} +b \cdot (a-c)+c\cdot (b-d).
\end{eqnarray}

We are now ready to approximate $\mathbf{x}^{\mathbf{1}_r}$. For the first $2^2$ elements of $\pmb{\ell}^\dag$, we have

\begin{eqnarray*} 
&& \mathcal{N}_{\pmb{\ell}^m} (\pmb{\ell}_{1:4}  \, |\, \mathbf{W}^{\star}_{m+3}) - \pmb{\ell}_{1:4}^{\mathbf{1}_4}  \nonumber \\
&= & \mathcal{N}_{\pmb{\ell}^m} (\pmb{\ell}_{1:4} \, |\, \mathbf{W}^{\star}_{m+3}) -\nn (\pmb{\ell}^m_{\pmb{\ell}_1, \pmb{\ell}_2 } \, |\,  \mathbf{W}^{\star}_{m+3})\nn (\pmb{\ell}^m_{\pmb{\ell}_3 , \pmb{\ell}_4 } \, |\,  \mathbf{W}^{\star}_{m+3})  \nonumber \\
&&+ \nn (\pmb{\ell}^m_{\pmb{\ell}_1 , \pmb{\ell}_2 } \, |\,  \mathbf{W}^{\star}_{m+3})\nn (\pmb{\ell}^m_{\pmb{\ell}_3 , \pmb{\ell}_4 } \, |\,  \mathbf{W}^{\star}_{m+3}) -\pmb{\ell}_{1:4}^{\mathbf{1}_4} .
\end{eqnarray*}
Using \eqref{defineq1} and Lemma \ref{LM1}.2, we can further obtain that

\begin{eqnarray}\label{eqx4}
0 \le  \mathcal{N}_{\pmb{\ell}^m} (\pmb{\ell}_{1:4}  \, |\, \mathbf{W}^{\star}_{m+3}) - \pmb{\ell}_{1:4}^{\mathbf{1}_4} \le 3\cdot 2^{-m}.
\end{eqnarray}
Similarly, for the first $2^3$ elements of $\pmb{\ell}$, we have

\begin{eqnarray}
0\le  \mathcal{N}_{\pmb{\ell}^m} (\pmb{\ell}_{1:8}  \, |\, \mathbf{W}^{\star}_{m+3}) - \pmb{\ell}_{1:8}^{\mathbf{1}_8} \le 3^2\cdot 2^{-m},
\end{eqnarray}
where we have used \eqref{defh}, \eqref{eqx4}, and Lemma \ref{LM1}.2.

We can keep doing this for the first $2^\zeta$ elements of $\pmb{\ell}$ with $\zeta \ge 2$. By induction we obtain that for $\pmb{\ell}$ defined in \eqref{defX2}

\begin{eqnarray*}
0\le \mathcal{N}_{\pmb{\ell}^m} (\pmb{\ell} \, |\, \mathbf{W}^{\star}_{m+3})   - \pmb{\ell}^{ \mathbf{1}_{2^{\lceil \log_2 r\rceil}}} =  \mathcal{N}_{\pmb{\ell}^m} (\pmb{\ell}\, |\, \mathbf{W}^{\star}_{m+3})   -\mathbf{x}^{\mathbf{1}_r}  \le 3^{\lceil \log_2 r\rceil-1} 2^{-m},
\end{eqnarray*}
where $\pmb{\ell}^{\mathbf{1}_{2^{\lceil \log_2 r\rceil}}}=\mathbf{x}^{\mathbf{1}_r}$ is obvious by the construction. Thus, the second result follows.

\medskip

(3). Consider $\frac{\partial}{\partial x_i} \mathcal{N}_{\pmb{\ell}^m} (\pmb{\ell} \, |\, \mathbf{W}^{\star}_{m+3})$, where $x_i$ stands for the $i^{th}$ element of $\mathbf{x}$. Note that

\begin{eqnarray*}
\frac{\partial}{\partial x_i} \mathcal{N}_{\pmb{\ell}^m} (\pmb{\ell} \, |\, \mathbf{W}^{\star}_{m+3}) =\frac{\partial}{\partial x_i} (\mathbf{x}^{\mathbf{1}_r}) +\frac{\partial}{\partial x_i}[ \mathcal{N}_{\pmb{\ell}^m} (\pmb{\ell} \, |\, \mathbf{W}^{\star}_{m+3}) -\mathbf{x}^{\mathbf{1}_r}].
\end{eqnarray*}
Thus, we focus on $\frac{\partial}{\partial x_i}[ \mathcal{N}_{\pmb{\ell}^m} (\pmb{\ell} \, |\, \mathbf{W}^{\star}_{m+3}) -\mathbf{x}^{\mathbf{1}_r}]$ below, and without loss of generality let $i=1$.

Write

\begin{eqnarray*}
&&\frac{\partial}{\partial x_1}[ \mathcal{N}_{\pmb{\ell}^m} (\pmb{\ell}_{1:4} \, |\, \mathbf{W}^{\star}_{m+3}) -\pmb{\ell}_{1:4}^{\mathbf{1}_4}] \nonumber \\
&=&\frac{\partial}{\partial x_1}\Big[\mathcal{N}_{\pmb{\ell}^m} (\pmb{\ell}_{1:4} \, |\, \mathbf{W}^{\star}_{m+3}) -\nn (\pmb{\ell}^m_{\pmb{\ell}_1 , \pmb{\ell}_2 } \, |\,  \mathbf{W}^{\star}_{m+3})\nn (\pmb{\ell}^m_{\pmb{\ell}_3 , \pmb{\ell}_4 } \, |\,  \mathbf{W}^{\star}_{m+3})   \nonumber \\
&&+\nn (\pmb{\ell}^m_{\pmb{\ell}_1 , \pmb{\ell}_2 } \, |\,  \mathbf{W}^{\star}_{m+3})\nn (\pmb{\ell}^m_{\pmb{\ell}_3 , \pmb{\ell}_4 } \, |\,  \mathbf{W}^{\star}_{m+3}) -\pmb{\ell}_{1:4}^{\mathbf{1}_4} \Big]  .
\end{eqnarray*}
By \eqref{defh}, it is easy to obtain that

\begin{eqnarray*}
&&\left\| \frac{\partial}{\partial x_1}\Big[\mathcal{N}_{\pmb{\ell}^m} (\pmb{\ell}_{1:4} \, |\, \mathbf{W}^{\star}_{m+3}) -\nn (\pmb{\ell}^m_{\pmb{\ell}_1 , \pmb{\ell}_2 } \, |\,  \mathbf{W}^{\star}_{m+3})\nn (\pmb{\ell}^m_{\pmb{\ell}_3 , \pmb{\ell}_4 } \, |\,  \mathbf{W}^{\star}_{m+3})  \Big]\right\|_{\infty}^{[0,h]^4} \le  2^{-m},
\end{eqnarray*}
and

\begin{eqnarray*}
&&\left\| \frac{\partial}{\partial x_1}\big[\nn (\pmb{\ell}^m_{\pmb{\ell}_1 , \pmb{\ell}_2 } \, |\,  \mathbf{W}^{\star}_{m+3})\nn (\pmb{\ell}^m_{\pmb{\ell}_3 , \pmb{\ell}_4 } \, |\,  \mathbf{W}^{\star}_{m+3}) -\pmb{\ell}_{1:4}^{ \mathbf{1}_4} \big] \right\|_{\infty}^{[0,h]^4}\nonumber \\
&=&\Big\|\frac{\partial}{\partial x_1}\big[\nn (\pmb{\ell}^m_{\pmb{\ell}_1 , \pmb{\ell}_2 } \, |\,  \mathbf{W}^{\star}_{m+3})\big(\nn (\pmb{\ell}^m_{\pmb{\ell}_3 , \pmb{\ell}_4 } \, |\,  \mathbf{W}^{\star}_{m+3})-  \pmb{\ell}_{3:4}^{ \mathbf{1}_2}\big)\nonumber \\
&&+\big(\nn (\pmb{\ell}^m_{\pmb{\ell}_1, \pmb{\ell}_2 } \, |\,  \mathbf{W}^{\star}_{m+3}) -\pmb{\ell}_{1:2}^{ \mathbf{1}_2} \big)\pmb{\ell}_{3:4}^{\mathbf{1}_2}\big] \Big\|_{\infty}^{[0,h]^4}\nonumber \\
&\le &  \Big\|\frac{\partial}{\partial x_1}\big[\nn (\pmb{\ell}^m_{\pmb{\ell}_1 , \pmb{\ell}_2 } \, |\,  \mathbf{W}^{\star}_{m+3})\big]\big(\nn (\pmb{\ell}^m_{\pmb{\ell}_3 , \pmb{\ell}_4 } \, |\,  \mathbf{W}^{\star}_{m+3})-  \pmb{\ell}_{3:4}^{\mathbf{1}_2}\big)\Big\|_{\infty}^{[0,h]^4}\nonumber \\
&&+\Big\| \frac{\partial}{\partial x_1}\big[\nn (\pmb{\ell}^m_{\pmb{\ell}_1 , \pmb{\ell}_2 } \, |\,  \mathbf{W}^{\star}_{m+3}) -\pmb{\ell}_{1:2}^{\mathbf{1}_2} \big]\pmb{\ell}_{3:4}^{ \mathbf{1}_2}\big] \Big\|_{\infty}^{[0,h]^4}\nonumber \\
&=&2^{-m}+ 2^{-m} =2\cdot 2^{-m},
\end{eqnarray*}
where we have used Lemma \ref{LM1} of the main text and the chain rule. Thus, we can conclude that

\begin{eqnarray*}
\left\| \frac{\partial}{\partial x_1}[ \mathcal{N}_{\pmb{\ell}^m} (\pmb{\ell}_{1:4} \, |\, \mathbf{W}^{\star}_{m+3}) -\pmb{\ell}_{1:4}^{ \mathbf{1}_4}]\right\|_{\infty}^{[0,h]^4} \le 3 \cdot 2^{-m}.
\end{eqnarray*}

Repeat the above procedure as in the second step of this lemma. Then we are able to conclude the validity of the third result. The proof is now completed.  \hspace*{\fill}{$\blacksquare$}

\bigskip

\noindent \textbf{Proof of Lemma \ref{LM.A7}:}

Without loss of generality, we assume all elements of $\pmb{\alpha}$ are greater than 0. If one element is 0 (say $\alpha_1=0$), we just rearrange $\pmb{\ell}_{\mathbf{x}\, |\, {\pmb{\alpha}} }$ as follows:

\begin{eqnarray*}
\pmb{\ell}_{\mathbf{x}\, |\, {\pmb{\alpha}} } = ( x_2^{\alpha_2},\ldots, x_r^{\alpha_r}, \mathbf{1}_{q}^\top, x_1^{\alpha_1})^\top=( x_2^{\alpha_2},\ldots, x_r^{\alpha_r}, \mathbf{1}_{q}^\top, 1)^\top,
\end{eqnarray*}
which has no impact on the proof at all.

Apparently, we have

\begin{eqnarray*}
\mathbf{x}^{\pmb{\alpha}} =\pmb{\ell}_{\mathbf{x}\,| \, {\pmb{\alpha}}}^{\mathbf{1}_{2^{\lceil \log_2 r\rceil}}} =\prod_{i=1}^r x_i^{\alpha_i},
\end{eqnarray*}
which in connection with Lemma \ref{LM.A6} immediately yields the three results of this lemma. Here the third result follows from the chain rule of the derivative. The proof is now completed. \hspace*{\fill}{$\blacksquare$}

\bigskip

\noindent \textbf{Proof of Lemma \ref{LM2}:}

Recall that we have defined the necessary monomials in Section \ref{Sec.1}. By Lemma \ref{LM.A2}, we can find a $p_{\vartheta} (\mathbf{x}\, | \, \mathbf{x}_0)$ admitting a form of $p_{\vartheta} (\mathbf{x}\, | \, \mathbf{x}_0) := \pmb{\psi}_{r_\vartheta}(\mathbf{x}\, |\, \mathbf{x}_0)^\top \pmb{\beta}_{\star}$ such that

\begin{eqnarray*}
\sup_{C_{\mathbf{x}_0,h} }| f_\star (\mathbf{x}) - p_{\vartheta}  (\mathbf{x}\, | \, \mathbf{x}_0) | =O(h^p ) ,\nonumber
\end{eqnarray*}
where $\pmb{\beta}_{\star}$ is an $r_\vartheta\times 1$ vector depending on $\mathbf{x}_0$.

To proceed, we note that as $h\to 0$, it is guaranteed that  

\begin{eqnarray*}
\mathbf{x}-\mathbf{x}_0\in [0,h]^r \subseteq  [0,1]^r\nonumber
\end{eqnarray*}
in the definition of $C_{\mathbf{x}_0,h}$. Also, recall that we have defined $\pmb{\psi}_{r_\vartheta} (\mathbf{x}|\mathbf{x}_0)$ in the end of Section \ref{Sec.1}. Then we write

\begin{eqnarray*}
&&\sup_{C_{\mathbf{x}_0,h} }| f_\star (\mathbf{x}) - \mathbf{N}  (\mathbf{x} \, | \, \mathbf{x}_0)^\top  \pmb{\beta}_{\star}  |\nonumber \\
&\le & \sup_{C_{\mathbf{x}_0,h} }| f_\star (\mathbf{x}) - p_{\vartheta} (\mathbf{x}\, | \, \mathbf{x}_0)|+\sup_{C_{\mathbf{x}_0,h} }| p_{\vartheta}  (\mathbf{x}\, | \, \mathbf{x}_0)  -\mathbf{N}  (\mathbf{x} \, | \, \mathbf{x}_0)^\top  \pmb{\beta}_{\star}|\nonumber \\
&\le &O(h^p)  + \|\pmb{\beta}_\star \|\cdot \sup_{C_{\mathbf{x}_0,h} } \| \pmb{\psi}_{r_\vartheta} (\mathbf{x}\, |\, \mathbf{x}_0)- \mathbf{N} (\mathbf{x} \, | \, \mathbf{x}_0)\|  \nonumber \\
&\le &O(h^p) + \|\pmb{\beta}_\star \|\cdot \sqrt{r_{\vartheta}} \cdot  3^{\lceil \log_2 \vartheta\rceil-1} 2^{-m} = O(h^p +2^{-m}),
\end{eqnarray*}
where the second inequality follows from Lemma \ref{LM.A2}, and the third inequality follows from Lemma \ref{LM.A7} and the definitions of $\pmb{\psi}_{r_\vartheta} (\mathbf{x}\, |\, \mathbf{x}_0)$ and $\mathbf{N} (\mathbf{x} \, | \, \mathbf{x}_0) $. 

The proof is now completed. \hspace*{\fill}{$\blacksquare$}

\bigskip

\noindent \textbf{Proof of Theorem \ref{TM1}:}

By construction of $C_{\mathbf{x}_{\mathbf{i}}, h}$'s, for $\forall\mathbf{x}\in [-a,a]^r$, we can always find a $C_{\mathbf{x}_{\mathbf{i}}, h}$ to ensure 

\begin{eqnarray}\label{eqmainlem1}
\mathbf{x}\in C_{\mathbf{x}_{\mathbf{i}}, h}.
\end{eqnarray}
Also, we note that $I_{\mathbf{i}}(\mathbf{x})\cdot I_{\mathbf{j}}(\mathbf{x})=0$ when $\mathbf{i}\ne \mathbf{j}$. Therefore, for any $\mathbf{x}\in [-a,a]^r$, there is only one $C_{\mathbf{x}_{\mathbf{i}}, h}$ to ensure \eqref{eqmainlem1}. Finally, invoking Lemma \ref{LM2}, the result  follows.   \hspace*{\fill}{$\blacksquare$}

\bigskip

\noindent \textbf{Proof of Lemma \ref{LM3}:}

Write

\begin{eqnarray*}
&&Q_T (\pmb{\theta}, \mathbf{B})=\frac{1}{T}\sum_{t=1}^T [y_t - \mathscr{N} ( \mathbf{z}_t \,\pmb{\theta} \, | \, ,\mathbf{B}) ]^2 \nonumber \\
&=& \frac{1}{T}\sum_{t=1}^T \big[ f_\star(\mathbf{z}_{t} \, \pmb{\theta}_\star)-\mathscr{N} (\mathbf{z}_t \,\pmb{\theta}_\star \, | \, \mathbf{B}_\star) + \mathscr{N} (\mathbf{z}_t \,\pmb{\theta}_\star \, | \, \mathbf{B}_\star) - \mathscr{N} (\mathbf{z}_t \,\pmb{\theta} \, | \, \mathbf{B}_\star) \nonumber \\
&&+ \mathscr{N} (\mathbf{z}_t \,\pmb{\theta} \, | \,\mathbf{B}_\star) - \mathscr{N} (\mathbf{z}_t \, \pmb{\theta} \, | \,\mathbf{B}) + \varepsilon_t\big]^2\nonumber \\
&=& \frac{1}{T}\sum_{t=1}^T [ f_{\star}(\mathbf{z}_{t} \, \pmb{\theta}_\star)-\mathscr{N} (\mathbf{z}_t \,\pmb{\theta}_\star \, | \, \mathbf{B}_\star)  ]^2 + \frac{1}{T}\sum_{t=1}^T[\mathscr{N} (\mathbf{z}_t \,\pmb{\theta}_\star \, | \,\mathbf{B}_\star) - \mathscr{N} (\mathbf{z}_t\, \pmb{\theta}  \, | \,\mathbf{B}_\star)]^2 \nonumber \\
&&+ \frac{1}{T}\sum_{t=1}^T[\mathscr{N} (\mathbf{z}_t \,  \pmb{\theta}\, | \, \mathbf{B}_\star) -\mathscr{N} (\mathbf{z}_t\, \pmb{\theta} \, | \,\mathbf{B})]^2+\frac{1}{T}\sum_{t=1}^T \varepsilon_t^2 \nonumber \\
&&+ \frac{2}{T}\sum_{t=1}^T [ f_\star(\mathbf{z}_{t} \, \pmb{\theta}_\star)- \mathscr{N} (\mathbf{z}_t \,\pmb{\theta}_\star \, | \, \mathbf{B}_\star)  ][\mathscr{N}(\mathbf{z}_t \, \pmb{\theta}_\star\, | \, \mathbf{B}_\star) - \mathscr{N} (\mathbf{z}_t \, \pmb{\theta} \, | \, \mathbf{B}_\star)]\\
&&+ \frac{2}{T}\sum_{t=1}^T [ f_\star(\mathbf{z}_{t} \, \pmb{\theta}_\star)-\mathscr{N} (\mathbf{z}_t\, \pmb{\theta}_\star  \, | \, \mathbf{B}_\star) ][\mathscr{N} (\mathbf{z}_t \,\pmb{\theta} \, | \, \mathbf{B}_\star) -\mathscr{N} (\mathbf{z}_t\, \pmb{\theta}  \, | \, \mathbf{B})]\\
&&+ \frac{2}{T}\sum_{t=1}^T [ f_\star(\mathbf{z}_{t} \, \pmb{\theta}_\star)-\mathscr{N} (\mathbf{z}_t \,\pmb{\theta}_\star \, | \, \mathbf{B}_\star)  ]\varepsilon_t \\
&&+\frac{2}{T}\sum_{t=1}^T[\mathscr{N} (\mathbf{z}_t \, \pmb{\theta}_\star \, | \, \mathbf{B}_\star) - \mathscr{N} (\mathbf{z}_t \, \pmb{\theta}  \, | \, \mathbf{B}_\star)][\mathscr{N} (\mathbf{z}_t\, \pmb{\theta}  \, | \, \mathbf{B}_\star) - \mathscr{N} (\mathbf{z}_t \, \pmb{\theta} \, | \,\mathbf{B})]\\
&&+\frac{2}{T}\sum_{t=1}^T[\mathscr{N} (\mathbf{z}_t \, \pmb{\theta}_\star  \, | \, \mathbf{B}_\star) - \mathscr{N} (\mathbf{z}_t\, \pmb{\theta}  \, | \, \mathbf{B}_\star)]\varepsilon_t \\
&&+\frac{2}{T}\sum_{t=1}^T[\mathscr{N} (\mathbf{z}_t\, \pmb{\theta}  \, | \, \mathbf{B}_\star) -\mathscr{N} (\mathbf{z}_t\,\pmb{\theta}  \, | \,\mathbf{B}) ]\varepsilon_t \\
&:=&Q_{T,1} + \cdots + Q_{T,10},
\end{eqnarray*}
where the definitions of $Q_{T,j}$ for $j\in[10]$ are obvious. In what follows, we consider these terms one by one.

\medskip

Among $Q_{T,j}$'s, some of them can be studied quite easily. For example, by Theorem \ref{TM1}, it is straightforward to obtain 

\begin{eqnarray*}
Q_{T,1} =O(h^{2p}+2^{-2m}),
\end{eqnarray*}
which in connection with Cauchy-Schwarz inequality yields that

\begin{eqnarray*}
\sup_{\pmb{\theta} ,\mathbf{B}}|Q_{T,j}|=O_P( h^{p}+2^{-m})\quad\text{for}\quad j=5,6,7.
\end{eqnarray*}
Also, by Assumption \ref{As2}, it is obvious 

\begin{eqnarray*}
Q_{T,4}=\sigma_\varepsilon^2 +o_P(1).
\end{eqnarray*}

We next consider $\frac{1}{T}\sum_{t=1}^T \mathscr{N} (\mathbf{z}_t \, \pmb{\theta} \, | \, \mathbf{B}) \varepsilon_t$. Note that by Theorem \ref{TM1} we can write

\begin{eqnarray*}
\frac{1}{T}\sum_{t=1}^T \mathscr{N} (\mathbf{z}_t \, \pmb{\theta} \, | \, \mathbf{B}) \varepsilon_t &=& \sum_{\mathbf{i}\in [M]^2}\pmb{\beta}_{\mathbf{i}}^\top \frac{1}{T}\sum_{t=1}^T I_{\mathbf{i}} (\mathbf{z}_t)\, \mathbf{N} (\mathbf{z}_t\, \pmb{\theta}\, |\, \mathbf{x}_{\mathbf{i}})\varepsilon_t\nonumber \\
&\asymp & \sum_{\mathbf{i}\in [M]^2}\pmb{\beta}_{\mathbf{i}}^\top \frac{1}{T}\sum_{t=1}^T I_{\mathbf{i}} (\mathbf{z}_t)\, \pmb{\psi}_{r_\vartheta} (\mathbf{z}_t\,\pmb{\theta} \, | \, \mathbf{x}_{\mathbf{i}})\varepsilon_t ,
\end{eqnarray*}
where the last step follows from Lemma \ref{LM.A7}, and the proof of Lemma \ref{LM2}. Here, it is easy to show that

\begin{eqnarray*}
\sup_{ \pmb{\theta}}\left\|\frac{1}{T}\sum_{t=1}^T I_{\mathbf{i}} (\mathbf{z}_t)\, \pmb{\psi}_{r_\vartheta} (\mathbf{z}_t\,\pmb{\theta} \, | \, \mathbf{x}_{\mathbf{i}})\varepsilon_t \right\|=o_P(1)
\end{eqnarray*}
using the facts that $r_\vartheta$ is fixed, and $\pmb{\psi}_{r_\vartheta} (\mathbf{z}\,\pmb{\theta} \, | \,\mathbf{x}_{\mathbf{i}})$ is formed by $(\mathbf{z}\,\pmb{\theta}-\mathbf{x}_{\mathbf{i}})^{\mathbf{J}}$ with $0\le |\mathbf{J}|\le \vartheta$. Thus, simple algebra shows that 

\begin{eqnarray*}
\sup_{\pmb{\theta},\mathbf{B}}\frac{1}{M^r}|Q_{T,j}| =o_P(1) \quad\text{for}\quad j=9, 10.
\end{eqnarray*}

\medskip

With the above results in hand, we only need to pay attention to $Q_{T,2}$, $Q_{T,3}$ and $Q_{T,8}$. To study $Q_{T,3}$, we note that

\begin{eqnarray}\label{cont2}
&&\frac{1}{h^r}\sum_{\mathbf{i}\in [M]^r}E\left[ I_{\mathbf{i}}(\mathbf{z}_t)\mathbf{H}^{-1} \pmb{\psi}_{r_\vartheta}(\mathbf{z}_t\,\pmb{\theta}\, |\, \mathbf{x}_{\mathbf{i}}) \pmb{\psi}_{r_\vartheta}(\mathbf{z}_t\,\pmb{\theta}\, |\,\mathbf{x}_{\mathbf{i}})^\top\mathbf{H}^{-1}\right] \nonumber \\
&=&\frac{1}{h^r} \sum_{\mathbf{i}\in [M]^r}\int_{\mathbf{w}\in C_{\mathbf{x}_\mathbf{i}}} \mathbf{H}^{-1} \pmb{\psi}_{r_\vartheta}( \mathbf{w} \, |\,\mathbf{x}_{\mathbf{i}}) \pmb{\psi}_{r_\vartheta}(\mathbf{w} \, |\,\mathbf{x}_{\mathbf{i}})^\top\mathbf{H}^{-1} \phi_{\pmb{\theta}} (\mathbf{w}) \mathrm{d}\mathbf{w} \nonumber \\
&=& \sum_{\mathbf{i}\in [M]^r}\phi_{\pmb{\theta}} (\mathbf{x}_{\mathbf{i}})\int_{[0,1]^r}  \pmb{\psi}_{r_\vartheta}(\mathbf{w} )\, \pmb{\psi}_{r_\vartheta}(\mathbf{w} )^\top \mathrm{d}\mathbf{w}\cdot (1+o(1)),
\end{eqnarray}
where the first equality follows from the construction of $C_{\mathbf{i}}$ in \eqref{defcnew}, and the second equality follows from integration by substitution and Assumption \ref{As2}. Then we can write

\begin{eqnarray}\label{der.qt3}
Q_{T,3}&=&\frac{1}{T}\sum_{t=1}^T[\mathscr{N} (\mathbf{z}_t\, \pmb{\theta} \, | \, \mathbf{B}_\star) -\mathscr{N} (\mathbf{z}_t \, \pmb{\theta} \, | \,\mathbf{B})]^2 \nonumber \\
&\asymp & \sum_{\mathbf{i}\in [M]^r} \frac{1}{T}\sum_{t=1}^TI_{\mathbf{i}}(\mathbf{z}_t) \left[ \pmb{\psi}_{r_\vartheta}(\mathbf{z}_t\, \pmb{\theta} \, | \, \mathbf{x}_{\mathbf{i}})^\top \mathbf{H}^{-1}\mathbf{H} (\pmb{\beta}_{\star\mathbf{i}} - \pmb{\beta}_{\mathbf{i}})  \right]^2\nonumber  \\
&\asymp& h^r \sum_{\mathbf{i}\in [M]^r}\|\mathbf{H}(\pmb{\beta}_{\star\mathbf{i}} - \widehat{\pmb{\beta}}_{\mathbf{i}})  \|^2
\asymp \frac{1}{M^r}\sum_{\mathbf{i}\in [M]^r}\|\mathbf{H}(\pmb{\beta}_{\star\mathbf{i}} - \widehat{\pmb{\beta}}_{\mathbf{i}}) \|^2,
\end{eqnarray}
where the second step follows from Lemma \ref{LM.A7}, Theorem \ref{TM1} and $I_{\mathbf{i}}(\mathbf{z}\, \pmb{\theta})I_{\mathbf{j}}(\mathbf{z}\, \pmb{\theta})=0$ for $\mathbf{i}\ne \mathbf{j}$, the third step follows from \eqref{cont2} and Assumption \ref{As2}, and the last step follows from the definition of $h$. 

Note further that by \eqref{der.qt3}, $Q_{T,2}+Q_{T,3}+Q_{T,8}$ admits a quadratic form using matrix notation which reaches its minimum value (i.e., 0) at $(\pmb{\theta}_\star,\mathbf{B}_\star)$. From here,  using Assumption \ref{As2}, it is easy to see that $\|\widehat{\pmb{\theta}} -\pmb{\theta}_\star\|= o_P(1)$ and $\frac{1}{M^r}\sum_{\mathbf{i}\in [M]^r}\|\mathbf{H}(\pmb{\beta}_{\star\mathbf{i}} - \widehat{\pmb{\beta}}_{\mathbf{i}}) \|^2=o_P(1)$ must be satisfied.  Otherwise, in view of the continuity of the quadratic form, it is easy to know that $Q_{T,2}+Q_{T,3}+Q_{T,8} =c>0$ for some positive constant $c$ in probability one. The proof is now completed.\hspace*{\fill}{$\blacksquare$}

\bigskip

\noindent \textbf{Proof of Lemma \ref{LM.A8}:}

We consider the two asymptotic distributions one by one below.

(1). For notational simplicity, we let

\begin{eqnarray}\label{defA1}
\widetilde{\mathbf{A}}_{T} &=&\frac{1}{\sqrt{T}}\sum_{t=1}^T\varepsilon_t \cdot\mathbf{f}^{(1)}_\star (\mathbf{z}_t\, \pmb{\theta}_\star)\cdot \widetilde{\mathbf{z}}_t \cdot I_{a,t}:= \frac{1}{\sqrt{T}}\sum_{t=1}^T\varepsilon_t \mathbf{A}_{t},
\end{eqnarray}
where the definition of $\mathbf{A}_{t}$ is obvious.

We now proceed to prove the asymptotic normality for $\frac{1}{\sqrt{T}}\sum_{t=1}^T\varepsilon_t \mathbf{A}_{t}$. Obviously, we have

\begin{eqnarray}\label{defA2}
E[\widetilde{\mathbf{A}}_{T} \widetilde{\mathbf{A}}_{T}^\top] &=& \frac{1}{T}\sum_{t,s=1}^T E[\varepsilon_t \varepsilon_s \,I_{a,t} I_{a,s}\, \mathbf{f}^{(1)}_\star(\mathbf{z}_t\, \pmb{\theta}_\star)\, \widetilde{\mathbf{z}}_t  \widetilde{\mathbf{z}}_s^\top\, \mathbf{f}^{(1)}_\star(\mathbf{z}_t\, \pmb{\theta}_\star) ] \nonumber \\
&\to &\pmb{\Sigma}_{11} +\pmb{\Sigma}_{12}+\pmb{\Sigma}_{12}^\top.
\end{eqnarray}

Below, we use small-block and large-block to prove the normality. To employ the small-block and large-block arguments, we partition the set $\{1,\ldots, T \}$ into $2k_T+1$ subsets with large blocks of size $l_T$ and small blocks of size $s_T$ and the last remaining set of size $T-k_T(l_T+s_T)$, where $l_T$ and $s_T$ are selected such that

\begin{eqnarray*}
s_T\to \infty,\quad  \frac{s_T}{l_T}\to 0,\quad \frac{l_T^{1+\nu} }{ T^{\frac{\nu}{2}}}\to 0,\quad\text{and}\quad k_T\equiv \left\lfloor \frac{T}{l_T+s_T} \right\rfloor,
\end{eqnarray*}
and $\nu$ is defined in Assumption \ref{As2}.

For $j=1,\ldots, k_T$, define

\begin{eqnarray*}
&&\boldsymbol{\xi}_j = \sum_{t=(j-1)(l_T+s_T)+1}^{jl_T+(j-1)s_T} \varepsilon_t \mathbf{A}_t,\quad \boldsymbol{\eta}_j = \sum_{t=jl_T+ (j-1)s_T+1}^{j(l_T+s_T)} \varepsilon_t \mathbf{A}_t,\quad \boldsymbol{\zeta} =\sum_{t=k_T(l_T+s_T)+1}^T \varepsilon_t \mathbf{A}_t.
\end{eqnarray*}
Note that $\alpha(T) = o(1/T)$ and $k_Ts_T/T\to 0$. By direct calculation, we immediately obtain that

\begin{eqnarray*}
\frac{1}{T}E\left\|\sum_{j=1}^{k_T} \boldsymbol{\eta}_j \right\|^2\to 0\quad \text{and}\quad \frac{1}{T}E\left\| \boldsymbol{\zeta}\right\|^2\to 0.
\end{eqnarray*}
Therefore,

\begin{eqnarray*}
\frac{1}{\sqrt{T}}\sum_{t=1}^T\varepsilon_t \mathbf{A}_t = \frac{1}{\sqrt{T}}\sum_{j=1}^{k_T}\boldsymbol{\xi}_j +o_P(1).
\end{eqnarray*}
By Proposition 2.6 of \cite{FanYao}, we have as $T\to 0$

\begin{eqnarray*}
&&\left| E\left[\exp\left( \frac{iw}{\sqrt{T}}\sum_{j=1}^{k_T}\boldsymbol{\xi}_j\right) \right] -\prod_{j=1}^{k_T}E\left[\exp\left( \frac{iw \boldsymbol{\xi}_j}{\sqrt{T}} \right) \right]\right| \nonumber \\
&\le & 16(k_T-1) \alpha(s_T)\to 0,
\end{eqnarray*}
where $i$ is the imaginary unit. In connection with \eqref{defA1} and \eqref{defA2}, the Feller condition is fulfilled as follows:

\begin{eqnarray*}
\frac{1}{T}\sum_{j=1}^{k_T}E[\boldsymbol{\xi}_j\boldsymbol{\xi}_j^\top]\to \pmb{\Sigma}_{11} +\pmb{\Sigma}_{12}+\pmb{\Sigma}_{12}^\top.
\end{eqnarray*}

Also, we note that

\begin{eqnarray*}
E[\|\boldsymbol{\xi}_1\|^2 \cdot I(\|\boldsymbol{\xi}_1\| \ge \epsilon \sqrt{T})] &\le &\{E\|\boldsymbol{\xi}_1\|^{2\cdot \frac{2+\nu}{2}}\}^{\frac{2}{2+\nu}} \left\{ E[ I(\|\boldsymbol{\xi}_1\| \ge \epsilon \sqrt{T})] \right\}^{\frac{\nu}{2+\nu}}\nonumber \\
&\le  &\{E\|\boldsymbol{\xi}_1\|^{2+\nu}\}^{\frac{2}{2+\nu}} \left\{ \frac{E\|\boldsymbol{\xi}_1\|^{2+\nu}}{\epsilon^{2+\nu} T^{\frac{2+\nu}{2}}}\right\}^{\frac{\nu}{2+\nu}}\nonumber \\
&= &  \frac{1}{\epsilon^{\nu} T^{\frac{\nu}{2}}}\left\{ E\|\boldsymbol{\xi}_1\|^{2+\nu}\right\}^{\frac{1}{2+\nu} \cdot (2+\nu)} \nonumber \\
&=&  O(1) \frac{l_T^{2+\nu} }{\epsilon^{\nu} T^{\frac{\nu}{2}}}  E \|\varepsilon_1 \mathbf{A}_1 \|^{2+\nu}   \nonumber \\
&=&  O(1) \frac{l_T^{2+\nu} }{ T^{\frac{\nu}{2}}} ,
\end{eqnarray*}
where the first inequality follows from H\"older inequality, the second inequality follows from  Chebyshev's inequality, and the second equality follows from Minkowski inequality. Consequently,

\begin{eqnarray*}
\frac{1}{T}\sum_{j=1}^{k_T}E[\|\boldsymbol{\xi}_j\|^2 \cdot I(\|\boldsymbol{\xi}_j\| \ge \epsilon \sqrt{T})] =O\left( \frac{k_Tl_T^{2+\nu} }{ T T^{\frac{\nu}{2}}}\right) = O\left( \frac{l_T^{1+\nu} }{  T^{\frac{\nu}{2}}}\right)=o(1),
\end{eqnarray*}
where the last step follows from the choice of $l_T$ as specified above. Therefore, the Lindberg condition is justified. Using a Cram{\'e}r-Wold device, the first result follows immediately by the standard argument. 

\medskip

(2). Without loss generality, we suppose that $ \mathbf{x}_0\in C_{\mathbf{i}}$ for some $\mathbf{i}$. For notational simplicity, we further let

\begin{eqnarray}
\widetilde{\mathbf{C}}_T &= & \frac{1}{\sqrt{Th^r}}\sum_{t=1}^T\varepsilon_t I_{\mathbf{i}} (\mathbf{z}_t) \mathbf{H}^{-1}\pmb{\psi}_{r_\vartheta} (\mathbf{z}_t\, \pmb{\theta}_\star \, | \, \mathbf{x}_{\mathbf{i}}):=\frac{1}{\sqrt{Th^r}}\sum_{t=1}^T\varepsilon_t \mathbf{C}_t,
\end{eqnarray}
where the second equality follows from Lemma \ref{LM.A7}, and the definition of $\mathbf{C}_t$ is obvious.

We now proceed and write

\begin{eqnarray}\label{EqA.14}
&&E\left[ \left(\frac{1}{\sqrt{Th^r}}\sum_{t=1}^T\varepsilon_t \mathbf{C}_t \right) \left(\frac{1}{\sqrt{Th^2}}\sum_{t=1}^T\varepsilon_t \mathbf{C}_t\right)^\top \right]\nonumber \\
&=&\frac{1}{Th^r}\sum_{t=1}^T\sigma_\varepsilon^2 E [\mathbf{C}_t\mathbf{C}_t^\top  ] + \frac{1}{ h^r}\sum_{t=1}^{T-1} (1-t/T)E [\mathbf{C}_{1+t} \mathbf{C}_1^\top\varepsilon_1\varepsilon_{1+t} ] \nonumber \\
&&+\frac{1}{ h^r}\sum_{t=1}^{T-1} (1-t/T)E [\mathbf{C}_1\mathbf{C}_{1+t}^\top \varepsilon_1\varepsilon_{1+t} ],
\end{eqnarray}
where the last two terms are the same up to a transpose operation.

The term $\frac{1}{ h^r}\sum_{t=1}^{T-1} (1-t/T)E [\mathbf{C}_1\mathbf{C}_{1+t}^\top \varepsilon_1\varepsilon_{1+t} ]$ on the right hand side can be bounded as follows.

\begin{eqnarray*}
&& \| E [\mathbf{C}_1\mathbf{C}_{1+t}^\top\varepsilon_1\varepsilon_{1+t} ] \| \nonumber \\
&\le &O(1)  \alpha(t)^{\nu/(2+\nu)}\left\{E\|\mathbf{C}_1 \varepsilon_1 \|^{2+\nu} \right\}^{2/(2+\nu)}\nonumber \\
&\le &O(1) \alpha(t)^{\nu/(2+\nu)}\left\{E\|\mathbf{C}_1  \varepsilon_1 \|^{2+\nu} ] \frac{1}{h^r} \right\}^{2/(2+\nu)}\cdot (h^r)^{2/(2+\nu)}\nonumber \\
&=&O((h^r)^{  2/(2+\nu)})\alpha(t)^{\nu/(2+\nu)}.
\end{eqnarray*}
Thus, we have

\begin{eqnarray*}
&&\left\|\frac{1}{ h^r}\sum_{t=1}^{T-1} (1-t/T)E [\mathbf{C}_1\mathbf{C}_{1+t}^\top \varepsilon_1\varepsilon_{1+t} ] \right\| \nonumber \\
&\le & O(1) \frac{1}{ h^r}\sum_{t=1}^{d_T} \left\| E[\mathbf{C}_1\mathbf{C}_{1+t}^\top \varepsilon_1\varepsilon_{1+t}] \right\| +O(1) \frac{1}{ h^r}\sum_{t=d_T+1}^{T} \left\| E [ \mathbf{C}_1\mathbf{C}_{1+t}^\top \varepsilon_1\varepsilon_{1+t} ] \right\| \nonumber \\
&\le & O(1) h^r\sum_{t=1}^{d_T}  \left\| E[\frac{1}{ h^r} I_{\mathbf{i}}(\mathbf{z}_1)  \frac{1}{ h^r} I_{\mathbf{i}}(\mathbf{z}_{1+t}) \varepsilon_1\varepsilon_{1+t}] \right\| \\
&&  +O(1) \frac{1}{ h^r}\sum_{t=d_T+1}^{T} \left\| E[\mathbf{C}_1\mathbf{C}_{1+t}^\top  \varepsilon_1\varepsilon_{1+t} ] \right\| \nonumber \\
&\le &O(1)h^r d_T+O(1)\frac{(h^r)^{  2/(2+\nu)}}{h^r}\sum_{t=d_T+1}^T \alpha^{\nu/(2+\nu)}(t) \nonumber \\
&=&O(1)h^r d_T+O(1)\frac{1}{h^{\frac{r\nu}{2+\nu}}}\sum_{t=d_T+1}^T  \alpha^{\nu/(2+\nu)}(t) =o(1),
\end{eqnarray*}
where  $\nu>0$ is defined in Assumption \ref{As2}, and the last step follows from Assumption 2.1 that we can choose $d_T$ to ensure 
\begin{eqnarray}\label{newINEQ1}
d_T h^r \to 0\quad\text{and}\quad  \sum_{t= d_T+ 1}^{T} \alpha^{\nu/(2+\nu)}(t) = o\left(h^{\frac{r\nu}{2+\nu}}\right),
\end{eqnarray}
which can be achieved by choosing $d_T =\lfloor T^{c_1}\rfloor$, $h = T^{-c_2}$ and $\alpha(t) = t^{-c_3}$ for some suitable $c_1>0$, $c_2>0$ and $c_3>0$, for example.

Thus, we can conclude that

\begin{eqnarray}\label{EqA.17}
&&E\left[ \left(\frac{1}{\sqrt{Th^r}}\sum_{t=1}^T\varepsilon_t \mathbf{C}_t \right) \left(\frac{1}{\sqrt{Th^r}}\sum_{t=1}^T\varepsilon_t \mathbf{C}_t\right)^\top \right]\nonumber \\
&=&\frac{1}{Th^r}\sum_{t=1}^T\sigma_\varepsilon^2 E [\mathbf{C}_t\mathbf{C}_t^\top ] +o(1)\nonumber \\
&\to &\sigma_\varepsilon^2 \phi_{\pmb{\theta}_0} (\mathbf{x}_{0})\int_{[0,1]^r}  \pmb{\psi}_{r_\vartheta}(\mathbf{w} )\, \pmb{\psi}_{r_\vartheta}(\mathbf{w})^\top \mathrm{d}\mathbf{w},
\end{eqnarray}
where the last line follows from the development of \eqref{cont2} and the continuity of $\phi_{\pmb{\theta}_0}$ by Assumption \ref{As2}.

\medskip

Below, we further use small-block and large-block to prove the normality. To employ the small-block and large-block arguments, we partition the set $\{1,\ldots, T \}$ into $2k_T+1$ subsets with large blocks of size $l_T$ and small blocks of size $s_T$ and the last remaining set of size $T-k_T(l_T+s_T)$, where $l_T$ and $s_T$ are selected such that

\begin{eqnarray*}
s_T\to \infty,\quad  \frac{s_T}{l_T}\to 0,\quad \frac{l_T^{1+\nu} }{  (Th^r)^{\frac{\nu}{2}}}\to 0,\quad\text{and}\quad k_T\equiv \left\lfloor \frac{T}{l_T+s_T} \right\rfloor,
\end{eqnarray*}
and $\nu$ is defined in Assumption \ref{As2}.

For $j=1,\ldots, k_T$, define

\begin{eqnarray*}
&&\boldsymbol{\xi}_j = \sum_{t=(j-1)(l_T+s_T)+1}^{jl_T+(j-1)s_T} \frac{1}{\sqrt{h^r}}\mathbf{C}_t \varepsilon_t,\quad \boldsymbol{\eta}_j = \sum_{t=jl_T+ (j-1)s_T+1}^{j(l_T+s_T)} \frac{1}{\sqrt{h^r}}\mathbf{C}_t \varepsilon_t ,\nonumber \\
&&\boldsymbol{\zeta} =\sum_{t=k_T(l_T+s_T)+1}^T \frac{1}{\sqrt{h^r}}\mathbf{C}_t \varepsilon_t.
\end{eqnarray*}
Note that $\alpha(T) = o(1/T)$ and $k_Ts_T/T\to 0$. By direct calculation, we immediately obtain that

\begin{eqnarray*}
\frac{1}{T}E\left\|\sum_{j=1}^{k_T} \boldsymbol{\eta}_j \right\|^2\to 0\quad \text{and}\quad \frac{1}{T}E\left\| \boldsymbol{\zeta}\right\|^2\to 0.
\end{eqnarray*}
Therefore,

\begin{eqnarray*}
\frac{1}{\sqrt{Th^r}}\sum_{t=1}^T\mathbf{C}_t\varepsilon_t = \frac{1}{\sqrt{T}}\sum_{j=1}^{k_T}\boldsymbol{\xi}_j +o_P(1).
\end{eqnarray*}
By Proposition 2.6 of \cite{FanYao}, we have as $T\to 0$

\begin{eqnarray*}
&&\left| E\left[\exp\left( \frac{iw}{\sqrt{T}}\sum_{j=1}^{k_T}\boldsymbol{\xi}_j\right) \right] -\prod_{j=1}^{k_T}E\left[\exp\left( \frac{iw \boldsymbol{\xi}_j}{\sqrt{T}} \right) \right]\right| \nonumber \\
&\le & 16(k_T-1) \alpha(s_T)\to 0,
\end{eqnarray*}
where $i$ is the imaginary unit. In connection with \eqref{EqA.14}-\eqref{EqA.17}, the Feller condition is fulfilled as follows:

\begin{eqnarray*}
\frac{1}{T}\sum_{j=1}^{k_T}E[\boldsymbol{\xi}_j\boldsymbol{\xi}_j^\top]\to \sigma_\varepsilon^2 \phi_{\pmb{\theta}_0} (\mathbf{x}_{0})\int_{[0,1]^r}  \pmb{\psi}_{r_\vartheta}(\mathbf{w})\, \pmb{\psi}_{r_\vartheta}(\mathbf{w} )^\top \mathrm{d}\mathbf{w}.
\end{eqnarray*}

Also, we note that

\begin{eqnarray*}
E[\|\boldsymbol{\xi}_1\|^2 \cdot I(\|\boldsymbol{\xi}_1\| \ge \epsilon \sqrt{T})] &\le &\{E\|\boldsymbol{\xi}_1\|^{2\cdot \frac{2+\nu}{2}}\}^{\frac{2}{2+\nu}} \left\{ E[ I(\|\boldsymbol{\xi}_1\| \ge \epsilon \sqrt{T})] \right\}^{\frac{\nu}{2+\nu}}\nonumber \\
&\le  &\{E\|\boldsymbol{\xi}_1\|^{2+\nu}\}^{\frac{2}{2+\nu}} \left\{ \frac{E\|\boldsymbol{\xi}_1\|^{2+\nu}}{\epsilon^{2+\nu} T^{\frac{2+\nu}{2}}}\right\}^{\frac{\nu}{2+\nu}}\nonumber \\
&= &  \frac{1}{\epsilon^{\nu} T^{\frac{\nu}{2}}}\left\{ E\|\boldsymbol{\xi}_1\|^{2+\nu}\right\}^{\frac{1}{2+\nu} \cdot (2+\nu)} \nonumber \\
&=&  O(1) \frac{l_T^{2+\nu} }{\epsilon^{\nu} T^{\frac{\nu}{2}}}  E\| \mathbf{C}_1 \varepsilon_1 \|^{2+\nu} \cdot \frac{1}{h^{\frac{r}{2} \cdot (2+\nu) }} \nonumber \\
&=&  O(1) \frac{l_T^{2+\nu} }{\epsilon^{\nu} (Th^r)^{\frac{\nu}{2}}}  \cdot  \frac{1}{h^r} E\| \mathbf{C}_1 \varepsilon_1 \|^{2+\nu} \nonumber \\
&=&  O(1) \frac{l_T^{2+\nu} }{ (Th^r)^{\frac{\nu}{2}}} ,
\end{eqnarray*}
where the first inequality follows from H\"older inequality, the second inequality follows from  Chebyshev's inequality, and the second equality follows from Minkowski inequality. Consequently,

\begin{eqnarray*}
\frac{1}{T}\sum_{j=1}^{k_T}E[\|\boldsymbol{\xi}_j\|^2 \cdot I(\|\boldsymbol{\xi}_j\| \ge \epsilon \sqrt{T})] =O\left( \frac{k_Tl_T^{2+\nu} }{ T (Th^r)^{\frac{\nu}{2}}}\right) = O\left( \frac{l_T^{1+\nu} }{  (Th^r)^{\frac{\nu}{2}}}\right)=o(1),
\end{eqnarray*}
where the last step follows from the choice of $l_T$ as specified above. Therefore, the Lindberg condition is justified. Using a Cram{\'e}r-Wold device, the CLT follows immediately by the standard argument.

The proof is now completed.\hspace*{\fill}{$\blacksquare$}

\bigskip

\noindent \textbf{Proof of Theorem \ref{TM2}:}

We now start the investigation. 

%
%

By the first order condition, we have

\begin{eqnarray}\label{foc1}
0&=&\frac{\partial Q_T (\pmb{\theta}, \mathbf{B})}{\partial \pmb{\theta}}\big|_{(\pmb{\theta}, \mathbf{B})=(\widehat{\pmb{\theta}}, \widehat{\mathbf{B}})}\nonumber \\
&=&-\frac{2}{T}\sum_{t=1}^T[y_t - \mathscr{N}  (\mathbf{z}_t\, \widehat{\pmb{\theta}}  \, | \,  \widehat{\mathbf{B}}) ]\sum_{\mathbf{i}}   I_{\mathbf{i}} (\mathbf{z}_t)\, \frac{\partial  \mathscr{N} (\mathbf{z}_t\, \widehat{\pmb{\theta}}  \, | \, \widehat{\pmb{\beta}}_{\mathbf{i}})}{\partial \pmb{\theta}}\nonumber \\
&\asymp &-\frac{2}{T}\sum_{t=1}^T[y_t - \mathscr{N}(\mathbf{z}_t  \, | \, \widehat{\pmb{\theta}}, \widehat{\mathbf{B}}) ]\, \mathbf{f}^{(1)}_{\star}(\mathbf{z}_t\,\pmb{\theta}_\star) \widetilde{\mathbf{z}}_t I_{a,t}\nonumber \\
&:=&-2\mathbf{A}_{T,\pmb{\theta}}(\widehat{\pmb{\theta}}, \widehat{\mathbf{B}}) ,
\end{eqnarray}
where the third line follows from Lemma \ref{LM.A7} and Lemma \ref{LM3}. Similarly, we have

\begin{eqnarray}\label{foc2}
0 &=& \frac{\partial Q_T (\pmb{\theta}, \mathbf{B})}{\partial \pmb{\beta}_{\mathbf{i}}}\big|_{(\pmb{\theta}, \mathbf{B})=(\widehat{\pmb{\theta}}, \widehat{\mathbf{B}})} \nonumber \\
&=&-\frac{2}{T}\sum_{t=1}^T[y_t - \mathscr{N} (\mathbf{z}_t \, \widehat{\pmb{\theta}} \, | \,  \widehat{\mathbf{B}}) ] I_{\mathbf{i}} (\mathbf{z}_t)\, \frac{\partial  \mathscr{N} (\mathbf{z}_t \,\widehat{\pmb{\theta}} \, | \,  \widehat{\pmb{\beta}}_{\mathbf{i}})}{\partial \pmb{\beta}_{\mathbf{i}}}\nonumber \\
&\asymp & -\frac{2}{T}\sum_{t=1}^T[y_t - \mathscr{N} (\mathbf{z}_t \,\widehat{\pmb{\theta}} \, | \,  \widehat{\mathbf{B}}) ]\, I_{\mathbf{i}} (\mathbf{z}_t) \pmb{\psi}_{r_\vartheta} (\mathbf{z}_t\,\widehat{\pmb{\theta}} \, | \, \mathbf{x}_{\mathbf{i}}) \nonumber \\
&:=&-2\mathbf{A}_{T,\pmb{\beta}_{\mathbf{i}}} (\widehat{\pmb{\theta}}, \widehat{\pmb{\beta}}_{\mathbf{i}}),
\end{eqnarray}
where we have used Lemma \ref{LM.A7} again to obtain the third equality.

\medskip

To proceed, we further label $\pmb{\beta}_{\mathbf{i}}$'s and $C_{\mathbf{i}}$'s as 

\begin{eqnarray*}
\pmb{\beta}_{1},\ldots,\pmb{\beta}_{M^r} \quad\text{and}\quad C_{1},\ldots,C_{M^r} 
\end{eqnarray*}
respectively. Thus, we define

\begin{eqnarray}\label{foc}
&&\mathbf{V}(\pmb{\theta}, \mathbf{B}) = (\pmb{\theta}^\top,\pmb{\beta}_{1}^\top,\ldots, \pmb{\beta}_{ {M^r}}^\top)^\top,\nonumber \\
&&\mathbf{R}_T =\diag\{ \sqrt{T}\mathbf{I}_{d}, \sqrt{Th^r}  \mathbf{H}\}, \nonumber \\
&&\overline{\mathbf{I}}_{\mathbf{x}_0} = \diag\{I(\mathbf{x}_0\in [-a,a]^r)\mathbf{I}_d, \mathbf{I}_{\mathbf{x}_0}\},\nonumber \\
&&\mathbf{I}_{\mathbf{x}_0}= \left(I(\mathbf{x}_0\in C_1)  \mathbf{I}_{r_\vartheta},\ldots, I(\mathbf{x}_0\in C_{M^r})\mathbf{I}_{r_\vartheta} \right),\nonumber \\
&&\widetilde{\mathbf{I}}_{\mathbf{x}_0} = \diag\{I(\mathbf{x}_0\in [-a,a]^r)\mathbf{I}_d,I(\mathbf{x}_0\in C_1)  \mathbf{I}_{r_\vartheta},\ldots, I(\mathbf{x}_0\in C_{M^r})\mathbf{I}_{r_\vartheta} \},\nonumber \\
&&\mathbf{A}_T  (\pmb{\theta}, \mathbf{B}) =(\mathbf{A}_{T,\pmb{\theta}} (\pmb{\theta}, \mathbf{B})^\top,\mathbf{A}_{T,\pmb{\beta}_1} (\pmb{\theta}, \mathbf{B})^\top ,\ldots, \mathbf{A}_{T,\pmb{\beta}_{M^r}} (\pmb{\theta}, \mathbf{B})^\top )^\top ,\nonumber \\
&&\mathbf{C}_T (\pmb{\theta}, \mathbf{B}) =\begin{pmatrix}
\frac{\partial\mathbf{A}_{T,\pmb{\theta}}(\pmb{\theta}, \mathbf{B})}{\pmb{\theta}^\top}&  \frac{\partial\mathbf{A}_{T,\pmb{\theta}}(\pmb{\theta}, \mathbf{B})}{ \pmb{\beta}_{1}^\top} & \frac{\partial\mathbf{A}_{T,\pmb{\theta}}(\pmb{\theta}, \mathbf{B})}{ \partial \pmb{\beta}_{2}^\top}& \cdots & \frac{\partial\mathbf{A}_{T,\pmb{\theta}} (\pmb{\theta}, \mathbf{B})}{ \partial \pmb{\beta}_{ M^r}^\top}  \\
\frac{\partial \mathbf{A}_{T,\pmb{\beta}_{1}}(\pmb{\theta}, \mathbf{B})}{ \partial \pmb{\theta}^\top} & \frac{\partial\mathbf{A}_{T,\pmb{\beta}_{1}} (\pmb{\theta}, \mathbf{B})}{ \partial \pmb{\beta}_{1}^\top} & \mathbf{0} &\cdots & \mathbf{0} \\
\frac{\partial\mathbf{A}_{T,\pmb{\beta}_{2}}  (\pmb{\theta}, \mathbf{B})}{ \partial \pmb{\theta}^\top} & \mathbf{0} & \frac{\partial\mathbf{A}_{T,\pmb{\beta}_{2}} (\pmb{\theta}, \mathbf{B})}{ \partial \pmb{\beta}_{2}^\top} &\cdots & \mathbf{0} \\
\vdots &\vdots & \vdots & \ddots & \vdots\\
\frac{\partial \mathbf{A}_{T,\pmb{\beta}_{M^r}} (\pmb{\theta}, \mathbf{B})}{ \partial \pmb{\theta}^\top} & \mathbf{0} & \mathbf{0} &\cdots &  \frac{\partial \mathbf{A}_{T,\pmb{\beta}_{ M^r}}  (\pmb{\theta}, \mathcal{B})}{ \partial \pmb{\beta}_{ M^r}^\top} 
 \end{pmatrix}.
\end{eqnarray}

We establish a joint CLT. By Taylor expansion, simple algebra shows that

\begin{eqnarray*}
\widetilde{\mathbf{I}}_{\mathbf{x}_0}\mathbf{A}_T (\pmb{\theta}, \mathbf{B})\big|_{(\pmb{\theta}, \mathbf{B})=( \pmb{\theta}_\star, \mathbf{B}_\star)} &=&\widetilde{\mathbf{I}}_{\mathbf{x}_0} \mathbf{C}_T (\pmb{\theta}, \mathbf{B})\big|_{(\pmb{\theta}, \mathbf{B})=( \widetilde{\pmb{\theta}},\widetilde{\mathbf{B}})}(\mathbf{V}(\widehat{\pmb{\theta}}, \widehat{\mathbf{B}}) -\mathbf{V}(\pmb{\theta}_\star, \mathbf{B}_\star) ),
\end{eqnarray*}
where $( \widetilde{\pmb{\theta}},\widetilde{\mathbf{B}})$ lie between $(\widehat{\pmb{\theta}}, \widehat{\mathbf{B}})$ and $(\pmb{\theta}_\star, \mathbf{B}_\star)$. In connection with the fact that $[-a,a]^r= \cup_{i=1}^{M^r} C_i$ and $C_i\cap C_j=\emptyset$ for $i\ne j$, it is easy to see that

\begin{eqnarray*}
&&\mathbf{R}_T \overline{\mathbf{I}}_{\mathbf{x}_0}(\mathbf{V}(\widehat{\pmb{\theta}}, \widehat{\mathbf{B}}) -\mathbf{V}(\pmb{\theta}_\star, \mathbf{B}_\star) )\nonumber \\
&=& \left(\mathbf{R}_T^{-1} \overline{\mathbf{I}}_{\mathbf{x}_0} \cdot \mathbf{C}_T (\pmb{\theta}, \mathbf{B})\big|_{(\pmb{\theta}, \mathbf{B})=( \widetilde{\pmb{\theta}},  \widetilde{\mathbf{B}})} \cdot \overline{\mathbf{I}}_{\mathbf{x}_0}^\top \mathbf{R}_T^{-1}\right)^{-1}\cdot \mathbf{R}_T^{-1} \cdot \overline{\mathbf{I}}_{\mathbf{x}_0}\mathbf{A}_T (\pmb{\theta}, \mathbf{B})\big|_{(\pmb{\theta}, \mathbf{B})=( \pmb{\theta}_\star, \mathbf{B}_\star)}.
\end{eqnarray*}

Note that the evaluation of $\mathbf{R}_T^{-1} \overline{\mathbf{I}}_{\mathbf{x}_0} \cdot \mathbf{C}_T (\pmb{\theta}, \mathbf{B})\big|_{(\pmb{\theta}, \mathbf{B})=( \widetilde{\pmb{\theta}},  \widetilde{\mathbf{B}})} \cdot \overline{\mathbf{I}}_{\mathbf{x}_0}^\top \mathbf{R}_T^{-1}$ is straightforward and similar to \eqref{defA2} and \eqref{EqA.14} after invoking Weak Law of Large Numbers and noting that $\mathbf{C}_T (\pmb{\theta}, \mathbf{B})$ is continuous with respect to $(\pmb{\theta}, \mathbf{B})$. Therefore, we focus on $\mathbf{R}_T^{-1} \overline{\mathbf{I}}_{\mathbf{x}_0}\cdot \mathbf{A}_T (\pmb{\theta}, \mathbf{B})\big|_{(\pmb{\theta}, \mathbf{B})=( \pmb{\theta}_\star, \mathbf{B}_\star)}$ below.

%

Write

\begin{eqnarray}\label{clt1}
&&\sqrt{T}\mathbf{A}_{T,\pmb{\theta}} (\pmb{\theta}, \mathbf{B})\big|_{(\pmb{\theta}, \mathbf{B})=( \pmb{\theta}_\star, \mathbf{B}_\star)} \nonumber \\
&=&\frac{1}{\sqrt{T}}\sum_{t=1}^T[f_\star(\mathbf{z}_t\,\pmb{\theta}_\star) +\varepsilon_t - \mathscr{N}  (\mathbf{z}_t \, \pmb{\theta}_\star \, | \, \mathbf{B}_\star) ]\cdot \mathbf{f}^{(1)}_\star (\mathbf{z}_t\, \pmb{\theta}_\star)\cdot \widetilde{\mathbf{z}}_t \cdot I_{a,t}\nonumber \\
&=&\frac{1}{\sqrt{T}}\sum_{t=1}^T\varepsilon_t \cdot \mathbf{f}^{(1)}_\star (\mathbf{z}_t\, \pmb{\theta}_\star)\cdot \widetilde{\mathbf{z}}_t \cdot I_{a,t}\nonumber \\
& &+ \frac{1}{\sqrt{T}}\sum_{t=1}^T[f_\star(\mathbf{z}_t\,\pmb{\theta}_\star)  - \mathscr{N} (\mathbf{z}_t \, \pmb{\theta}_\star \, | \, \mathbf{B}_\star) ]\cdot \mathbf{f}^{(1)}_\star (\mathbf{z}_t\, \pmb{\theta}_\star)\cdot \widetilde{\mathbf{z}}_t \cdot I_{a,t}.
\end{eqnarray}
Here, it is easy to see that the bias term converges to

\begin{eqnarray*}
c_{\pmb{\theta}}&=&  \sum_{\mathbf{i}}\int_{C_{\mathbf{i}}}\phi(\mathbf{z})\sum_{\|\mathbf{J}\|_1=\vartheta}\frac{\vartheta (\mathbf{z}\,\pmb{\theta}_\star-\mathbf{x}_{\mathbf{i}})^{\mathbf{J}}}{\mathbf{J}!} F_{\mathbf{i}}(\mathbf{z}\, \pmb{\theta}_\star)\mathbf{f}^{(1)}_\star (\mathbf{z}\, \pmb{\theta}_\star)\cdot \mathbf{z}^\top \mathbf{1}_r  d\mathbf{z},
\end{eqnarray*}
where

\begin{eqnarray*}
  F_{\mathbf{i}}(\mathbf{x} ) =\int_{0}^1\left[ (1-w)^{\vartheta-1} f^{(\mathbf{J})}(\mathbf{x}_{\mathbf{i}} + w(\mathbf{x}-\mathbf{x}_{\mathbf{i}})) - f^{(\mathbf{J})}(\mathbf{x}_{\mathbf{i}})   (\mathbf{x}-\mathbf{x}_{\mathbf{i}})^{\mathbf{J}} \right] \mathrm{d}w.
\end{eqnarray*}

Without loss of generality, we suppose that $\mathbf{x}_0\in C_{\mathbf{i}}$, and then write

\begin{eqnarray}\label{clt2}
&&\sqrt{Th^r}\mathbf{H}^{-1}\mathbf{A}_{T,\pmb{\beta}_{\mathbf{i}}} ( \pmb{\theta}, \pmb{\beta}_{\mathbf{i}})\big|_{(\pmb{\theta}, \mathbf{B})=(\pmb{\theta}_\star, \mathbf{B}_\star)} \nonumber \\
&=& \frac{1}{\sqrt{Th^r}}\mathbf{H}^{-1}\sum_{t=1}^T[y_t - \mathscr{N}  (\mathbf{z}_t \, \pmb{\theta}_\star \, | \,  \mathbf{B}_\star) ]I_{\mathbf{i}} (\mathbf{z}_t) \pmb{\psi}_{r_\vartheta} (\mathbf{z}_t\, \pmb{\theta}_\star \, | \, \mathbf{x}_{\mathbf{i}})\nonumber \\
&=& \frac{1}{\sqrt{Th^r}}\sum_{t=1}^T\varepsilon_t I_{\mathbf{i}} (\mathbf{z}_t) \mathbf{H}^{-1}\pmb{\psi}_{r_\vartheta} (\mathbf{z}_t\, \pmb{\theta}_\star \, | \, \mathbf{x}_{\mathbf{i}})\nonumber \\
&&+\frac{1}{\sqrt{Th^r}}\sum_{t=1}^T[f_\star(\mathbf{z}_t\,\pmb{\theta}_\star)  - \mathscr{N} (\mathbf{z}_t \, \pmb{\theta}_\star \, | \, \mathbf{B}_\star) ]I_{\mathbf{i}} (\mathbf{z}_t) \mathbf{H}^{-1}\pmb{\psi}_{r_\vartheta} (\mathbf{z}_t\, \pmb{\theta}_\star \, | \, \mathbf{x}_{\mathbf{i}}).
\end{eqnarray}
Here, the bias term converges to

\begin{eqnarray*}
c_{f} &=&\sum_{\mathbf{i}}\frac{I(\mathbf{x}_0\in C_{\mathbf{X}_\mathbf{i}})}{h^r}\int_{C_{\mathbf{X}_\mathbf{i}}}\phi_{\pmb{\theta}}(\mathbf{x})\sum_{\|\mathbf{J}\|_1=\vartheta}\frac{\vartheta (\mathbf{x}-\mathbf{x}_{\mathbf{i}})^{\mathbf{J}}}{\mathbf{J}!} F_{\mathbf{i}}(\mathbf{x})\mathbf{H}^{-1}\pmb{\psi}_{r_\vartheta} (\mathbf{x} \, | \, \mathbf{x}_{\mathbf{i}}) d\mathbf{x}.
\end{eqnarray*}

In addition, we note that

\begin{eqnarray}\label{clt3}
&& \left\|\frac{\sqrt{h^r}}{Th^r}\mathbf{H}^{-1}\frac{\partial\mathbf{A}_{T,\pmb{\beta}_{\mathbf{i}}}^{(1)} (\pmb{\theta}, \mathbf{B})}{ \partial \pmb{\theta}^\top} \big|_{(\pmb{\theta}, \mathbf{B})=( \widetilde{\pmb{\theta}},  \widetilde{\mathbf{B}})} \right\|\nonumber \\
&=&\left\| \mathbf{H}^{-1}\frac{\sqrt{h^r}}{Th^r}\sum_{t=1}^T I_{\mathbf{i}} (\mathbf{z}_t) \pmb{\psi}_{r_\vartheta} (\mathbf{z}_t\, \pmb{\theta}_\star \, | \, \mathbf{x}_{\mathbf{i}}) \widetilde{\mathbf{z}}_t^\top \mathbf{f}^{(1)}_\star ( \mathbf{z}_t\,\pmb{\theta}_\star )  \right\|   \nonumber \\
&\le &O(1)\frac{\sqrt{h^r}}{Th^r}\sum_{t=1}^T I_{\mathbf{i}} (\mathbf{z}_t)\| \mathbf{H}^{-1} \pmb{\psi}_{r_\vartheta} (\mathbf{z}_t\, \pmb{\theta}_\star \, | \, \mathbf{x}_{\mathbf{i}})\| \cdot \|  \mathbf{f}^{(1)}_\star ( \mathbf{z}_t\,\pmb{\theta}_\star )\| \nonumber \\
&=&\sqrt{h^r} \cdot \phi_{\pmb{\theta}_\star}(\mathbf{x}_0)\cdot \| \mathbf{f}^{(1)}_\star ( \mathbf{x}_0 )\| \int_{[0,1]^r}\| \pmb{\psi}_{r_\vartheta} (\mathbf{w}) \|\mathrm{d} \mathbf{w} \cdot (1+o_P(1))\nonumber \\
&=&O_P(\sqrt{h^r} ),
\end{eqnarray}
where the inequality follows from $\mathbf{z}_{t}\in C_{\mathbf{i}}$ (i.e., belonging to a bounded set by design),\ and the second equity follows from the integration by substitution and Assumption \ref{As2}.

Finally, invoking Lemma \ref{LM.A8} and in view of \eqref{clt1}-\eqref{clt3}, the result follows immediately. \hspace*{\fill}{$\blacksquare$}

\bigskip

\noindent \textbf{Proof of Corollary \ref{coroinf}:}

In view of the development of Theorem \ref{TM2}, we consider the following term only:  

\begin{eqnarray*}
\frac{1}{\sqrt{T}}\sum_{t=1}^T\varepsilon_t \cdot \mathbf{f}^{(1)}_\star (\mathbf{z}_t\, \pmb{\theta}_\star)\cdot \widetilde{\mathbf{z}}_t \cdot I_{a,t}\cdot \eta_{t}:=\frac{1}{\sqrt{T}}\sum_{t=1}^T\varepsilon_t \mathbf{A}_t\eta_t.
\end{eqnarray*}
For the rest of the terms, the development can be done similarly but much simpler.

Using the Cram{\'e}r-Wold device, let $\pmb{\ell}$ be a $d\times 1$ vector and $\|\pmb{\ell}\|=1$. Thus,  we consider

\begin{eqnarray*} 
B^*   = \frac{1}{\sqrt{T}}\sum_{t=1}^T\varepsilon_t \pmb{\ell}^\top\mathbf{A}_t\eta_t  .
\end{eqnarray*}
The goal is to show that

\begin{eqnarray}\label{EQA.10}
B^* \to_{D^*} N(0, \pmb{\ell}^\top( \pmb{\Sigma}_{11}+\pmb{\Sigma}_{12}+\pmb{\Sigma}_{21}^\top) \pmb{\ell}),
\end{eqnarray}
which in connection with Theorem \ref{TM2} immediately yields the result.  

We now consider

\begin{eqnarray*}
E^*[(B^*)^2] &=& \frac{1}{T}\sum_{t,s=1}^T\varepsilon_t\varepsilon_s \pmb{\ell}^\top\mathbf{A}_t\mathbf{A}_s\pmb{\ell} E[\eta_t\eta_s]\nonumber \\
&=& \frac{1}{T}\sum_{t,s=1}^T\varepsilon_t\varepsilon_s \pmb{\ell}^\top\mathbf{A}_t\mathbf{A}_s\pmb{\ell} + \frac{1}{T}\sum_{t,s=1}^T\varepsilon_t\varepsilon_s \pmb{\ell}^\top\mathbf{A}_t\mathbf{A}_s\pmb{\ell} (E[\eta_t\eta_s]-1)
\end{eqnarray*}
Note that

\begin{eqnarray*}
&&E\left| \frac{1}{T}\sum_{t,s=1}^T\varepsilon_t\varepsilon_s \pmb{\ell}^\top\mathbf{A}_t\mathbf{A}_s\pmb{\ell} (E[\eta_t\eta_s]-1) \right|\nonumber \\
&\le &\frac{1}{T}E\left| \sum_{t=1}^{d_T}\sum_{s=1}^{T-t}\varepsilon_t\varepsilon_s \pmb{\ell}^\top\mathbf{A}_t\mathbf{A}_s\pmb{\ell} (E[\eta_t\eta_s]-1) \right|+\frac{1}{T}E\left| \sum_{t=d_T+1}^{T}\sum_{s=1}^{T-t}\varepsilon_t\varepsilon_s \pmb{\ell}^\top\mathbf{A}_t\mathbf{A}_s\pmb{\ell} (E[\eta_t\eta_s]-1) \right|\nonumber \\
&=&O(1)\sum_{t=1}^{d_T}|a(t/\ell)-a(0)| + \sum_{t=d_T+1}^TE|\varepsilon_0 \varepsilon_t|\nonumber \\
&\le &O(1)d_T^2/\ell + \sum_{t=d_T+1}^TE|\varepsilon_0 \varepsilon_t| =o(1),
\end{eqnarray*}
where the second inequality follows from $a(w)$ being Lipschitz continuous on $[-1,1]$, and the last equality holds by letting $d_T^2/\ell \to 0$ and $d_T\to \infty$. In addition, using the property of $\ell$ dependent time series, we know that $E|(B^*)^2 -E[(B^*)^2]|=o(1)$ by the development of Theorem 1 of \cite{Hansen1992}. Then we can further obtain that $E^*[(B^* )^2] = \pmb{\ell}^\top (\sigma_\varepsilon^2\pmb{\Sigma}_1+\pmb{\Sigma}_3+\pmb{\Sigma}_3^\top) \pmb{\ell}+o_P(1)$.

We now rewrite $B^*$ as follows.

\begin{eqnarray} \label{EQA.11}
B^* = \sum_{j=1}^K \nu_{j}^*  + \sum_{j=1}^K \varpi_{j}^*,
\end{eqnarray}
where 

\begin{eqnarray*}
\nu_{j}^*  = \sum_{t=B_j+1}^{B_j+r_1} \frac{1}{\sqrt{T}}\varepsilon_t \pmb{\ell}^\top\mathbf{A}_t\eta_{\pmb{\theta},t},\quad \varpi_{j}^*  =  \sum_{t=B_j+r_1+1}^{B_j+r_1+r_2}  \frac{1}{\sqrt{T}}\varepsilon_t \pmb{\ell}^\top\mathbf{A}_t\eta_{\pmb{\theta},t}.
\end{eqnarray*}
 Moreover, $B_j = (j-1)(r_1+r_2)$, and without loss of generality we suppose that $K= T/(r_1+r_2) $ is an integer for simplicity. Otherwise, one needs to include the remaining terms in \eqref{EQA.11} which are negligible for an obvious reason. In addition, we let 

\begin{eqnarray}\label{EQA.12}
(r_1, r_2)\to (\infty,\infty), \quad \left(\frac{r_2}{r_1},\frac{r_1}{T}\right)\to (0,0),\quad r_1\ge \ell ,
\end{eqnarray}
so the blocks $\varpi_{j}^*$'s are mutually independent by the construction of $\xi_t$'s. Note that by $\frac{r_2}{r_1}\to 0$ of \eqref{EQA.12},

\begin{eqnarray*}
\frac{Kr_2}{T}\to 0\quad\text{and}\quad \frac{Kr_1}{T}\to 1.
\end{eqnarray*}

By construction, a direct calculation on the small blocks shows that

\begin{eqnarray*}
EE^* \left[ \left( \sum_{j=1}^K \varpi_{j}^* \right)^2\right] =  \sum_{j=1}^K EE^*[(\varpi_{j}^*)^2]  =O(1) \frac{Kr_2 }{T}=o(1).
\end{eqnarray*}
Therefore, the term $\sum_{j=1}^K \varpi_{j}^*$ of \eqref{EQA.11} is negligible.

\medskip

Next, we employ the Lindeberg CLT to establish the asymptotic normality of $ \sum_{j=1}^K \nu_{j}^*$. Recall that we have shown that $E^*[(B^* )^2] = \pmb{\ell}^\top (\sigma_\varepsilon^2\pmb{\Sigma}_1+\pmb{\Sigma}_3+\pmb{\Sigma}_3^\top) \pmb{\ell}+o_P(1)$ and  $\sum_{j=1}^K \varpi_{j}^*$ of \eqref{EQA.11} is negligible, so it is easy to know that 

\begin{eqnarray*}
E^*( \sum_{j=1}^K \nu_{j}^* )^2  =  \pmb{\ell}^\top (\sigma_\varepsilon^2\pmb{\Sigma}_1+\pmb{\Sigma}_3+\pmb{\Sigma}_3^\top) \pmb{\ell}+o_P(1).
\end{eqnarray*}
Similar arguments can be seen in (A.8)-(A.9) of \cite{CGL2012ET}. That said, we just need to verify that for $\forall \epsilon>0$

\begin{eqnarray}\label{EQA.13}
\sum_{j=1}^K E^* \left[(\nu_{j}^*)^2\cdot I\left(|\nu_{j}^*|>\epsilon\right) \right] =o_P(1).
\end{eqnarray}

Before proceeding further, we point out that the series $\frac{1}{\sqrt{T}}\varepsilon_t \pmb{\ell}^\top\mathbf{A}_t\eta_{t}$ is in fact  a mixingale sequence mentioned in Definition 1 of \cite{hansen_1991}, where the term $| \frac{1}{\sqrt{T}}\varepsilon_t \pmb{\ell}^\top\mathbf{A}_t\eta_{t} |$ is equivalent to $c_i$ in the notation of \cite{hansen_1991}. This is not hard to justify given $\{\xi_t\}$ is an $\ell$-dependent series. When $m$ in the notation of \cite{hansen_1991} is greater than $\ell$, all the requirements of Definition 1 of \cite{hansen_1991} are fulfilled. Thus, it allows us to invoke the asymptotic properties associated to the mixingale sequence in the following development.  

Write

\begin{eqnarray*} 
&&\sum_{j=1}^K \mathbb{E}^* [(\nu_{j}^* )^2\cdot I ( |\nu_{j}^*|>\epsilon ) ] \nonumber \\
&\le & \sum_{j=1}^K\{\mathbb{E}^* | (\nu_{j}^* )^2 |^{\delta/2} \}^{2/\delta}\cdot  \{ \mathbb{E}^* [ I( |\nu_{j}^*|>\epsilon ) ] \}^{(\delta-2)/\delta}\le  \sum_{j=1}^K\{\mathbb{E}^* |(\nu_{j}^* )^2|^{\delta/2} \}^{2/\delta}  \left\{\frac{\mathbb{E}^* |\nu_{j}^*|^{\delta}}{\epsilon^{\delta} }\right\}^{(\delta-2)/\delta}\nonumber \\
&=&\epsilon^{\delta-2} \sum_{j=1}^K \mathbb{E}^* | \nu_{j}^*|^{\delta}  =\epsilon^{\delta-2} \sum_{j=1}^K \left\{ \mathbb{E}^*\left[\left(\sum_{t=B_j+1}^{B_j+r_1} \frac{1}{\sqrt{T}}\varepsilon_t \pmb{\ell}^\top\mathbf{A}_t\eta_{t} \right)^{\delta}\right] \right\}^{\frac{1}{\delta} \cdot \delta}\nonumber \\
&\le &O(1)\epsilon^{\delta-2}\sum_{j=1}^K \left\{   \sum_{t=B_j+1}^{B_j+r_1} \left(  \frac{1}{\sqrt{T}}\varepsilon_t \pmb{\ell}^\top\mathbf{A}_t\eta_{t}  \right)^{2}  \right\}^{\frac{1}{2} \cdot \delta} \nonumber \\
&\le &O(1)\epsilon^{\delta-2}\sum_{j=1}^Kr_1^{\delta/2-1}     \sum_{t=B_j+1}^{B_j+r_1} \left(  \frac{1}{\sqrt{T}}\varepsilon_t \pmb{\ell}^\top\mathbf{A}_t\eta_{t} \right)^{\delta}\nonumber \\
&\le &O(1)\epsilon^{\delta-2}\frac{  r_1^{\delta/2-1} }{T^{\delta/2-1}} \cdot\frac{1}{T}\sum_{t=1}^T   \left( \varepsilon_t \pmb{\ell}^\top\mathbf{A}_t\eta_{t} \right)^{\delta} \nonumber \\
&=&O_P(1)\frac{  r_1^{\delta/2-1} }{T^{\delta/2-1}}=o_P(1),
\end{eqnarray*}
where the first inequality follows from the H\"older inequality, the second inequality follows from the Chebyshev's inequality, the third inequality follows from Lemma 2 of \cite{hansen_1991},  and the last equality follows from $r_1/(Th)\to 0$ and $\delta>2$ (say, letting $\delta=4$). Thus, we can conclude the validity of \eqref{EQA.13}.

Based on the above development, we are readily to conclude that \eqref{EQA.10} holds. The proof is now completed.  \hspace*{\fill}{$\blacksquare$}

\bigskip

\noindent \textbf{Proof of Corollary \ref{cor.nnvec}:}

By Lemma \ref{LM.A2}, we can find a $p_{\vartheta} (\mathbf{x}\, | \, \mathbf{x}_0)$ admitting a form of $p_{\vartheta} (\mathbf{x}\, | \, \mathbf{x}_0) := \pmb{\psi}_{r_\vartheta}(\mathbf{x}\, |\, \mathbf{x}_0)^\top \pmb{\beta}_{\star}$ such that

\begin{eqnarray*}
\sup_{C_{\mathbf{x}_0,h} }| f_\star^{(\pmb{\delta})}(\mathbf{x}) - p_{\vartheta}^{(\pmb{\delta})} (\mathbf{x}\, | \, \mathbf{x}_0) | =O(h^{p-\|\pmb{\delta}\|_1}) ,\nonumber
\end{eqnarray*}
where $\pmb{\beta}_{\star}$ is an $r_\vartheta\times 1$ vector depending on $\mathbf{x}_0$.

To proceed, we note that as $h\to 0$, it is guaranteed that  

\begin{eqnarray*}
\mathbf{x}-\mathbf{x}_0\in [0,h]^r \subseteq [0,1]^r\nonumber
\end{eqnarray*}
in the definition of $C_{\mathbf{x}_0,h}$. Also, for notational simplicity, define

\begin{eqnarray*}
\pmb{\psi}_{r_\vartheta}^{(\pmb{\delta})}(\mathbf{x}) =(\psi_1^{(\pmb{\delta})}(\mathbf{x}),\ldots, \psi_{r_\vartheta}^{(\pmb{\delta})}(\mathbf{x}))^\top.
\end{eqnarray*}
Then we write

\begin{eqnarray*}
&&\sup_{C_{\mathbf{x}_0,h} }| f_\star^{(\pmb{\delta})}(\mathbf{x}) - \mathbf{N}_{\pmb{\delta}}  (\mathbf{x} \, | \, \mathbf{x}_0)^\top  \pmb{\beta}_{\star}  |\nonumber \\
&\le & \sup_{C_{\mathbf{x}_0,h} }| f_\star^{(\pmb{\delta})}(\mathbf{x}) - p_{\vartheta}^{(\pmb{\delta})} (\mathbf{x}\, | \, \mathbf{x}_0)|+\sup_{C_{\mathbf{x}_0,h} }| p_{\vartheta}^{(\pmb{\delta})} (\mathbf{x}\, | \, \mathbf{x}_0)  - \mathbf{N}_{\pmb{\delta}}  (\mathbf{x} \, | \, \mathbf{x}_0)^\top  \pmb{\beta}_{\star}|\nonumber \\
&\le &O(h^{p-\|\pmb{\delta}\|_1})  + \|\pmb{\beta}_\star \|\cdot \sup_{C_{\mathbf{x}_0,h} } \| \pmb{\psi}_{r_\vartheta}^{(\pmb{\delta})}(\mathbf{x}\, |\, \mathbf{x}_0)- \mathbf{N}_{\pmb{\delta}}  (\mathbf{x} \, | \, \mathbf{x}_0)\|  \nonumber \\
&\le &O(h^{p-\|\pmb{\delta}\|_1}) + \|\pmb{\beta}_\star \|\cdot \sqrt{r_{\vartheta}} \cdot  3^{\lceil \log_2 \vartheta\rceil-1} 2^{-m} \nonumber \\
&= &O(h^{p-\|\pmb{\delta}\|_1}+2^{-m}),
\end{eqnarray*}
where the second inequality follows from Lemma \ref{LM.A2}, and the third inequality follows from Lemma \ref{LM.A7} and the definitions of $\pmb{\psi}_{r_\vartheta}^{(\pmb{\delta})}(\mathbf{x}\, |\, \mathbf{x}_0)$ and $\mathbf{N}_{\pmb{\delta}}  (\mathbf{x} \, | \, \mathbf{x}_0) $. 

The proof is now completed. \hspace*{\fill}{$\blacksquare$}

 \bigskip
 
\noindent \textbf{Proof of Lemma \ref{theorem44}:}

(1). By Theorem 9.42 of \citet{rudin2004}, we may take derivative under the integral.  Note that $\phi_s(x)=s\phi(sx)$, and
\begin{eqnarray*}
\sup_u|\sigma_s(u)-\sigma(u)|&=&\sup_u \left|\int \sigma(x)\phi_s(x-u)dx-\sigma(u)\right|\\
&\le&\sup_u\int |\sigma(x+u)-\sigma(u)|\phi_s(x)dx\\
&\le&c\int |x|\phi_s(x)dx =\frac{c}{s}\int |x|\phi(x)dx.
\end{eqnarray*}

\medskip

(2). Note that we can always write

\begin{eqnarray*}
\sigma_s(u)-\sigma(u) &=& \int [\sigma(x+u)-\sigma(u)]\phi_s(x)dx \nonumber \\
&=& \int [\sigma(x+u)-\sigma(u)]s\phi(sx)dx \nonumber \\
&=&\int [\sigma(x/s+u)-\sigma(u)] \phi(x)dx,
\end{eqnarray*}
where the third equality follows from integration by substitution.

In what follows, we consider two cases: (i) $u\ge 0$ and (ii) $u<0$. For case (i), write

\begin{eqnarray*}
\sigma_s(u)-\sigma(u) &=& \int_{-1}^{-su} [0-u] \phi(x)dx+\frac{1}{s}\int_{-su}^{1} x \phi(x)dx\nonumber \\
&=&-u\int_{-1}^{-su}  \phi(x)dx+\frac{1}{s}\int_{-su}^{1} x \phi(x)dx,
\end{eqnarray*}
where the first equality follows from the definition of $\sigma(\cdot)$. Note further that if $ su\ge 1$ (i.e., $u\ge s^{-1}$), we obtain that

\begin{eqnarray*}
\sigma_s(u)-\sigma(u) =0
\end{eqnarray*}
by the definition of $\phi(\cdot)$. Therefore, it remains to consider the case $0\le  su\le 1$ (i.e., $0\le u\le \frac{1}{s}$), then it is obvious that

\begin{eqnarray*}
\sigma_s(u)-\sigma(u) &=&-u\int_{-1}^{-su}  \phi(x)dx+\frac{1}{s}\int_{-su}^{su} x \phi(x)dx+\frac{1}{s}\int_{su}^{1} x \phi(x)dx\nonumber \\
&=&-u\int_{-1}^{-su}  \phi(x)dx +\frac{1}{s}\int_{su}^{1} x \phi(x)dx\nonumber \\
&=&u\int_{1}^{su}  \phi(x)dx +\frac{1}{s}\int_{su}^{1} x \phi(x)dx\nonumber \\
&=& \int_{su}^{1} \frac{x-su}{s} \phi(x)dx,
\end{eqnarray*}
where the second equality follows from $\phi(x)$ being symmetric, and the third equality follows from integration by substitution. As $\phi(x)$ is nonnegative by Assumption \ref{As4}, it immediately yields that

\begin{eqnarray*}
0\le  \sigma_s(u)-\sigma(u) \le  O(1)\frac{1}{s}.
\end{eqnarray*}

For case (ii), it is easy to know that

\begin{eqnarray*}
\sigma_s(u)-\sigma(u)&=&  \frac{1}{s}\int_{-su}^{1} x \phi(x)dx.
\end{eqnarray*} 
For $su\le -1$ (i.e., $u\le -s^{-1}$), we have

\begin{eqnarray*}
\sigma_s(u)-\sigma(u) =0
\end{eqnarray*}
by the definition of $\phi(\cdot)$. Thus, it remains to consider $-1\le su\le 0$ (i.e., $-\frac{1}{s}\le u\le 0$), then it is obvious that

\begin{eqnarray*}
0\le \sigma_s(u)-\sigma(u) \le  O(1)\frac{1}{s}.
\end{eqnarray*}

Collecting the results for both cases (i) and (ii), we conclude that for $|u|\le s^{-1}$

\begin{eqnarray*}
0\le \sigma_s(u)-\sigma(u) \le  O(1)s^{-1},
\end{eqnarray*}
and for $|u|> s^{-1}$

\begin{eqnarray*}
\sigma_s(u)-\sigma(u) =0.
\end{eqnarray*}

The proof is now completed. \hspace*{\fill}{$\blacksquare$}

}

\end{document}